\documentclass[a4paper,fleqn,usenatbib]{mnras}

\usepackage{newtxtext,newtxmath}   

\usepackage[T1]{fontenc}
\usepackage{ae,aecompl}

\usepackage{graphicx}	% Including figure files
\usepackage{amsmath}	% Advanced maths commands
\usepackage{amssymb}	% Extra maths symbols
\usepackage{indentfirst}
\usepackage{mathrsfs}
\usepackage{multicol}        % Multi-column entries in tables
\usepackage{bm}		% Bold maths symbols, including upright Greek
\usepackage{pdflscape}	% Landscape pages
\usepackage{mathrsfs}

\usepackage{color}      % RJL

\title[Discrete chemo-dynamical models of M87's GCs]{A discrete chemo-dynamical model of M87's globular clusters: Kinematics extending to $\sim$ 400 kpc}

\author[Li et al.]{Chao Li,$^{1,5}$
Ling Zhu,$^{3}$\thanks{lzhu@shao.ac.cn}
R. J. Long,$^{2,1,4}$
Shude Mao,$^{2,1}$
Eric W. Peng,$^{6,7}$
Marc Sarzi,$^{8,16}$\and
Glenn van de Ven,$^{9}$
Hongxin Zhang,$^{13}$
Rui Guo,$^{1,5}$
Xiangxiang Xue,$^{1}$
Alessia Longobardi,$^{15}$\and
Patrick C$\mathrm{\hat{o}}$t\'e,$^{10}$
Laura Ferrarese,$^{10}$
Chengze Liu,$^{11, 12}$
Stephen Gwyn,$^{10}$
Sungsoon Lim,$^{7}$\and
Youkyung Ko$^{6,7,14}$
\\
$^{1}$National Astronomical Observatories, Chinese Academy of Sciences, A 20 Datun Road, Chaoyang District, Beijing 100012, China\\
$^{2}$Department of Astronomy and Tsinghua Center for Astrophysics, Tsinghua University, Beijing 100084, China\\
$^{3}$Shanghai Astronomical Observatory, Chinese Academy of Sciences, 80 Nandan Road, Shanghai 200030, China\\
$^{4}$Jodrell Bank Centre for Astrophysics, Department of Physics and Astronomy, The University of Manchester, Oxford Road, Manchester M13 9PL, UK\\
$^{5}$University of Chinese Academy of Sciences, Beijing 100049, China\\
$^{6}$Kavli Institute for Astronomy and Astrophysics, Peking University, Beijing 100871, China\\
$^{7}$Department of Astronomy, Peking University, Beijing 100871, China\\
$^{8}$Armagh Observatory and Planetarium, College Hill, Armagh, BT61 9DG, UK \\
$^{9}$Department of Astrophysics, University of Vienna, T\"urkenschanzstrasse 17, 1180 Vienna, Austria\\
$^{10}$National Research Council of Canada, Herzberg Astronomy and Astrophysics Program, 5071 West Saanich Road, Victoria, BC V9E 2E7, Canada\\
$^{11}$Center for Astronomy and Astrophysics, Department of Physics and Astronomy, Shanghai Jiao Tong University, Shanghai 200240, China\\
$^{12}$Shanghai Key Lab for Particle Physics and Cosmology, Shanghai Jiao Tong University, Shanghai 200240, China \\
$^{13}$CAS Key Laboratory for Research in Galaxies and Cosmology, Department of Astronomy, University of Science and Technology of China, 
Hefei 230026, China\\
$^{14}$Astronomy Program, Department of Physics and Astronomy, Seoul National University, 1 Gwanak-ro, Gwanak-gu, Seoul 08826, Republic of Korea\\
$^{15}$Aix Marseille Universit\'e, CNRS, CNES, Laboratoire d'Astrophysique de Marseille UMR 7326, F-13388, Marseille, France \\
$^{16}$Centre for Astrophysics Research, University of Hertfordshire, Hatfield, Herts AL1 9AB, UK \\
}

\date{Accepted XXX. Received YYY; in original form ZZZ}

\pubyear{2019}

\begin{document}
\label{firstpage}
\pagerange{\pageref{firstpage}--\pageref{lastpage}}
\maketitle

\begin{abstract}
We study the mass distribution and kinematics of the giant elliptical galaxy M87 (NGC 4486) using discrete chemo-dynamical, axisymmetric Jeans equation modelling. Our catalogue comprises 894 globular clusters (GCs) extending to a projected radius of $\sim 430$ kpc with line-of-sight velocities and colours, and  Multi Unit Spectroscopic Explorer (MUSE) integral field unit data within the central $2.4$ kpc of the main galaxy. The gravitational potential for our models is a combination of a luminous matter potential with a varying mass-to-light ratio for the main galaxy, a supermassive black hole and a dark matter (DM) potential with a cusped or cored DM halo. The best-fitting models with either a cusped or a cored DM halo show no significant differences and both are acceptable. We obtain a total mass of $(2.16 \pm 0.38) \times 10^{13} M_{\odot}$ within $\sim$ 400 kpc. By including the stellar mass-to-light ratio gradient, the DM fraction increases from $\sim$ 26 percent (with no gradient) to $\sim$ 73 percent within $1\,R_e^{\rm maj}$ (major axis of half-light isophote, 14.2 kpc), and from $\sim$ 84 percent to $\sim$ 94 percent within $5\,R_e^{\rm maj}$ (71.2 kpc). Red GCs have moderate rotation with $V_{\rm max}/\sigma \sim$ 0.4, and blue GCs have weak rotation with $V_{\rm max}/\sigma \sim$ 0.1. Red GCs have tangential velocity dispersion anisotropy, while blue GCs are consistent with being nearly isotropic. Our results suggest that red GCs are more likely to be born in-situ, while blue GCs are more likely to be accreted.
\end{abstract}


%%%%%%%%%%%%%%%%%%%%%%%%%%%%%%%%%%%%%%%%%%%%%%%%%%

%%%%%%%%%%%%%%%%% BODY OF PAPER %%%%%%%%%%%%%%%%%%

\section{Introduction}

In the outer haloes of galaxies, the observed stellar surface brightness is too low for stellar kinematics to be determined. Bright globular clusters (GCs), considered as discrete point sources, are powerful tracers of dark matter distributions, and fossil records of the halo's formation history.

M87 is a giant elliptical galaxy, located in the dynamical centre of the Virgo cluster. It has the largest GC system in the local supercluster with a total of 15,000 GCs \citep{Peng2008}, and the number of M87 GCs with line-of-sight (LOS) velocities has been growing rapidly in recent years \citep{Strader2011, Durrell2014}. M87's GCs have been used previously to estimate its mass profile (e.g. \citealt{Murphy2011}; \citealt{Gebhardt2009}; \citealt{Strader2011}), but only recently, has its data coverage been significantly increased, reducing the uncertainty of mass profiles extending to $\sim 150$ kpc \citep{Zhu2014}. Recent observations of M87's GC LOS velocities from NGVS have reached the boundary of the galaxy's potential. A population of intra-cluster GCs, which is gravitationally unbound to M87 but bound to the cluster potential, has been identified with a velocity dispersion of $\sim 640\,{\rm km\,s^{-1}}$ \citep{Longobardi2018}. GCs bound to M87 have a velocity dispersion of $\sim 350\,{\rm km\,s^{-1}}$.

M87 has diffuse X-ray emission from hot gas, which has been observed extending to $\sim 1.2$ Mpc \citep{Urban2011}.  A mean temperature of $kT = 2.3$ keV of the hot gas implies $M_{\rm 200} = 1.4\times 10^{14} M_{\odot}$ and $R_{\rm 200} = 1.08$ Mpc (\citealt{Urban2011}; \citealt{Pointecouteau2005}). A detailed mass profile has also been obtained from X-ray emission extending to a few hundred kpc (\citealt{Das2010}; \citealt{Nulsen1995}). The virial mass obtained from X-ray emission is generally consistent with the estimates from the kinematics of dwarf galaxies extending to $\sim 2$ Mpc \citep{McLaughlin1999}. There is evidence that the masses obtained from X-ray gas are larger than those obtained from GC kinematics (e.g., \citealt{Zhu2014}; \citealt{Strader2011}) at $R \gtrsim 100$ kpc, although these previous mass profiles were obtained from GCs not extending beyond $\sim 150$ kpc. Due to the fact that Virgo may not be fully relaxed (\citealt{Urban2011}; \citealt{Binggeli1987}), it is possible that neither X-ray gas, dwarf galaxies, nor GCs are fully relaxed at all regions from the inner radii of M87 to the outermost cluster. X-ray gas and dwarf galaxies may be better tracers of the mass profile of the outer regions of the Virgo cluster, while GCs, which are more spatially concentrated, may be a better mass tracer for M87 within its dynamical boundary.

How the outer haloes of giant elliptical galaxies are assembled is still under debate. Cosmological simulations suggest that massive early type galaxies undergo a two-phase formation process: an early phase of rapid in-situ star formation driven by gas accretion and dissipative mergers (the so called wet phase), followed by a dry merger-dominated phase with markedly smaller star formation rates \citep{Oser2010}. These two phases result in stars with different chemical and kinematic properties, which are usually difficult to determine from observations alone. 

Velocity dispersion anisotropy is a good indicator of underlying orbit distributions, and is widely used to discriminate between stars from the different galaxy formation phases. Galaxy formation simulations show that accreted stars contribute mainly to isotropic to radial anisotropy, while stars formed in-situ could result in tangential anisotropy \citep{Rottgers2014, Wu2014}.

The colour distributions within GC systems are typically bimodal, with GCs with different colours being associated with different spatial and kinematic distributions. GC colour bimodality is thought to be related to the two phase formation of galaxies (\citealt{Cote1998}; \citealt{Brodie2006}). Metal-rich (red) GCs are believed to have been created in the dissipational construction of their parent galaxies, while metal-poor (blue) GCs are considered to have been created in the early universe's low-mass dark matter halos and became part of their current galaxies by accretion.

The relationship, however, between the properties of GCs and their host galaxies is not straightforward to determine: different haloes contain different numbers of GCs, and GCs themselves are potentially subject to disruption by a variety of processes. Considering only GC abundance in different haloes, a purely accreted formation history results in GC systems that are mainly radially biased, with velocity anisotropy that is mildly radial in the centre but is more strongly radial in the outer region \citep{Amorisco2018}. Taking different GC formation schemes and disruption into account, the GC systems become nearly isotropic or mildly tangential in the inner region and more radial in the outer region \citep{Prieto2008}. The efficiency of GC disruption depends on their orbits within the halo. GCs that survive disruption could as a system be much more tangentially anisotropic than the original system, if the disruption occurs as a result of tidal effects \citep{Brockamp2014, Vesperini2003}. To understand the similarities and differences between the GCs and their host galaxies, as before, robust, observationally based estimates of the velocity anisotropy profiles of each population of GCs and central galaxy stars are important.

The velocity anisotropy of GCs in M87 has been provided by several dynamical models. The velocity dispersion anisotropy of GCs as a single population was assessed as near isotropic in \citet{Zhu2014} using a \citet{Syer1996} made-to-measure model as implemented by \citet{Long2010}. From binning the GC data, \citet{Zhang2015} using a two population spherical Jeans model found that the red GC population is highly radially anisotropic while the blue GC population is tangentially anisotropic.
In comparison, \citet{Oldham2018} used a discrete chemo-dynamical model, without binning data and also describing population kinematics by spherical Jeans models. They obtained medium radial anisotropies for both red and blue GCs.
The central galaxy has been observed with integral field units (IFUs) on a number of occasions (for example, Spectrographic Areal Unit for Research on Optical Nebulae \citep[SAURON]{deZeeuw2001} and Multi Unit Spectroscopic Explorer \citep[MUSE]{Emsellem2014}) and stellar kinematic data are readily available.

Our objectives in this investigation are to

1. produce a mass estimate for M87 including both stellar and dark matter,

2. determine the velocity dispersion anisotropy of its two populations of GCs, and

3. consider what the anisotropy may reveal about the formation of M87 and its GC system.

The paper is arranged as follows. In Section 2, we introduce the approach we take in our investigations. In Section 3, we describe the main characteristics of M87 and the observational data we use. In Section 4, we provide details on our modelling methods, and then, in Section 5, we present results from our modelling runs. Finally, we discuss what our investigation has revealed in Section 6, and draw our conclusions in Section 7.

\section{Approach}
\label{sec:approach}

Section 1 was concerned with describing the scientific background and objectives for our investigation. This section defines at a top level the approach we are going to take to meet our objectives.

No new modelling scheme is created for our investigation. The existing axisymmetric modelling scheme utilized with NGC 5846 and its globular clusters \citep{Zhu2016a} is used as the basis of our models. In essence, the scheme is Jeans Anisotropic Modelling (JAM) \citep{Cappellari2008} plus additions for discrete data \citep{Watkins2013}, and attributes such as metallicity \citep{Zhu2016a}. A Markow Chain Monte Carlo (MCMC) toolkit is used for parameter estimation purposes. Note that within this paper, we use colour as a surrogate for metallicity.

The modelling scheme we have chosen \citep{Zhu2016a} is not the only scheme we considered. Other schemes include using higher-order velocity moments \citep{Napolitano2014}, and two population spherical Jeans models \citep{Zhang2015,Pota2015b,Wasserman2018}. These schemes perform a hard cut on colour to separate the two populations of GCs and use different levels of information from their discrete data, and results are not always consistent with each other. The Gauss-Hermite (GH) kurtosis equivalent $h_4$ is taken as a direct indication of underlying velocity anisotropy in spherical systems. Used with our M87 data, it would suffer from poor sampling of data points and the need to bin across a large radius which would potentially affect any results. In the two population spherical Jeans models, the mass-anisotropy degeneracy is supposed to be partially removed.
Without a hard cut on colour, there are a few approaches to chemo-dynamical modelling in the literature which try to model the colour and kinematic distributions of multiple GC populations simultaneously. By adopting Gaussian distributions of colour and velocity for each GC population, \citet{Agnello2014} found evidence of three GC populations (red, yellow, and blue) in M87, rather than two. The resulting velocity dispersion profiles of the red and yellow populations are rather noisy with low number statistics. The data can also be explained by combining the red and yellow GCs into one population with a metallicity gradient from the inner to outer regions. In estimating the mass profile, \citet{Oldham2018} described the kinematics of each population by spherical Jeans models, with both populations sharing the same potential. As indicated in Section 1, the two population spherical models of \citet{Zhang2015} result in a red GC system which is highly radially anisotropic while the blue system has tangential anisotropy. \citet{Oldham2018} found very different results in that both populations have moderate radial anisotropy. Unfortunately in these models, the discrete points are modelled spherically and some of the information on anisotropy is lost. GC systems extending to medium radii with elongated shapes and weak rotation may be better modelled as axisymmetric.

To obtain good constraints from the kinematic data for each population, we consider two populations of GCs. Each model we run is made up of three components in total - one for the central M87 galaxy and one for each of the two colour groups of GCs. Our best-fitting model comes from maximizing a log likelihood function involving contributions from all three component models. In terms of our objectives, once we have a best-fitting model, we can estimate the mass of M87, and construct velocity dispersion anisotropy profiles for the two GC populations.

The role of the GCs is to constrain the mass distribution at large radii. In modelling the central galaxy, we use MUSE IFU data \citep{Emsellem2014} and a radially varying mass-to-light ($M/L$) ratio \citep{Sarzi2018}. This ratio has a negative gradient with increasing radius and is caused by a variation in the stellar initial mass function and the star formation history. The net effect of the varying ratio should be improved mass estimates with a decrease in the stellar matter fraction and an increase in the dark matter fraction, particularly at  radii close to the galactic centre.

We avoid binning GC data during modelling in order to utilize fully the GC information content. This should lead to more robust velocity anisotropy estimates. We do use binning as part of the analysis process after modelling is complete, to assess the reproduction of the GC kinematics.

Our gravitational potential has the usual three components arising from a stellar mass central galaxy, a central black hole (modelled as a point mass), and dark matter. The stellar potential comes from a Multi-Gaussian expansion (MGE) \citep{Cappellari2002, Emsellem1994} of the central galaxy's surface brightness together with the varying $M/L$ ratio. The dark matter potential is a generalised NFW profile and is used in both cusped and cored forms.

The MCMC toolkit we use to execute our models and arrive at our best-fitting model is the EMCEE package \citep{Foreman-Mackey2013}.
In total 15 parameters have to be determined - 3 for the gravitational potential and 2 for central galaxy, plus a further 5 for each GC population. Note that rotation and anisotropy parametrisation are specific to the individual component models and are not handled globally.

The GC density profiles required for modelling with the Jeans equations come from a red/blue colour separation on the full population of GCs, not just those with LOS velocities. After contamination removal, the resulting number profiles are converted to MGEs for modelling purposes. This separation is only an approximation and may require updating after model analysis, with models then being rerun. GCs with LOS velocities are regarded as a single data set with any split if required into red and blue being performed once modelling is complete, using relative probability values. Typically, our calculations for specific GC populations are weighted calculations using these probability values as weights.

Having described the overall approach in this section, in Section 3 we describe our data in more detail, and expand on specific aspects of our modelling scheme in Section 4.

\section{Data}

M87 (also known as Virgo A or NGC 4486) is a giant elliptical galaxy and is the second most massive and brightest galaxy of the Virgo Cluster. The distance to M87 is 16.5 Mpc \citep{Blakeslee2009, Mei2007}, and thus 1 kpc = 12.5 arcsec at this distance. Calculating the major axis of half-light isophote $R_e^{\rm maj}$  \cite[as defined in][]{Cappellari2013} with the stellar image MGEs shown in Table~\ref{tab:mge} gives 178 arcsec. The position angle, measured from the image Y-axis (North) over East to the photometric major axis, is $162^{\circ}$.

Throughout the paper, $(x,y)$ represent the coordinates on the projected plane, $x$ is along the major axis of galaxy, $z$ is along the line-of-sight, $R$ indicates the projected elliptical radius along the $x$ axis ($R= \sqrt{x^2+(y/q)^2}$, we call it elliptical $x$-axis radius for simplicity), $R'$ indicates the signed value ($R' = {\rm sign}(x) \times R$), and $r$ indicates the intrinsic 3D radius ($r = \sqrt{x^2+y^2+z^2}$). $q$ is the flattening of surface brightness on the 2D observational plane.

\subsection{Photometric data}

\begin{figure}
	\includegraphics[width=\columnwidth]{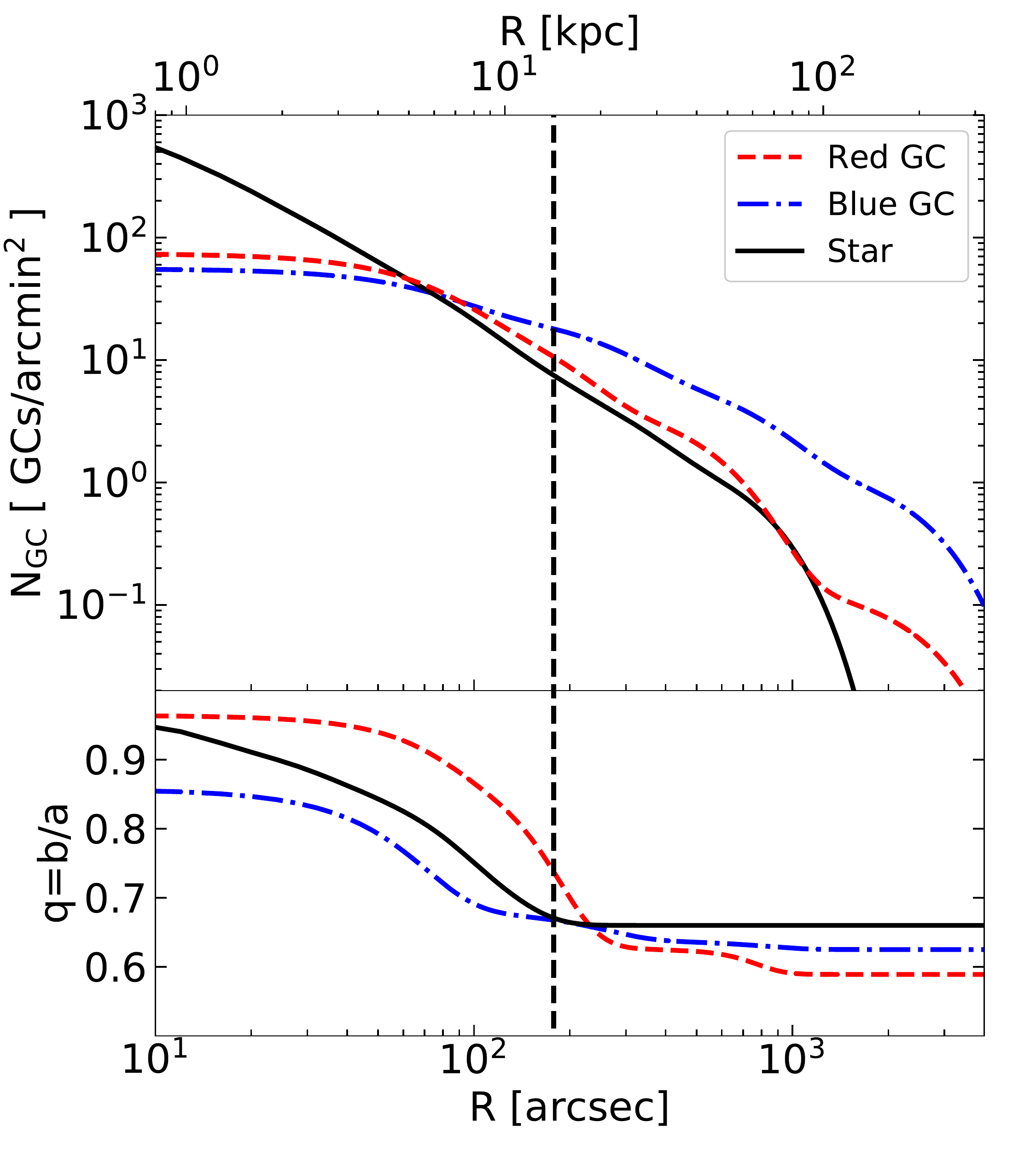}
    \caption{Surface Density profiles of different tracers for M87. \textbf{Top panel: }The GC number density profiles and the stellar surface brightness along the major axis. The dashed red and dash-dotted blue curves represent the number density profiles of the red and blue GCs respectively and they come from the 2D MGE fits to the number density of GCs. The flatness of the GC surface number density profiles in the inner 30 arcsec could partly be caused by catalogue incompleteness and smoothing. The solid black curve is from the MGE fit to the stellar surface brightness of NGVS $g$-band image and it has been arbitrarily scaled for comparison with GCs.  \textbf{Bottom panel: }The flattening profiles. The dashed red and dash-dotted blue curves represent the red and blue GCs respectively and the solid black curve is for the stars. The black vertical dashed line indicates the position of $1\,R_e^{\rm maj}$.}
    \label{fig:sfprofile}
\end{figure}

We take the photometric data from NGVS (The Next Generation Virgo Cluster Survey), including a $g$-band stellar surface brightness image of the main galaxy \citep{Ferrarese2016}, and GC photometric data. The stellar image covers a region of nearly 25 $\times$ 25 arcmin$^2$, which means the region $R$ < 60 kpc for M87. We fit the image with a two-dimensional (2D) MGE and plot the surface brightness profile along the projected major axis in Fig.~\ref{fig:sfprofile} (the solid black curve). For comparison purposes, we scale the surface brightness arbitrarily. We use this MGE to construct the luminous matter gravitational potential arising from the main galaxy.

The GC photometric data cover the $120 \times 120$ arcmin$^2$ region around M87, and contain all the photometrically-selected GC candidates with $g$-band magnitude smaller than 25. 
Following \citet{Durrell2014}, GCs are classified as red if meeting the conditions $0.85<g-i<1.15$, $20<g<24$ and $R<60$ arcmin, and are classified as blue with $0.55<g-i<0.85$, $20<g<24$ and $R<60$ arcmin.
To construct the individual projected surface number density maps for red and blue GCs, we take the following three steps.
(1) we construct a smoothed 2D surface density map, replacing the position of each source by a circular Gaussian with $\sigma = r5$, where $r5$ is the radial distance from each point to its 5th nearest neighbour \citep{Durrell2014}. In the inner region, $r5$ is very small, so we use $\sigma = 30$ arcsec instead.
(2) Background and foreground contaminations are determined by calculating the average surface number density for the fields with $80<R<100$ arcmin and where there is no obvious contamination from other galaxies. We subtract this constant value of contamination from the smoothed 2D map to form a cleaned 2D map. (3) we fit a 2D MGE to the cleaned surface density map. 

The best-fitting MGEs to the cleaned 2D surface density maps of the blue and red GCs are shown in Fig.~\ref{fig:sd2D} in Appendix B. In Fig.~\ref{fig:sd1D}, we show how well the MGE model matches the surface density profile along the major and minor axes. For the blue GCs, the MGE model matches the surface density profile from the centre out to $R\sim 5000$ arcsec (400 kpc), along both the major and minor axes. For the red GCs, there is an asymmetric structure to the north around $R=1000-2000$ arcsec in the 2D surface density map. Our MGE model does not match this  asymmetric structure, but does match the south side data. Thus the MGE constructed surface density for the red GCs may have some uncertainties at $R>1000$ arcsec (80 kpc). In the kinemetic data, the red GCs with velocity measurements are mostly located within 1000 arcsec. Note that the surface density profile beyond the coverage of the kinematic data points does not have any effect on our dynamical model. 
The number density profiles of red and blue GCs along the major axis and the flattening profiles from the MGE models are shown in Fig.~\ref{fig:sfprofile} (the dashed red and dash-dotted blue curves). These two MGEs are used as density functions in our Jeans modelling.

The parameters of the MGEs for the $g$-band stellar image, red GC number density and blue GC number density are shown in Table~\ref{tab:mge}. The MGE of the stellar image is composed of 9 Gaussians and the MGEs of both the red and blue GCs are composed of 4 Gaussians. For each Gaussian of the MGEs, $\Sigma$ is the peak surface brightness or surface number density, $\sigma$ is the dispersion and $q$ is the flattening. 

\begin{table*}
 \caption{The parameters of the MGEs for the $g$-band stellar image, red GC surface number density and blue GC surface number density. $\Sigma$ is the peak surface brightness or surface number density, $\sigma$ is the dispersion and $q$ is the flattening of each Gaussian.}
 \label{tab:mge}
 \renewcommand\arraystretch{1.25}
 \begin{tabular*}{16cm}{@{\extracolsep{\fill}}|c|ccc|ccc|ccc|}

   \hline
    && $g$-band stellar image &&& red GCs &&& blue GCs&\\
    & $\Sigma$ [${\rm L_{\odot}\,pc^{-2}}$] & $\sigma$ [arcsec] & $q$ & $\Sigma$ [${\rm N\,arcmin^{-2}}$] & $\sigma$ [arcsec]& $q$ & $\Sigma$ [${\rm N\,arcmin^{-2}}$] & $\sigma$ [arcsec] & $q$ \\
   \hline
    1 & 1.4597e+04 & 0.54 & 0.972 & 48.6816 & 51.2923 & 0.999 & 30.7399 & 53.4130 & 0.999\\
    2 & 1.6965e+03 & 4.89 & 0.945 & 20.6292 & 115.4347 & 0.959 & 16.7829 & 179.9053 & 0.702\\
    3 & 1.1315e+03 & 9.57 & 0.999 & 4.5376 & 382.9437 & 0.628 & 6.5417 & 512.8271 & 0.641\\
    4 & 4.8415e+02 & 17.77 & 0.945 & 0.1514 & 1729.8027 & 0.589 & 1.4576 & 1721.5215 & 0.625\\
    5 & 1.8451e+02 & 28.22 & 0.944 &&&&&&\\
    6 & 1.3272e+02 & 51.23 & 0.895 &&&&&&\\
    7 & 4.2822e+01 & 87.47 & 0.863 &&&&&&\\
    8 & 1.7848e+01 & 186.75 & 0.660 &&&&&&\\
    9 & 5.7816e+00 & 515.90 & 0.660 &&&&&&\\
   \hline
 \end{tabular*}
\end{table*}

Comparing the density profiles in the top panel of Fig.~\ref{fig:sfprofile}, the stellar surface brightness is steeper than the GC surface number density profiles in the inner $\sim$ 50 arcsec. The flatness of the GC surface number density profiles in the inner 30 arcsec could partly be caused by catalogue incompleteness and smoothing. At larger radii, the slope of the red curve follows the stellar light while the slope of the blue curve is shallower.

We also show the projected flattening profiles along the major axis in the bottom panel of Fig.~\ref{fig:sfprofile}. The solid black curve represents the surface brightness flattening profile, which varies from $q \sim 0.96$ to $q \sim 0.65$ from the centre to larger radii. The dashed red and dash-dotted blue curves represent red and blue GCs. The dashed red curve decreases from $q \sim 0.95$ to $q \sim 0.59$, while the blue from $q \sim 0.85$ to $q \sim 0.63$ from the inner $\sim 100$ arcsec to larger radii, after which they remain constant.

\subsection{Kinematic data}

We have two sets of kinematic data, MUSE IFU data for the central galaxy and LOS velocity data for the GCs. We describe the MUSE data first and then the GC data.

The MUSE data from \citet{Emsellem2014} cover the central region of nearly 1 $\times$ 1 arcmin$^2$ of M87 and comprise 523 Voronoi data bins. These bins correspond to stellar-kinematical measurements constructed from spectra binned to reach a minimum $S/N=300$. In our models, we use the MUSE LOS velocity and velocity dispersion values together with their error estimates.

We have 911 M87 GCs with LOS velocities from three different sources, a systematic spectroscopic survey using 2dF/AAOmega multi-fiber spectrograph on the 3.9m Anglo-Australian Telescope, an extensive spectroscopic survey using Hectospec multi-fibre spectrograph on the 6.5m Multiple Mirror Telescope \citep{Zhang2015}, and the radial velocity catalogues of Virgo GCs compiled by \citet{Strader2011}. Our GC catalogue is exactly the same as \citet{Zhang2015} but is slightly different from \citet{Zhu2014} (the catalogues are from the same source but there are a few more data points in the Zhu catalogue). We intend to publish this data set with the final catalogue paper in the near future (Youkyung Ko+ 2019 in preparation).

Although most GCs are located within 500 kpc, a small number of GCs are positioned as far out as 1000 kpc, and may be the intra-cluster GCs \citep{Longobardi2018}. We expect to be able to discover the intra-cluster GCs hidden in our catalogue and remove them. To do so, we adopt the friends-of-friends algorithm \citep{Huchra1982, Geller1983, Eke2004} and a dimensionless distance between two GCs as the linking length defined by
\begin{equation}
    l^2 = w_{\phi}\phi_{ij}^2 + w_{v}(v_i-v_j)^2 + w_{d}(d_i - d_j)^2~,
	\label{eq:4d}
\end{equation}
where $\phi_{ij} = {\rm atan2}(x_j-x_i,y_j-y_i)$ is the angular separation of two GCs relative to the centre of galaxy, $v$ is the LOS velocity and $d = \sqrt{x^2+y^2/q^2}$ is the elliptical $x$-axis radius. $(w_{\phi}, w_{v}, w_{d})$ are three weighting factors used to normalize the quantities (the largest possible angular separation is $\pi$ , distance 1064 kpc and velocity $3150\,{\rm km\,s^{-1}}$) and take into account uncertainties. The definition of the linking length $l$ in this paper is similar to the 4-distance in 
\citet{Starkenburg2009}, but we only possess information on three of the six components in phase space: the azimuth angle $\phi$ ($\phi = {\rm atan2}(y/x)$), the elliptical $x$-axis radius $d$ and the LOS velocity $v$.

\begin{equation}
    \begin{aligned}
    w_{\phi} &= \left(\frac{1}{\pi} \right)^2~,\\
    w_d &= \left(\frac{1}{1064}\right)^2 \frac{\left(\frac{d_{\rm err}(i)}{d(i)}\right)^2+\left(\frac{d_{\rm err}(j)}{d(j)}\right)^2}{2\left \langle \frac{d_{\rm err}}{d} \right \rangle^2}~,\\
    w_v &= \left(\frac{1}{3150}\right)^2 \frac{v_{\rm err}^{2}(i)+v_{\rm err}^{2}(j)}{2\left \langle v_{\rm err} \right \rangle^2}~.
	\label{eq:weight}
    \end{aligned}
\end{equation}
where $d_{\rm err}$ and $v_{\rm err}$ are the uncertainties on the elliptical $x$-axis radius and LOS velocity respectively, and the relative uncertainty $d_{\rm err}/d$ is used as distance uncertainties scale with the distance itself. Quantities within $\left \langle \right \rangle$ denote the average uncertainties over the whole sample.

\begin{figure}
	\includegraphics[width=\columnwidth]{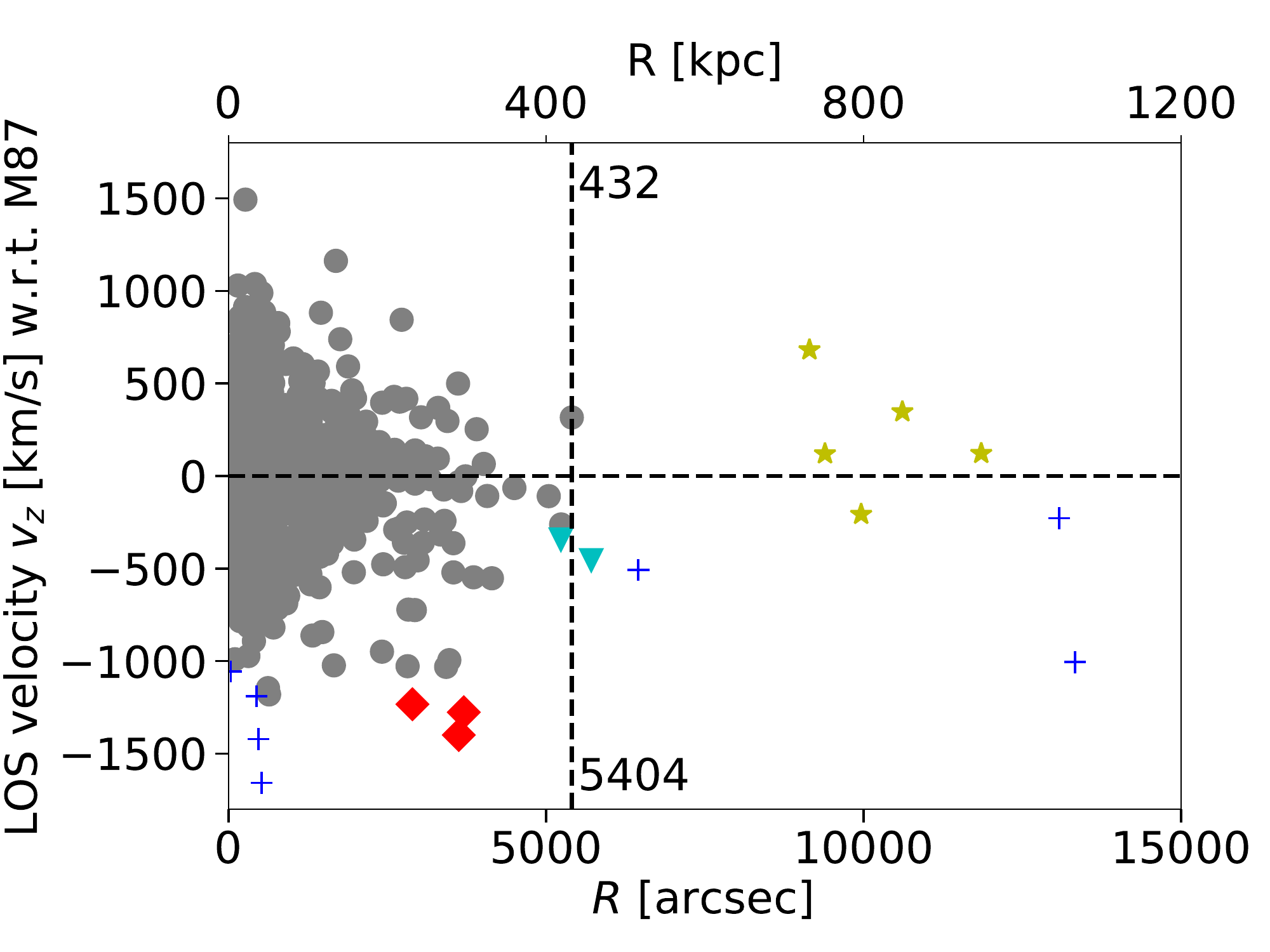}
    \caption{Groups in the GCs. There is a big group marked with grey points and several other small groups are marked with red diamonds, yellow asterisks and cyan triangles. The GCs which are isolated and have no neighborhood are marked with blue pluses. The vertical dashed line marks the maximum radius of the grey group.}
    \label{fig:groups}
\end{figure}

We set the linking length of the friends-of-friends algorithm as $l = 0.14$ and plot the groups in Fig.~\ref{fig:groups}. There is a big group marked with grey points and several other small groups are marked with red diamonds, yellow asterisks and cyan triangles. The GCs which are isolated and have no neighbours are marked with blue pluses. Most of the GCs are in the grey group. We treat GCs belonging to the non-grey groups as the intra-cluster GCs (17 in total) and remove them, leaving 894 GCs in our catalogue. These GCs are considered bound by the potential of M87 and trace the dynamical history of the galaxy. 
The 17 GCs excluded have a velocity dispersion of $620\,{\rm km\,s^{-1}}$, which is consistent with the intra-cluster GCs identified in \citet{Longobardi2018}. The remaining 894 GCs still show some asymmetry in velocity distribution in the region with $R>2000$ arcsec. Having checked the points with $R>2000$ arcsec and $V_z < -700\,{\rm km\,s^{-1}}$, we find that their photometric properties ($u-i$ vs. $i-K_s$ diagram) are consistent with GCs, and are unlikely to be Milky Way stars \citep{Munoz2014}. We presume this asymmetry is caused by low number statistics and we keep all the grey points in Fig~.\ref{fig:groups} as model constraints.

Regarding the non-symmetric structure in the north according to the surface number density maps, we do not mask out any red GCs that might be associated with the structure. At this radius, the number of red GCs is small compared to the number of blue GCs, and any impact on the modelling will be small.

\begin{figure*}
	\includegraphics[width=2\columnwidth]{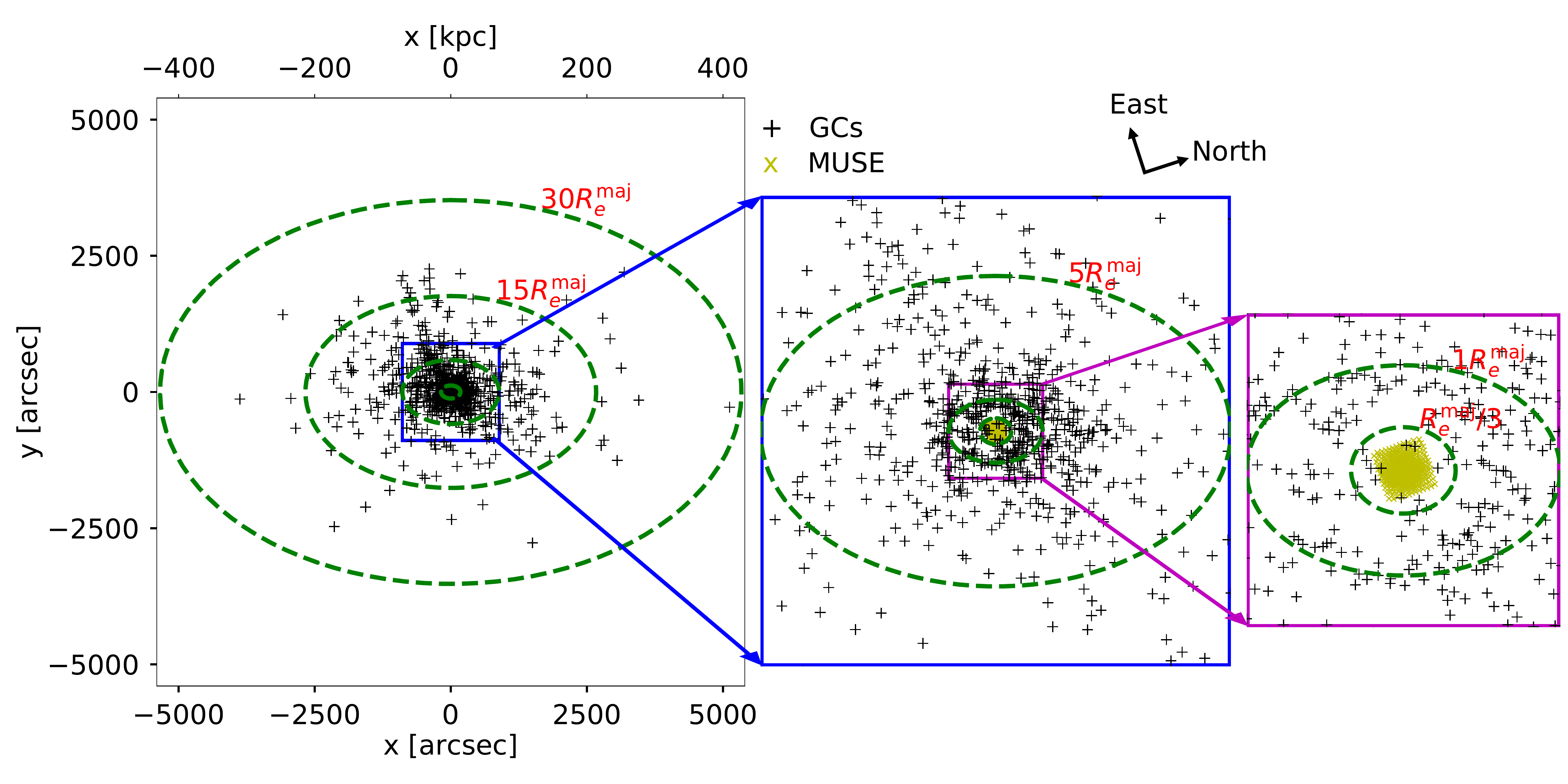}
    \caption{The kinematic data in the observational plane. $x$ and $y$ indicate the major and minor axes of the galaxy and the figure orientation is shown in the top-right corner of the middle panel. The black plus symbols represent the positions of GCs and the yellow multiplication signs represent the integrated stellar light from MUSE. The five green ellipses have $x$-axis radii of $(1/3,1,5,15,30)\,R_e^{\rm maj}$. The middle panel is a zoom-in of the central $5\,R_e^{\rm maj}$ region of the left panel and the right panel is a zoom-in of the central $1\,R_e^{\rm maj}$ region of the middle panel.}
    \label{fig:datamap}
\end{figure*}

The kinematic data in the projected sky plane are shown in Fig.~\ref{fig:datamap}. Each black plus symbol represents one GC, while the yellow square in the centre indicates the area covered by MUSE IFU data.
The five dashed green ellipses have $x$-axis radii of $(1/3,1,5,15,30)\,R_e^{\rm maj}$.

\section{Method}

Much of the material in this section just describes an application of the method
in \citet{Zhu2016a} and so we present it in summarised form.

\subsection{Gravitational potential}

The total potential is a combination of the contributions from stars in the central galaxy, the central supermassive black hole and the dark matter halo. 

\subsubsection{Central galaxy}

\begin{figure}
	\includegraphics[width=\columnwidth]{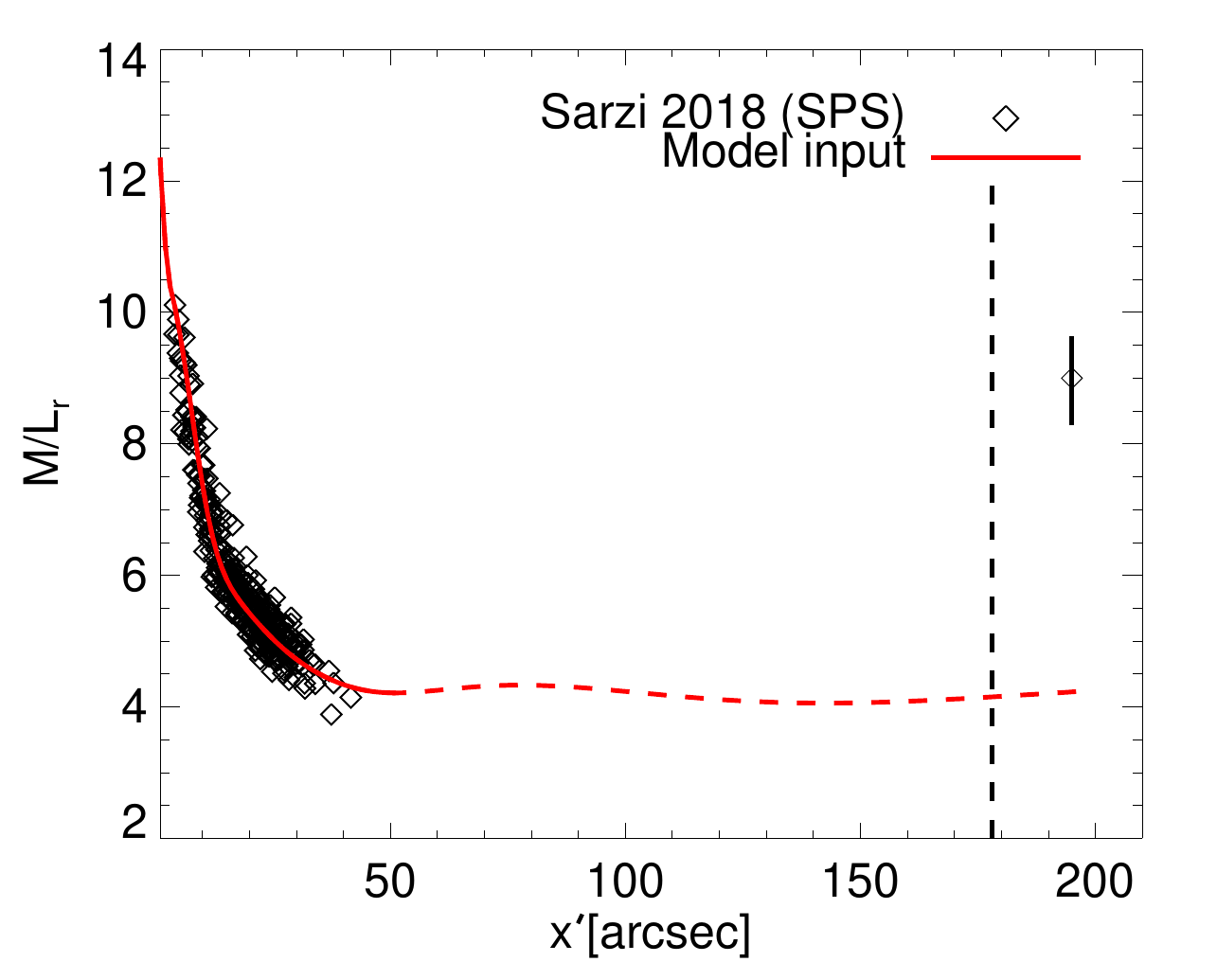}
    \caption{Stellar $M/L_r$ ratio gradient of M87. $x$ is along the major axis. The black diamonds are the $M/L_r$ ratio gradient from Sarzi et al. (2018). The diamond with error bar in the middle right presents the mean of errors. The solid red curve is our best-fitting $M/L_r$ ratio gradient with Eq.~\ref{eq:mlgrad}. The dashed red curve is the extension of our best-fitting $M/L_r$ ratio to large radii. The vertical dashed line indicates a radius of $1\,R_e^{\rm maj}$.}
    \label{fig:mlfit}
\end{figure}

\begin{figure}
	\includegraphics[width=\columnwidth]{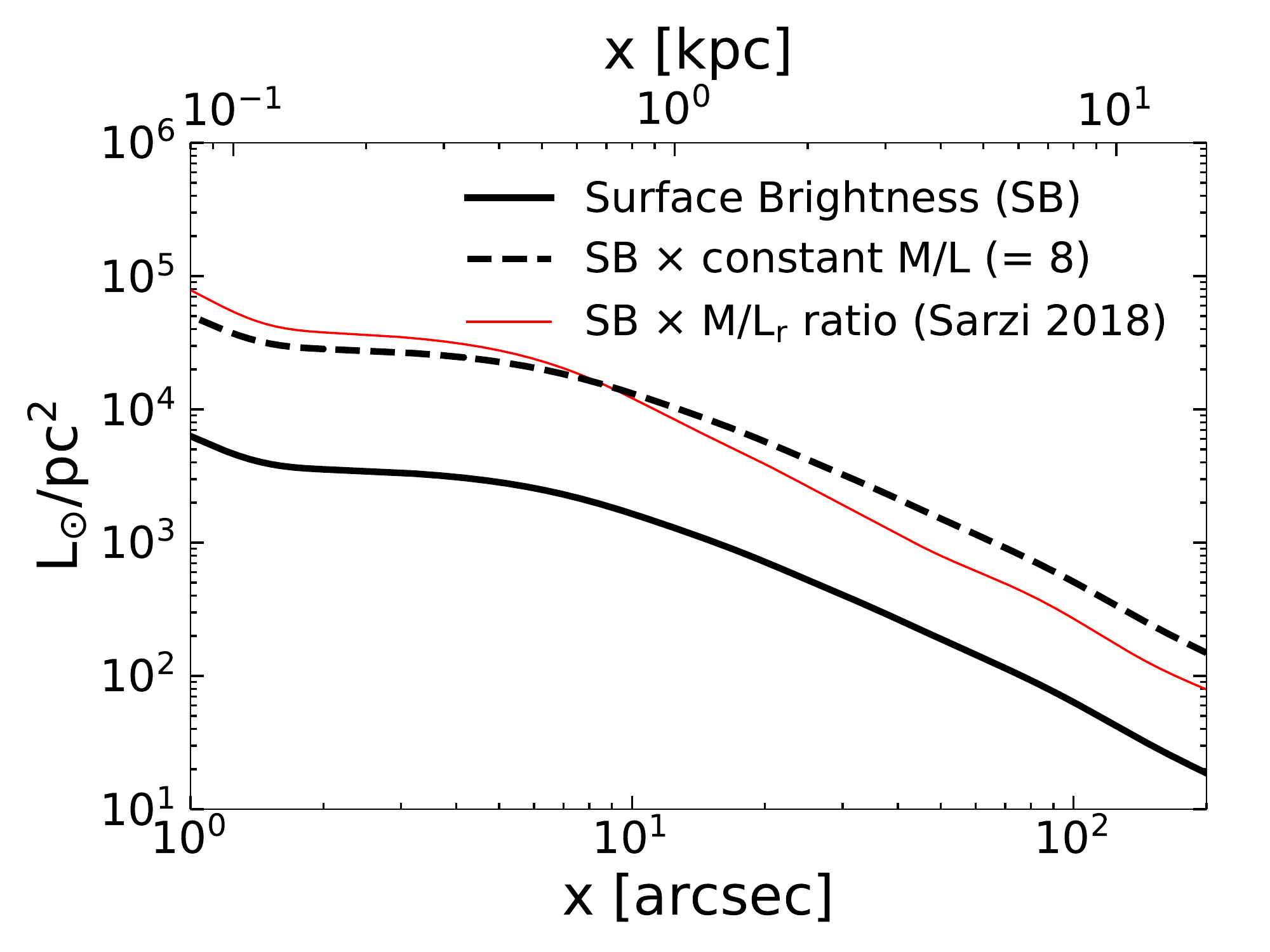}
    \caption{Stellar surface brightness profiles of M87. $x$ is along the major axis. The solid black, dashed black and thin red curves represent the stellar surface brightness profile, the stellar surface mass density profile using a constant $M/L$ ratio = 8, and the surface mass density profile using the $M/L_r$ profile from Fig.~\ref{fig:mlfit}.}
    \label{fig:ml_sbp}
\end{figure}

To create the potential from the stars in the central galaxy, we construct a 2D MGE from the galaxy's NGVS $g$-band surface brightness and apply the MGE formalism to the 2D MGE to arrive at the potential. In so doing we need to accommodate the varying $M/L$ ratio identified in Section 2 and the inclination of the galaxy to the line-of-sight. We proceed as follows.

Recent studies show that the stellar $M/L$ ratio for M87 has a gradient from the centre outwards \citep{Sarzi2018, Oldham2018}. The black diamonds in Fig.~\ref{fig:mlfit} show the $M/L$ ratio gradient from stellar population synthesis \citep{Sarzi2018}. Using MUSE observations, \citet{Sarzi2018} found a strong negative stellar $M/L$ ratio gradient in M87. To impose the stellar $M/L$ ratio variation, we assign each Gaussian component of the 2D MGE its own $M/L$ ratio $\gamma_i$ and define
\begin{equation}
    \gamma^{*}(x)=\frac{\sum_{i}^{N}\gamma_{i}l_{i}{\rm exp}\left(-\frac{x^2}{2\sigma_i^2}\right)}{\sum_{i}^{N}l_{i}{\rm exp}\left(-\frac{x^2}{2\sigma_i^2}\right)}~, 
	\label{eq:mlgrad}
\end{equation}
where $x$ is along the major axis, $N$ is the number of the MGE components, having peak surface brightness $l_i$ and dispersion $\sigma_i$ along the the major axis. We perform a minimum $\chi^2$ fit to the $M/L$ ratio gradient points constructed from Eq.~\ref{eq:mlgrad}, and obtain a solution for the $\gamma_i$. As shown in Fig.~\ref{fig:mlfit}, the resulting best-fitting $\gamma^{*}(x)$ profile (solid red curve) matches the data points well. The data points from \citet{Sarzi2018} only extend from $\sim$ 4 arcsec to $\sim$ 40 arcsec. The interpolation inward is consistent with the $M/L$ profile used in \citet{Oldham2018}. There is evidence that M87 has constant age and metallicity outside $\sim$ 60 arcsec \citep{Montes2014}, and so we extrapolate the $M/L$ ratio gradient from \citet{Sarzi2018} to large radii by assuming there is no strong variation in the outer region (dashed red curve).

Fig.~\ref{fig:ml_sbp} shows the comparison between the stellar surface brightness profile and $M/L$ ratio gradient revised surface mass density profile along the major axis. The solid black curve is the stellar surface brightness profile. The dashed black curve is the stellar surface mass density profile using constant $M/L$ = 8, and the solid red thin curve is the surface mass density profile constructed with the $M/L$ profile as shown in Fig.~\ref{fig:mlfit}.

Note that the $M/L$ radial profile from \citet{Sarzi2018} is in the SDSS $r$-band whereas the projected stellar surface brightness used in our models is from the NGVS $g$-band photometric image, thus we have used $L_g * M/L_r$ to impose the $M/L$ gradient in constructing the stellar mass model. M87 has a rather constant ratio of $L_r/L_g$ with increasing radius \citep{Montes2014}, thus the above assumption works for including the gradient itself, but has a mismatch in the absolute normalization of stellar mass.
Kinematics can be independent, and potentially stronger, constraints on the scale of stellar $M/L$ ratio \citep{Lyubenova2016}. Although \citet{Sarzi2018} has tried to pin down the effect of a varying IMF, we still allow a constant scale factor $\Upsilon$ as a free parameter in our models.

In order to deproject the 2D mass density to give a 3D density and then the potential, we need to know the inclination of M87 to the line-of-sight. Inclination angle is poorly constrained by kinematic data for a slow-rotating galaxy \citep{Remco2009}. 
We tried an initial model which showed that an inclination angle of $i = 90^{\circ}$ is preferred but it has large uncertainties (shown in Appendix C). Taking this into account, we choose to fix the inclination angle $i = 90^{\circ}$ (edge-on) for the rest of the models.

\subsubsection{Black hole}

M87 is believed to have a supermassive black hole. We adopt a black hole with a mass of $(6.6 \pm 0.4) \times 10^9 M_{\odot}$ \citep{Gebhardt2011} and model it as a point mass. The sphere of influence radius of the black hole is about 4 arcsec at the M87 distance of 16.5 Mpc. Note that stars anywhere in the galaxy, depending on their orbits, may have been affected by the black hole.

\subsubsection{Dark matter halo}

We adopt a generalized \citet*{Navarro1997} (NFW) dark matter (DM) density distribution 
\begin{equation}
    \rho(r) = \frac{\rho_s}{(r/r_s)^\gamma(1+r/r_s)^{3-\gamma}}~,
    \label{eq:nfw}
\end{equation}
with $r^2 = x^2 + y^2 + z^2/p_z^2$ for an axisymmetric model. The scale density $\rho_s$, the scale radius $r_s$ and the density slope $\gamma$ are the three DM halo parameters. Cosmological simulations show that the shape of a dark matter halo may be correlated with, but is usually rounder than, the luminous tracers \citep{Wu2014}. In the case of M87, it is unlikely to have a very flattened dark matter halo. The flattening $p_z$ of a dark matter halo is highly degenerate with its radial profile, and it is hard for it to be constrained by the kinematics of an external galaxy. To simplify matters, we just adopt a spherical dark matter halo ($p_z = 1$). We construct two models with different radial profiles in this paper, one with a classic cusped dark matter density profile ($\gamma = 1$), and the other with a cored dark matter density profile ($\gamma = 0$). In each case, $\rho_s$ and $r_s$ are the two free parameters for the DM halo.

\subsection{The likelihood function}

As indicated earlier we follow the approach in \citet{Zhu2016a}. We include, for completeness, a short description of the likelihood calculations.

\subsubsection{Discrete GCs}
The likelihood function for the GCs can be constructed by considering three probabilities involving chemistry, spatial position and kinematics. We describe these three probabilities individually and then combine them together to form the GCs' likelihood function.

We have a catalogue of $N_{\rm GC}$ discrete GCs where the $i^{\rm th}$ GC has sky coordinates $(x_i, y_i)$, LOS velocity $v_{z,i} \pm \delta v_{z,i}$ and measured metallicity $Z_i \pm \delta Z_i$. As stated earlier we adopt the colour $g-i$ as a proxy for metallicity. For this paper, we consider two GC populations, red and blue, corresponding to the metal rich and metal poor.

For each population $k$ ($k$ = red or blue), we adopt a Gaussian metallicity distribution with mean metallicity $Z_0^k$ and dispersion $\sigma_Z^k$ as free parameters. The metallicity probability of the $i^{\rm th}$ GC in population $k$ is 

\begin{equation}
    P_{{\rm chm},i}^k = \frac{1}{\sqrt{2\pi[(\sigma_Z^k)^2 + \delta Z_i^2]}}{\rm exp}\left[-\frac{1}{2}\frac{(Z_i-Z_0^k)^2}{(\sigma_Z^k)^2+\delta Z_i^2}\right]~.
    \label{eq:chm}
\end{equation}

Taking the observed projected number density for population $k$ as $\Sigma^k(x, y)$, the spatial probability of the $i^{\rm th}$ GC in population $k$ is 

\begin{equation}
    P_{{\rm spa},i}^k = \frac{\Sigma^k(x_i,y_i)}{\Sigma_{\rm obj}(x_i,y_i)}~,
    \label{eq:spa}
\end{equation}
where $\Sigma_{\rm obj} = \Sigma^{\rm red} + \Sigma^{\rm blue}$ is the combined density in our two population model.

Assuming a Gaussian velocity distribution with mean value and dispersion predicted by an axisymmetric JAM model for population $k$, the kinematic probability of the $i^{\rm th}$ GC in population $k$ is

\begin{equation}
    P_{{\rm dyn},i}^k = \frac{1}{\sqrt{[(\sigma_i^k)^2 + (\delta v_{z,i})^2]}}{\rm exp}\left[-\frac{1}{2}\frac{(v_{z,i}-\mu_i^k)^2}{(\sigma_i^k)^2+(\delta v_{z,i})^2}\right]~,
    \label{eq:dyn}
\end{equation}
where $\mu_i^k$ and $\sigma_i^k$ are the LOS mean velocity and velocity dispersion predicted by  the model at the position $(x_i, y_i)$. Axisymmetric JAM models assume: (i) the velocity ellipsoid is aligned with the cylindrical coordinates; and (ii) the velocity anisotropy in the meridional plane $\beta_z^k = 1 - \overline{v_z^2}/\overline{v_R^2}$ is constant \citep{Cappellari2008}.

In JAM modelling, the number density of population $k$ is expressed in terms of an MGE. In principle, for each Gaussian $j$ in the MGE (see Table~\ref{tab:mge}), the anisotropy parameters $\beta_{z,j}^k$ and rotation parameters $\kappa_j^k$ can take different values, and thus it is possible to model the velocity anisotropy in the meridional plane and intrinsic rotation at different radii. To investigate the rotation properties at inner and outer radii, we assume two different constant values for the first three and the last Gaussians in the MGE, and so we include two free parameters $\kappa_{\rm inner}^k$ and $\kappa_{\rm outer}^k$ for population $k$. Under the assumptions of axisymmetric JAM modelling, the velocity anisotropy in the meridional plane $\beta_z^k$ is constant and we include $\beta_z^k$ of population $k$ as another free parameter so we can investigate anisotropy.

Having defined the three probabilities individually, we now combine them to give a likelihood value. Given the chemical, spatial and kinematical probabilities, the likelihood for the $i^{\rm th}$ GC in the model is 
\begin{equation}
    L_i = \sum_k P_i^k = \sum_k P_{\rm chm,i}^k P_{\rm spa,i}^k P_{\rm dyn,i}^{k}~,
    \label{eq:gc}
\end{equation}
where $k$ = red or blue. The summation is over the red and blue populations in our model. $P_i^{'k} = P_i^k/L_i$ is the relative probability of the $i^{\rm th}$ GC belonging to population $k$.

The total log likelihood of all the GCs is 
\begin{equation}
    \mathcal{L} = \sum_i^{N_{\rm GC}} \log L_{i}~.
    \label{eq:tot}
\end{equation} 

\subsubsection{Stellar kinematics}

The MUSE IFU data provide the LOS mean velocities and velocity dispersions $(\mu_j^s \pm \delta \mu_j^s, \sigma_j^s \pm \delta \sigma_j^s)$ of stars in Voronoi bins with centroids $(x_j, y_j)$. We use another axisymmetric JAM model with free velocity anisotropy and rotation parameters to describe the stellar kinematics. We calculate $\chi_{\rm star}^2$ between the observed stellar mean velocities and velocity dispersions and the model predictions as
\begin{equation}
    \chi_{\rm star}^2 = \sum_{j=1}^{N_s}\left(\frac{\mu_j^s - \mu_j^{\rm star}}{\delta \mu_j^s}\right)^2 + \sum_{j=1}^{N_s}\left(\frac{\sigma_j^s - \sigma_j^{\rm star}}{\delta \sigma_j^s}\right)^2~,
    \label{eq:chi}
\end{equation} 
where $N_s$ is the number of Voronoi bins and $(\mu_j^{\rm star}, \sigma_j^{\rm star})$ are the model predicted mean velocities and velocity dispersions. 

\subsubsection{Total likelihood}
 
We obtain the best-fitting parameters by maximizing the log likelihood of the GCs and minimizing $\chi_{\rm star}^2$. This can be achieved by maximizing the combined likelihood as defined by

\begin{equation}
    \mathcal{L}_{\rm tot} = \mathcal{L} - \frac{1}{2}\alpha_s\chi_{\rm star}^2~,
    \label{eq:combined}
\end{equation} 
where $\alpha_s$ is a weight parameter which is used to balance the relative importance of the GCs and the stellar kinematics and also to balance the two terms numerically. The value of $\alpha_s$ is determined by experimentation. We set $\alpha_s = 0.004$ in our model. With this choice of $\alpha_s$, the kinematics of GCs and stars are both matched well by our model.

\subsection{Model parameters}

We summarize the 15 free parameters in our model. Three of them are related to the gravitational potential as described in Section 4.1:
\par
(1) $\rho_s$, DM scale density;
\par
(2) $d_s \equiv \log(\rho_s^2r_s^3)$, instead of DM scale radius $r_s$ for the purpose of reducing the strong degeneracy between $\rho_s$ and $r_s$;
\par
(3) $\Upsilon$, a constant scale factor of stellar mass.
\par
For the stars in the central galaxy, the projected stellar density profile is set by the surface brightness profile as shown in Fig.~\ref{fig:sfprofile} and the dynamical distribution is described by an axisymmetric JAM model, leaving two dynamical free parameters for the stars:
\par
(4) $\lambda^{\rm star} \equiv - {\rm ln}(1 - \beta_z^{\rm star})$, a symmetric re-casting of the constant velocity anisotropy parameter of the stars;
\par
(5) $\kappa^{\rm star}$, the rotation parameter of the stars.
\par
We consider two GC populations, red and blue, each with a Gaussian metallicity (colour) distribution, a fixed spatial density profile as shown in Fig.~\ref{fig:sfprofile} and a dynamical distribution described by an axisymmetric JAM model, leaving two chemical and three dynamical free parameters for each population, which have been introduced in detail in Section 4.2. So for each population, there are five free parameters:
\par
(6) $Z_0^{\rm k}$, mean of the Gaussian metallicity (colour) distribution for population k (red or blue);
\par
(7) $\sigma_Z^{\rm k}$, sigma of the Gaussian metallicity (colour) distribution;
\par
(8) $\lambda^{\rm k} \equiv - {\rm ln}(1 - \beta_z^{\rm k})$, a symmetric re-casting of the constant velocity anisotropy in the meridional plane;
\par
(9) $\kappa_{\rm inner}^{\rm k}$, the inner rotation parameter;
\par
(10) $\kappa_{\rm outer}^{\rm k}$, the outer rotation parameter.

\section{Results}

Both the cusped and cored models described in Section 4.1.3 are acceptable. For simplicity, we show the results from the cored model in this section and the corresponding results from the cusped model in the appendix A. The results from the cusped model hardly change relative to the cored model.

\subsection{Best-fitting parameters}

\begin{figure*}
	\includegraphics[width=2\columnwidth]{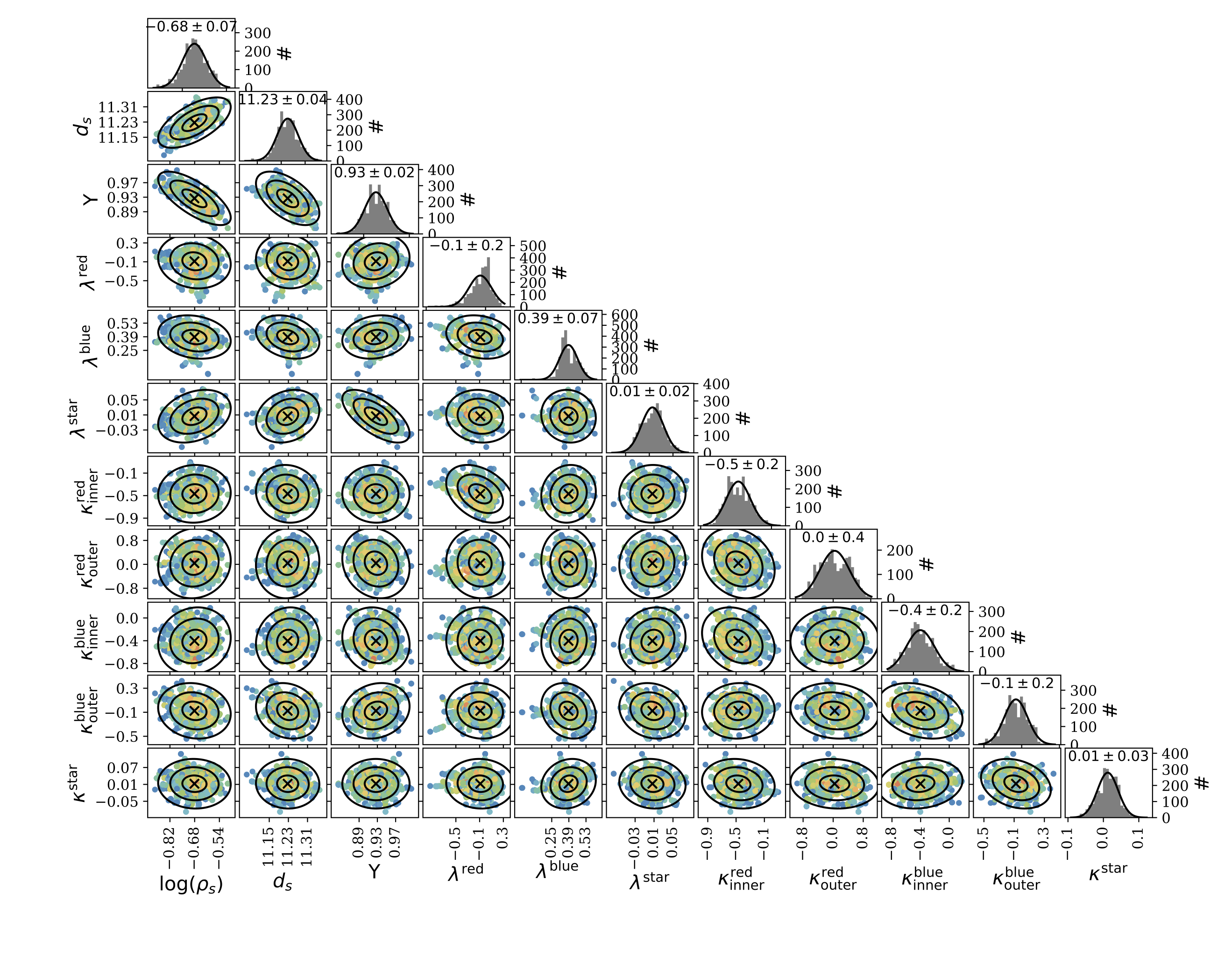}
    \caption{The posterior marginalized distributions for the model with a cored DM halo. The points in each scatter plot show the 2D posterior marginalized distribution of the two corresponding parameters and the colour of each point indicates the relative probabilities from the low (blue) to high (red). The three ellipses in each scatter plot show the 1, 2 and $3\,\sigma$ regions of the covariance ellipses of the distribution. The histogram in each row shows the 1D posterior marginalized distribution of the parameter and the curve is a Gaussian curve plotted with the mean and standard deviation of the distribution; the values are shown in the top. The parameters we show here, from left to right, are logarithm of the DM scale density $\log \rho_s$; $d_s={\rm log}(\rho_s^2r_s^3)$ where $r_s$ is the DM scale radius; the stellar $M/L$ ratio scale factor $\Upsilon$; the velocity anisotropy parameters for the red GCs $\lambda^{\rm red}$, blue GCs $\lambda^{\rm blue}$ and the central stellar light $\lambda^{\rm star}$; the inner and outer region rotation parameters for the red GCs $\kappa_{\rm inner}^{\rm red}$, $\kappa_{\rm outer}^{\rm red}$, blue GCs $\kappa_{\rm inner}^{\rm blue}$, $\kappa_{\rm outer}^{\rm blue}$; the rotation parameter for the central stellar light, $\kappa^{\rm star}$.}
    \label{fig:dbns}
\end{figure*}

We create two sets of models, one with a cusped DM halo and the other with a cored DM halo. As indicated in Section \ref{sec:approach}, we use the EMCEE package to run our models. For each model, we use 100 walkers and run for 600 steps to ensure that the parameter space is sufficiently explored and the parameter values have converged. The best-fitting parameters for the two models are shown in Table ~\ref{tab:landscape}. Note that we have converted $d_s$ back to $r_s$, and $\lambda^{\rm star}$, $\lambda^{\rm red}$ and $\lambda^{\rm blue}$ back to $\beta_z^{\rm star}$, $\beta_z^{\rm red}$ and $\beta_z^{\rm blue}$.  

Fig.~\ref{fig:dbns} shows the posterior marginalized distributions of the last 50 steps for 11 of the 15 parameters, which are related directly to the dynamical properties. The distributions are from the model with a cored DM halo. The points in each scatter plot show the 2D posterior marginalized distribution of the two corresponding parameters and the colour of each point indicates the relative probabilities from low (blue) to high (red). The three ellipses in each scatter plot show the 1, 2 and $3\,\sigma$ regions of the covariance ellipses of the distribution. When the covariance ellipse is tilted, the ellipticity indicates the degree of correlation of the two parameters such that the larger the ellipticity, the greater the correlation, indicating a stronger degeneracy. The histogram in each row shows the 1D posterior marginalized distribution of the parameter and the curve is a Gaussian curve plotted with the mean and standard deviation of the distribution. 

The cored and cusped models lead to similar maximum likelihoods, and so we have little ability to constrain the dark matter density slope. For the cored/cusped models, there are still some degeneracies between the DM scale density $\rho_s$, the DM scale radius $d_s$ and the stellar $M/L$ ratio scale factor $\Upsilon$ (see Fig.~\ref{fig:dbns}) even for the fixed DM density slope $\gamma$. There are no noticeable degeneracy concerns between the anisotropy and rotation parameters.

The $\rho_s$ and $r_s$ of the best-fitting cored halo correspond to virial values of $M_{200} = (2.2 \pm 0.4) \times 10^{13} \, M_{\odot}$ and $r_{200} = 573 \pm 30$ kpc, while those for the cusped halo correspond to $M_{200} = (3.1 \pm 0.4) \times 10^{13}\, M_{\odot}$ and $r_{200} = 647 \pm 30$ kpc. $M_{200}$ and $r_{200}$ of the cored and cusped models are roughly consistent with each other. 
Note that for the cusped model, its concentration is $C = 13 \pm 3$. The concentration uncertainty is relatively high, but consistent within $2\sigma$ scatter, compared to the prediction of the $M_{200} \sim r_{200}$ relation from cold dark model \citep{Dutton2014}. 

In using the stellar $M/L_r$ gradient from SPS \citep{Sarzi2018} and $g$-band surface brightness, we allow a scale parameter $\Upsilon$ to be free. We obtain $\Upsilon = 0.85\pm0.03$ for the cusped model, and $\Upsilon = 0.93\pm0.02$ for the cored model. The ratio of stellar mass obtained by our dynamical model to that obtained by SPS in \citet{Sarzi2018} is $M^{\rm dyn} / M^{\rm SPS} = L_g * \Upsilon/ L_r$. We have $g-r \sim 0.8$ for M87 \citep{Chen2010} and $g-r = 0.46$ for the Sun, thus $M^{\rm dyn} / M^{\rm SPS} \sim 0.73 \Upsilon$ which is $\sim 0.62$ for the cusped model and $\sim 0.68$ for the cored model. The total stellar mass from our dynamical model is $\sim 4.34 \times 10^{11}\, M_{\odot}$, which leads to $M_{\rm star}/ M_{\rm tot} \sim 0.017$.  This value is a little higher than might be expected from the stellar-to-halo mass relation \citep[e.g.][]{Behroozi2010}.

M87 is at the centre of the Virgo cluster, which is clearly not relaxed according to the non-continuous variation in velocity dispersions between the GCs within the dynamical boundary and the intra-cluster members. We would like to emphasize that the virial mass we have estimated here does not represent the virial mass of the Virgo cluster.

\begin{table*}
 \caption{The best-fitting parameters for the cusped and cored DM halo models. The parameters are presented in two rows for each model. The first row from left to right: DM scale density $\rho_s\,[10^{-3}\,M_{\odot}\,{\rm pc^{-3}}]$, DM scale radius $r_s$ [kpc], the $M/L$ ratio scale factor $\Upsilon$, the velocity anisotropy and the rotation parameters of the stars $\beta_z^{\rm star}$, $\kappa^{\rm star}$. The second row from left to right:  the mean and dispersion of the colour $g-i$ Gaussian distributions of the red GCs $Z_0^{\rm red}$ and $\sigma_Z^{\rm red}$, the velocity anisotropy parameter for the red GCs $\beta_z^{\rm red}$, the inner and outer rotation parameters for the red GCs $\kappa_{\rm inner}^{\rm red}$, $\kappa_{\rm outer}^{\rm red}$, the maximum likelihood $\mathcal{L}_{\rm max}$.  The third row from left to right: the mean and dispersion of the colour $g-i$ Gaussian distributions of the blue GCs $Z_0^{\rm blue}$ and $\sigma_Z^{\rm blue}$, the velocity anisotropy parameter for the blue GCs $\beta_z^{\rm blue}$, the inner and outer rotation parameters for the blue GCs $\kappa_{\rm inner}^{\rm blue}$, $\kappa_{\rm outer}^{\rm blue}$ and the maximum likelihood $1\,\sigma$ confidence level $\delta \mathcal{L}$. Note that we have converted $d_s$ back to $r_s$, and $\lambda^{\rm star}$, $\lambda^{\rm red}$ and $\lambda^{\rm blue}$ back to $\beta_z^{\rm star}$, $\beta_z^{\rm red}$ and $\beta_z^{\rm blue}$.}
 \label{tab:landscape}
 \renewcommand\arraystretch{1.75}
 \begin{tabular*}{17cm}{@{\extracolsep{\fill}}ccccccc}
   \hline
   \hline
   DM & $\rho_s$ & $r_s$ & $\Upsilon$  & $\beta_z^{\rm star}$ & $\kappa^{\rm star}$\\
   & $Z_0^{\rm red}$ & $\sigma_Z^{\rm red}$ & $\beta_z^{\rm red}$ & $\kappa_{\rm inner}^{\rm red}$ & $\kappa_{\rm outer}^{\rm red}$ & ${\rm \mathcal{L}_{max}}$\\ 
   & $Z_0^{\rm blue}$ & $\sigma_Z^{\rm blue}$ & $\beta_z^{\rm blue}$ & $\kappa_{\rm inner}^{\rm blue}$ & $\kappa_{\rm outer}^{\rm blue}$ & $\delta \mathcal{L} (1\,\sigma)$\\
   \hline
   Cusped & $10.8_{-2.2}^{+2.8}$ & $51_{-9}^{+11}$ & $0.85 \pm 0.03$ & $0.013_{-0.023}^{+0.022}$ & $0.005 \pm 0.032$\\
          & $0.946 \pm 0.009$ & $0.111 \pm 0.008$ & $-0.06_{-0.15}^{+0.13}$ & $-0.53 \pm 0.16$ & $0.22 \pm 0.32$ & -19806.5\\
          & $0.751 \pm 0.004$ & $0.072 \pm 0.004$ & $0.36_{-0.05}^{+0.05}$ & $-0.35 \pm 0.21$ & $0.10 \pm 0.18$ 
          & 6.9 \\
   \hline
   Cored & $208_{-30}^{+36}$ & $15.7_{-2.0}^{+2.3}$ & $0.93 \pm 0.02$ & $0.007_{-0.023}^{+0.023}$ & $0.01 \pm 0.03$ \\
         & $0.944 \pm 0.008$ & $0.112 \pm 0.008$ & $-0.09_{-0.23}^{+0.19}$ & $-0.47 \pm 0.17$ & $0.04 \pm 0.39$ & -19806.3 \\
         & $0.751 \pm 0.004$ & $0.071 \pm 0.003$ & $0.32_{-0.05}^{+0.05}$ & $-0.40 \pm 0.20$ & $-0.08 \pm 0.15$ & 7.4 \\
   \hline
   \hline
 \end{tabular*}
\end{table*}

\subsection{Models}

\subsubsection{GCs separation into two populations}

\begin{figure}
	\includegraphics[width=\columnwidth]{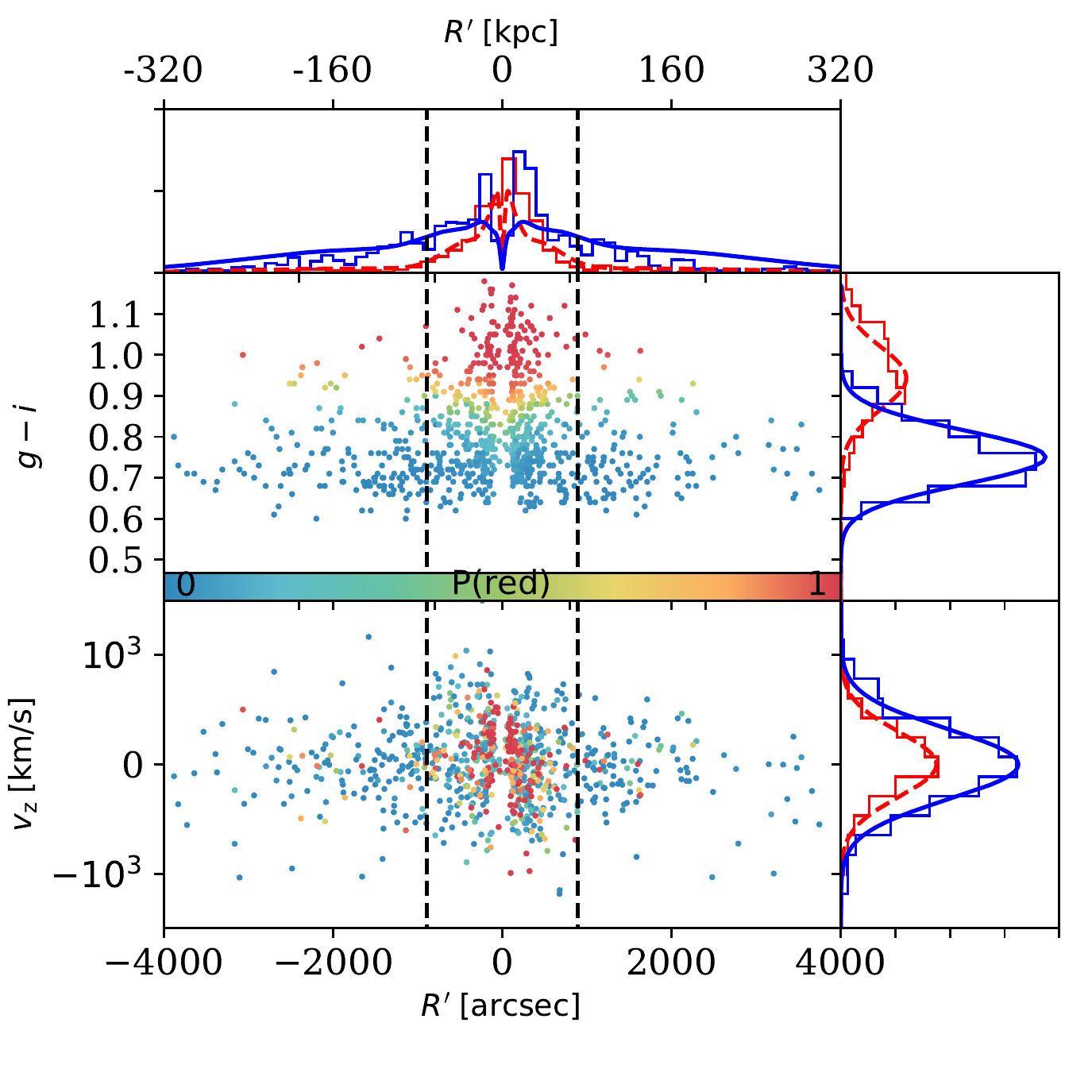}
    \caption{Chemo-dynamical modelling results for the spatial, colour and kinematic distributions of GCs. \textbf{Top scatter panel: }$R'$ vs. the colour $g-i$. \textbf{Bottom scatter panel: }$R'$ vs. LOS velocity $v_z$. GCs are plotted with points coloured by $P^{'\rm red}$, the relative probabilities of belonging to the red population, from blue (low) to red (high). In the top and right panels, we show the overall spatial, colour and velocity distributions for the red and blue populations. The histograms of red and blue populations are calculated from all kinematic GC data points weighted by $P_i^{'\rm red}$, and $P_i^{'\rm blue}$. The curves for the density distributions are from the best-fitting MGEs for the photometric red (dashed) and blue (solid) GCs. The curves for the colour and velocity distributions of red/blue GCs are from the best-fitting chemo-dynamical models. The histograms and the corresponding curves of density and colour distribution constructed from the kinematic data are generally consistent with the photometric data. The density curves and histograms are depressed in the inner region due to incompleteness in the photometric and kinematic data. The two vertical dashed lines indicate the position with elliptical $x$-axis radii of $-5\,R_e^{\rm maj}$ and $5\,R_e^{\rm maj}$.}
    \label{fig:discr_chemo}
\end{figure}

The relative probability of the $i^{\rm th}$ GC being red ($P_i^{'\rm red}$) or blue ($P_i^{'\rm blue}$) is calculated from the best-fitting model. With the posterior likelihood, the spatial, colour and velocity distribution of population $k$ can be calculated from the overall observed GC data values by weighting each by $P_i^{'k}$. For convenience of analysis of the two populations after modelling, we separate the GCs based on the relative probability of the $i^{\rm th}$ GC being red or blue in Section 5.3 (Figs.~\ref{fig:redGCs} and \ref{fig:azimuth_red}) and Section 5.4 (Figs.~\ref{fig:betaprofile} and \ref{fig:spherical_model}). The criteria used are that the $i^{\rm th}$ GC is identified as red if $P_i^{'\rm red}$ > 0.5, or blue if $P_i^{'\rm blue}$ > 0.5.

Fig.~\ref{fig:discr_chemo} shows the spatial, colour, and kinematic distributions of the red and blue populations from our best-fitting model with a cored DM halo. The top and bottom scatter panels show the $g-i$ and LOS velocity distributions along $R'$. The points are coloured by the relative probabilities of belonging to the red GCs ($P^{'\rm red}$) from low (blue) to high (red), which means the red points have higher probability of belonging to the red GCs and the blue points have higher probability of belonging to the blue GCs. In the top and right panels, we show the overall spatial, colour and velocity distributions for the red and blue populations. The histograms of red and blue populations are calculated from all the kinematic GC data points weighted by $P_i^{'\rm red}$, and $P_i^{'\rm blue}$. The curves for the density distributions are from the best-fitting MGEs for the photometric red (dashed) and blue (solid) GCs. The curves for the colour and velocity distributions of red/blue GCs are from the best-fitting chemo-dynamical models. The histograms and the corresponding curves of density and colour distribution constructed by the kinematic data are generally consistent with the photometric data. The density curves and histograms in the inner region are depressed due to incompleteness in the photometric and kinematic data. The histograms of the $g-i$ distributions for red and blue GCs show an overlap centred at $g-i \sim 0.85$, which is consistent with the hard cut colour value used in Section 3.1.

\citet{Zhang2015} used a double-Gaussian fitting to the NGVS $g-i$ bimodal colour distribution of the photometrically selected GCs in M87. This Gaussian fitting indicates that the blue and red components overlap at $g-i=0.89$ mag. We use this value of $g-i=0.89$ to separate initially the red and blue photometrical GCs so that we are able to construct MGEs for each component. Having run our model, which utilises the kinematical GC data, using these MGEs, we find that the overlap is now centred at $g-i \sim 0.85$ mag. We reconstruct our MGEs using this revised overlap value, and rerun our model. This second modelling run returns the consistent results we have shown. The photometrically selected GCs may contain other objects, such as stars, which may cause a bias in the $g-i$ separation from the double-Gaussian fitting. Our approach ensures we achieve consistency between the photometrical and kinematical GCs, which may eliminate this bias.

\subsubsection{Kinematic data reproduction}

\begin{figure}
	\includegraphics[width=\columnwidth]{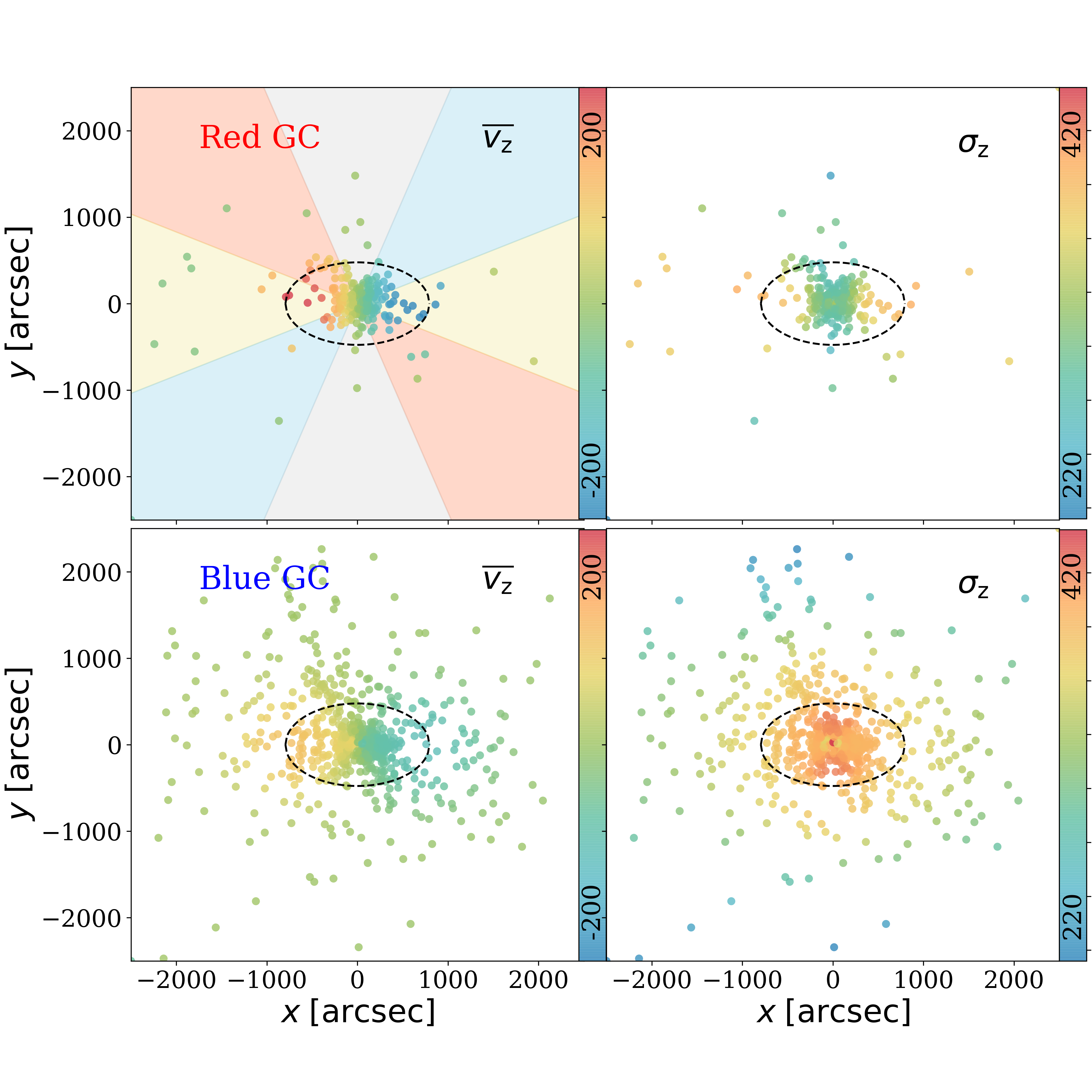}
    \caption{Mean velocity ($\overline{V_{z}}$) and velocity dispersion ($\sigma_{z}$) maps of red (upper panels) and blue (lower panels) GCs from the best-fitting model with a cored DM halo. Each point represents the position of a GC in the data sample, colours indicate the value of $\overline{V_{z}}$ and $\sigma_{z}$ as scaled by the colour bars. On the left-top panel, we illustrate the binning scheme we use for a direct comparison of the data and model predictions. The projected plane is divided into eight sectors and the regions along the major axis, minor axis and the $45^{\circ}$ and $135^{\circ}$ diagonals are filled with yellow, grey, blue and orange. The data located in each coloured region are binned along the radial radius from the left to right, or bottom to top. The x axis is along the major axis and y the minor axis. The dashed ellipse in each panel has an $x$-axis radius of $5\,R_e^{\rm maj}$.}
    \label{fig:binscheme}
\end{figure}

\begin{figure}
	\includegraphics[width=\columnwidth]{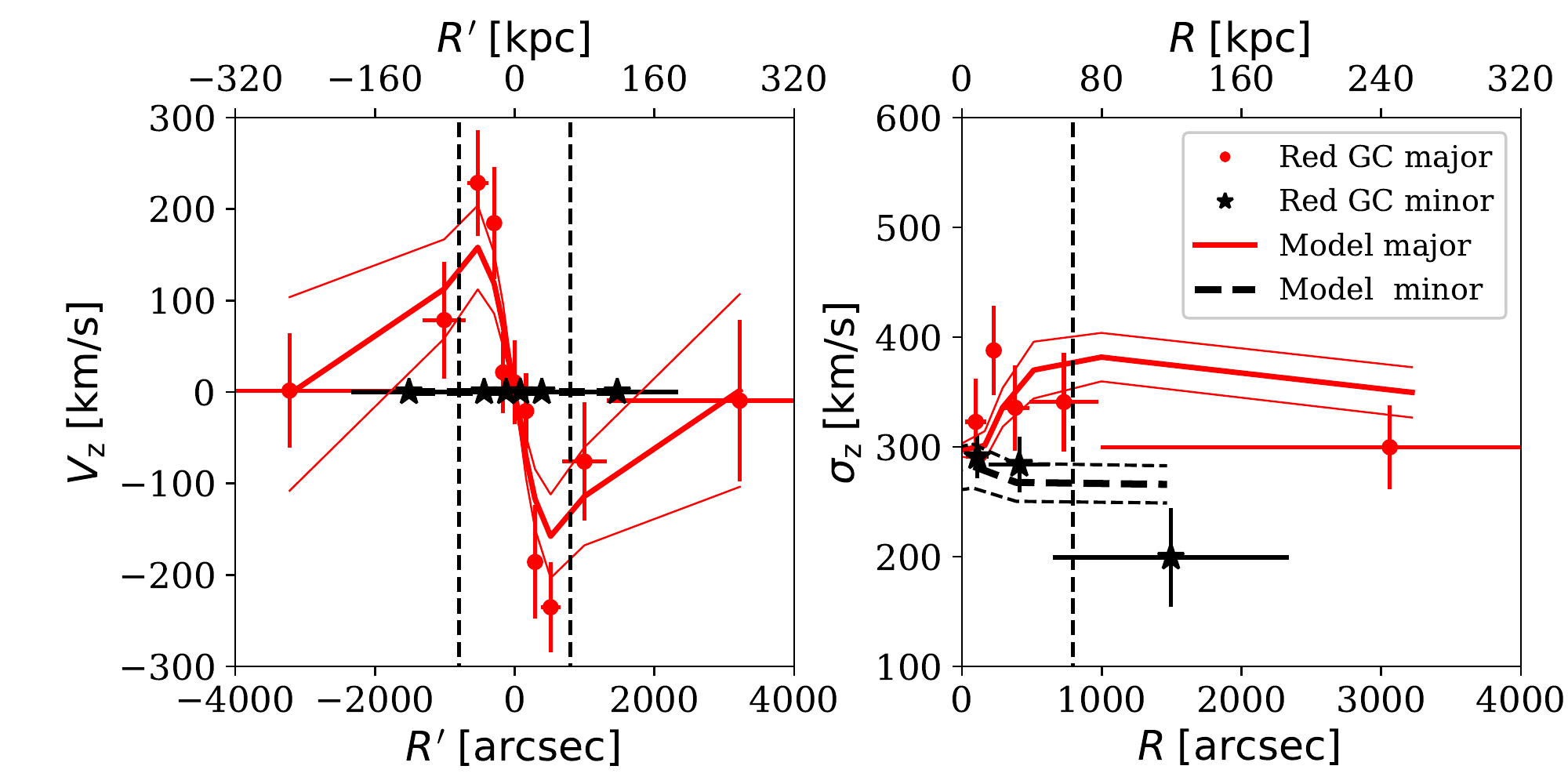}
	\includegraphics[width=\columnwidth]{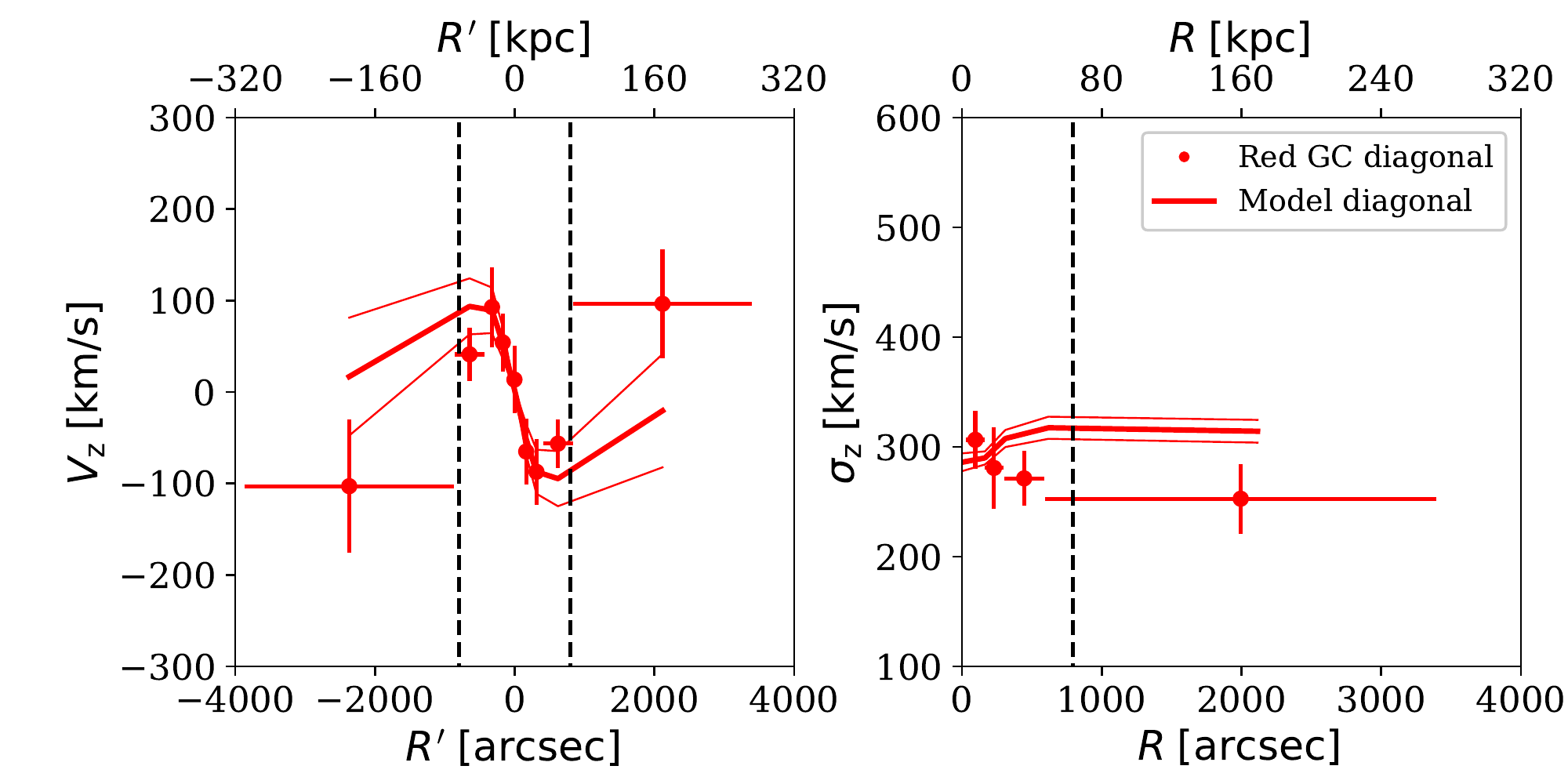}
	\includegraphics[width=\columnwidth]{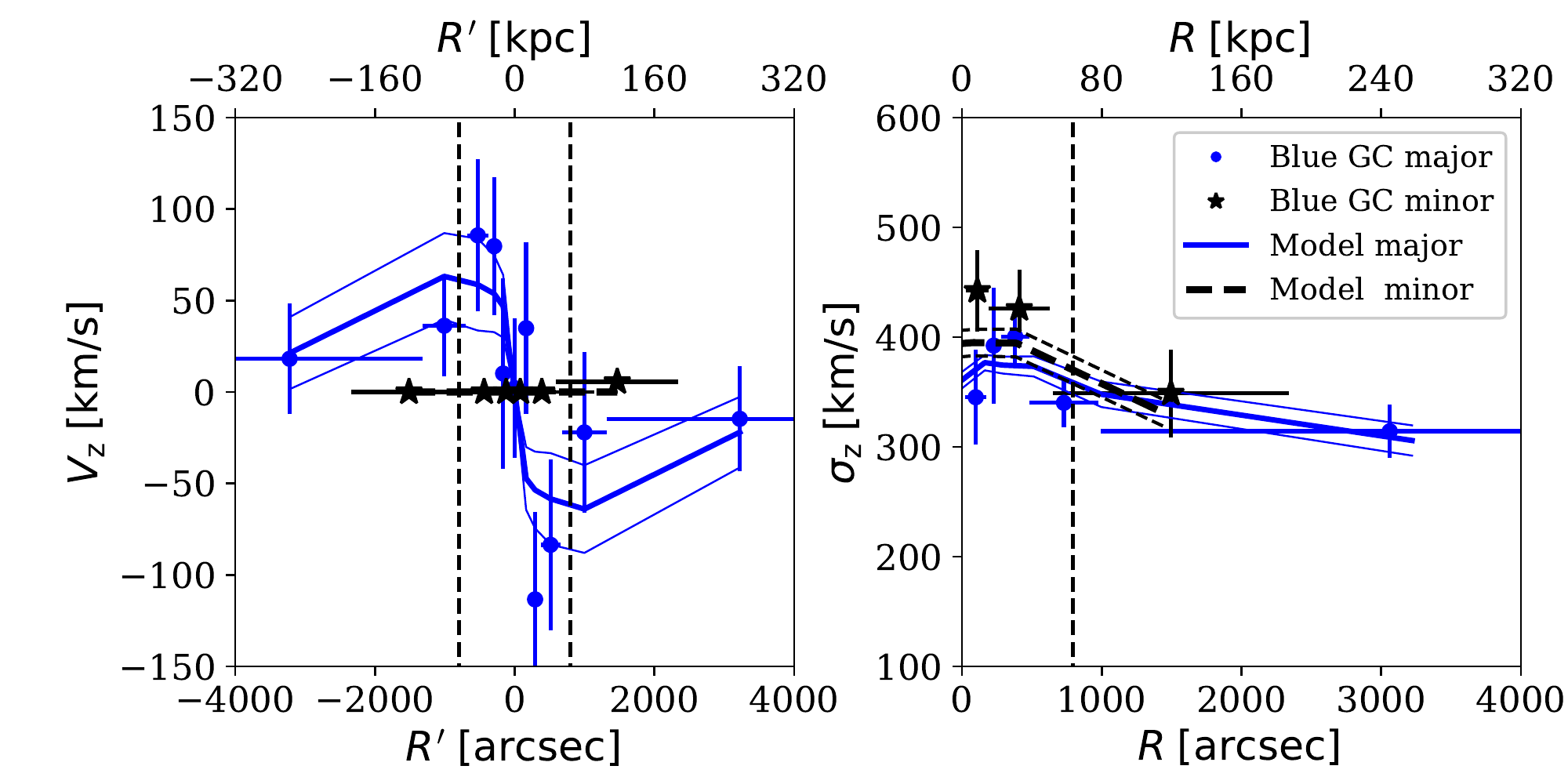}
	\includegraphics[width=\columnwidth]{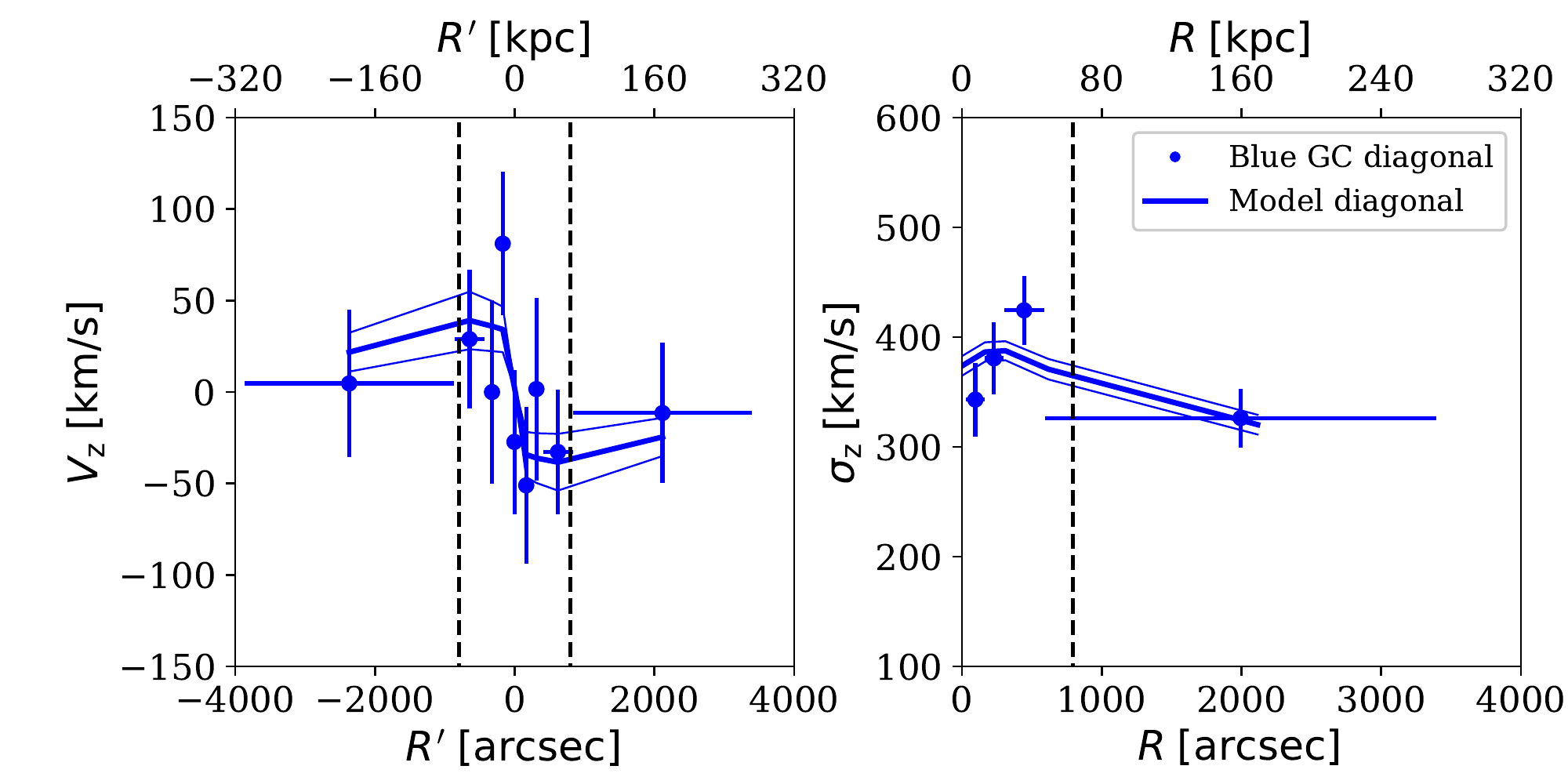}
    \caption{Comparison of the observed GC data with the best-fitting predictions from the cored DM halo model. In each panel, points with error bars are the data, thick and thin solid curves are the mean and $1\,\sigma$ uncertainties of best-fitting models. $R'$ is increasing from the left to right, or bottom to top for each binning region as shown in Fig.~\ref{fig:binscheme}. The vertical dashed lines in the panels mark the elliptical $x$-axis radius of $5\,R_e^{\rm maj}$.}
    \label{fig:GCkin}
\end{figure}

\begin{figure}
	\includegraphics[width=\columnwidth]{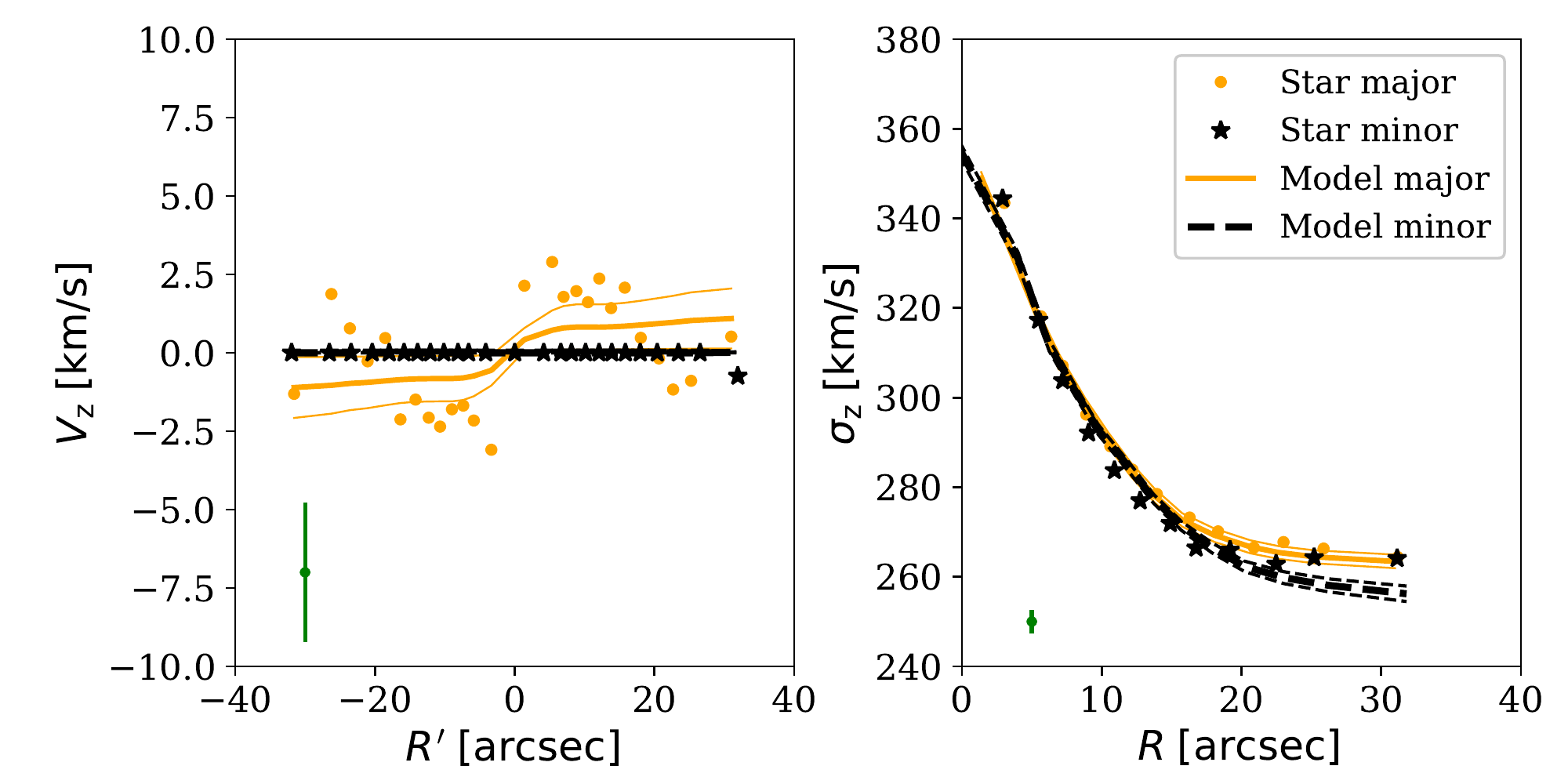}
	\includegraphics[width=\columnwidth]{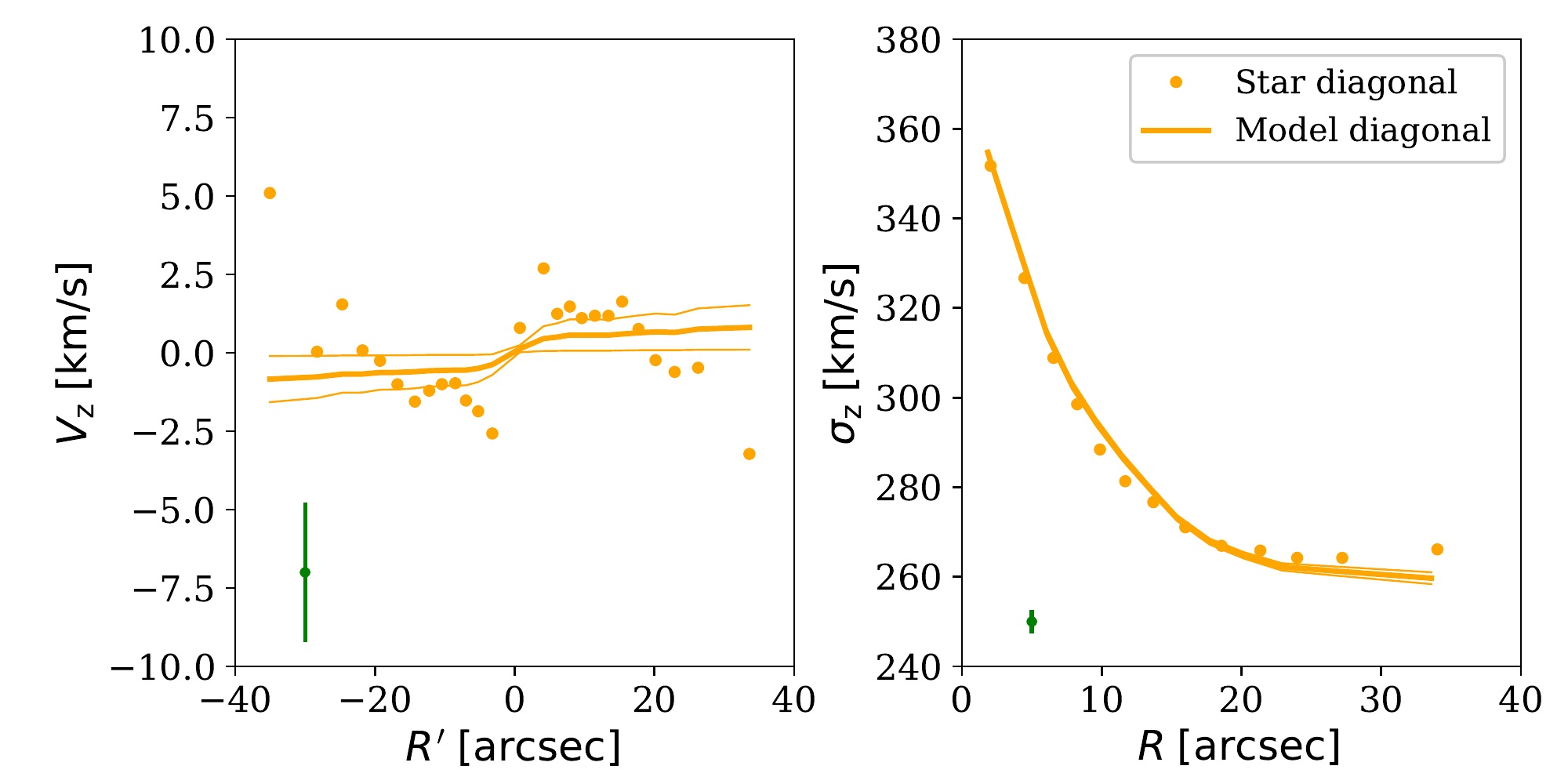}
    \caption{Comparison of the stellar data in the central galaxy with the best-fitting predictions from the cored DM halo model. In each panel, points are the data, thick and thin solid curves are the mean and $1\,\sigma$ uncertainties from the best-fitting model. The green points with vertical error bars in the bottom left hand corners of the panels indicate the mean $1\,\sigma$ uncertainties of the binned observed data.}
    \label{fig:starkin}
\end{figure}

In this section, we aim to quantify how well our model can recover the kinematics for both the discrete GCs and the stars in the central galaxy. In Fig.~\ref{fig:binscheme}, for the best-fitting cored model, we show the mean velocity and velocity dispersion maps for the red and blue GCs separately. Red GCs have moderate rotation while the rotation of the blue GCs is much weaker. On the velocity dispersion maps, red GCs have larger dispersions along the major axis than those along the minor axis, while conversely, blue GCs have larger dispersions along the minor axis, which are typical features for near edge-on systems with negative $\beta_z^{\rm red} (\sim -0.08$) and positive $\beta_z^{\rm blue} (\sim 0.35$) \citep{Cappellari2008}.

We maximize the likelihood of the discrete data points within a model. For the discrete data points with their single LOS velocity $V_z$ at a given position, we must bin the data points on the 2D plane to obtain the mean velocity and velocity dispersion values for direct comparison with the model predictions.
In order to give a visual impression of how well our model matches the observed kinematic data, we bin the model predictions and observed data with the same binning scheme. The binning scheme is shown in the top-left panel of Fig.~\ref{fig:binscheme}. Firstly, we divide the projected plane into eight sectors equally and the regions along the major axis, minor axis and the $45^{\circ}$ and $135^{\circ}$ diagonals are filled with yellow, grey, blue and orange. Secondly, we bin the observed data and model predictions located in each coloured region along the radial radius from the left to right, or bottom to top. We use axisymmetry and point symmetry for both the data and model predictions to increase the number of points by a factor of 4 before binning. After the symmetrization, the data and model predictions in the blue and orange regions along the diagonals are the same. 

For the discrete GCs, we take equal number of points in each bin and the bin size $n = 102$, but the size of the last bin is bigger because it contains all the remaining points. The observed mean velocity and dispersion values in each bin for the red and blue populations are calculated as below:

\begin{equation}
    \mu = \frac{\sum_{i}^n P_i^{'k} v_i}{\sum_{i}^n P_i^{'k}}~,
    \label{eq:v}
\end{equation}

\begin{equation}
    \sigma^2 = \frac{\sum_{i}^n P_i^{'k} (v_i-\mu)^2}{\sum_{i}^n P_i^{'k}}~,
    \label{eq:sigma}
\end{equation}
where $n$ is the bin size and $k$ represents the red or blue population. For the model, we just calculate the mean values of velocity and dispersion predicted at the different positions in that bin. Then we can calculate the mean $\chi^2$ per data point of the best-fitting model by 
\begin{equation}
    \overline{\chi^2} = \frac{\sum_{j=1}^{N}\left(\frac{\mu'_j - \mu_j}{\delta \mu'_j}\right)^2 + \sum_{j=1}^{N}\left(\frac{\sigma'_j - \sigma_j}{\delta \sigma'_j}\right)^2}{2N}~,
    \label{eq:chi2}
\end{equation}
where $N$ is the number of bins, $(\mu'_j, \sigma'_j)$ represent the mean velocity and dispersion of the observed GC data in the $j^{\rm th}$ bin and $(\mu_j, \sigma_j)$ represent that of the model predicted data. Our model matches the GC data well with a mean $\chi^2$ value of 0.94 for the red population and 0.49 for the blue population. 
Fig.~\ref{fig:GCkin} shows the resulting comparison of mean velocity and velocity dispersion profiles from the observed GC data and the model predictions along the major axis, the minor axis and the diagonals. In each panel, points with error bars are the data, thick and thin solid curves are the mean and $1\,\sigma$ uncertainties of the best-fitting models. 

For the stars, $\chi^2$ comes directly from the model as shown in Eq.~\ref{eq:chi}. The mean stellar $\chi^2$ value from the best-fitting cored model is 0.31. The visual comparison is shown in Fig.~\ref{fig:starkin}. We also bin the stellar data and model predictions as in Fig.~\ref{fig:binscheme} and the number of points per bin is 20. 
Note that we have symmetrized the data points to reduce noise, and so the data bins in Figs.~\ref{fig:GCkin} and \ref{fig:starkin} are not independent. $\chi^2$ will increase by a factor of 2 if we model without data symmetrization, and are then using independent bins.

For the velocity profiles, the model matches the data to within $1\,\sigma$ uncertainties for both the GCs and the stars in the central galaxy. As shown in the first and third left panels of Fig.~\ref{fig:GCkin}, red GCs have moderate rotation around the minor axis with $V_{\rm max}/\sigma \sim 0.4$ at $R \sim 40$ kpc, and blue GCs have a weak rotation around the minor axis with $V_{\rm max}/\sigma \sim 0.1$, and our model matches the rotation for each of the GC populations. The observed integrated starlight has no significant rotation around the minor axis, and our model prediction agrees with this.

For the velocity dispersion profiles, the model generally matches the radial profiles of both the GCs and the stars in the central galaxy, except for the red GCs along the diagonals. This explains why we obtained a higher mean $\chi^2$ for the red GCs. The dispersions along the major and minor axes are different for the two populations, with the red GCs having a larger dispersion along the major axis than that along the minor axis and the blue GCs behaving in just the opposite way. Our model matches the trend for both GC populations.

\subsection{Internal rotation}

\begin{figure}
	\includegraphics[width=\columnwidth]{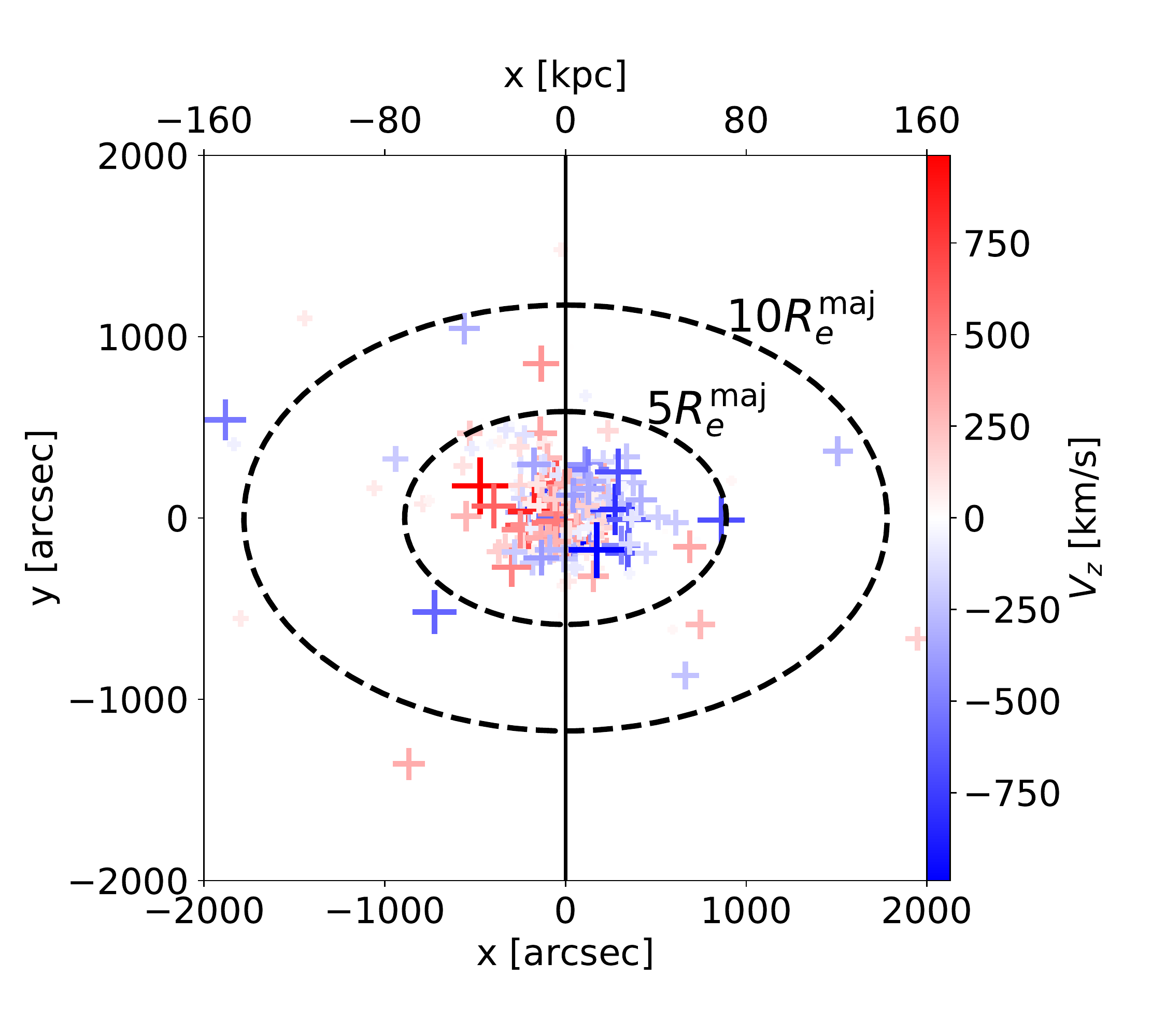}
    \caption{The positions of the red GCs identified by the cored model. $x$ is along the major axis and $y$ the minor axis. The plus symbols represent the red GCs coloured by the measured LOS velocity, while the size of each symbol is scaled with the absolute value of its LOS velocity. The two dashed ellipses have $x$-axis radii of $5\,R_e^{\rm maj}, 10\,R_e^{\rm maj}$. The vertical black line is the photometric minor axis of the galaxy.}
    \label{fig:redGCs}
\end{figure}

\begin{figure}
	\includegraphics[width=\columnwidth]{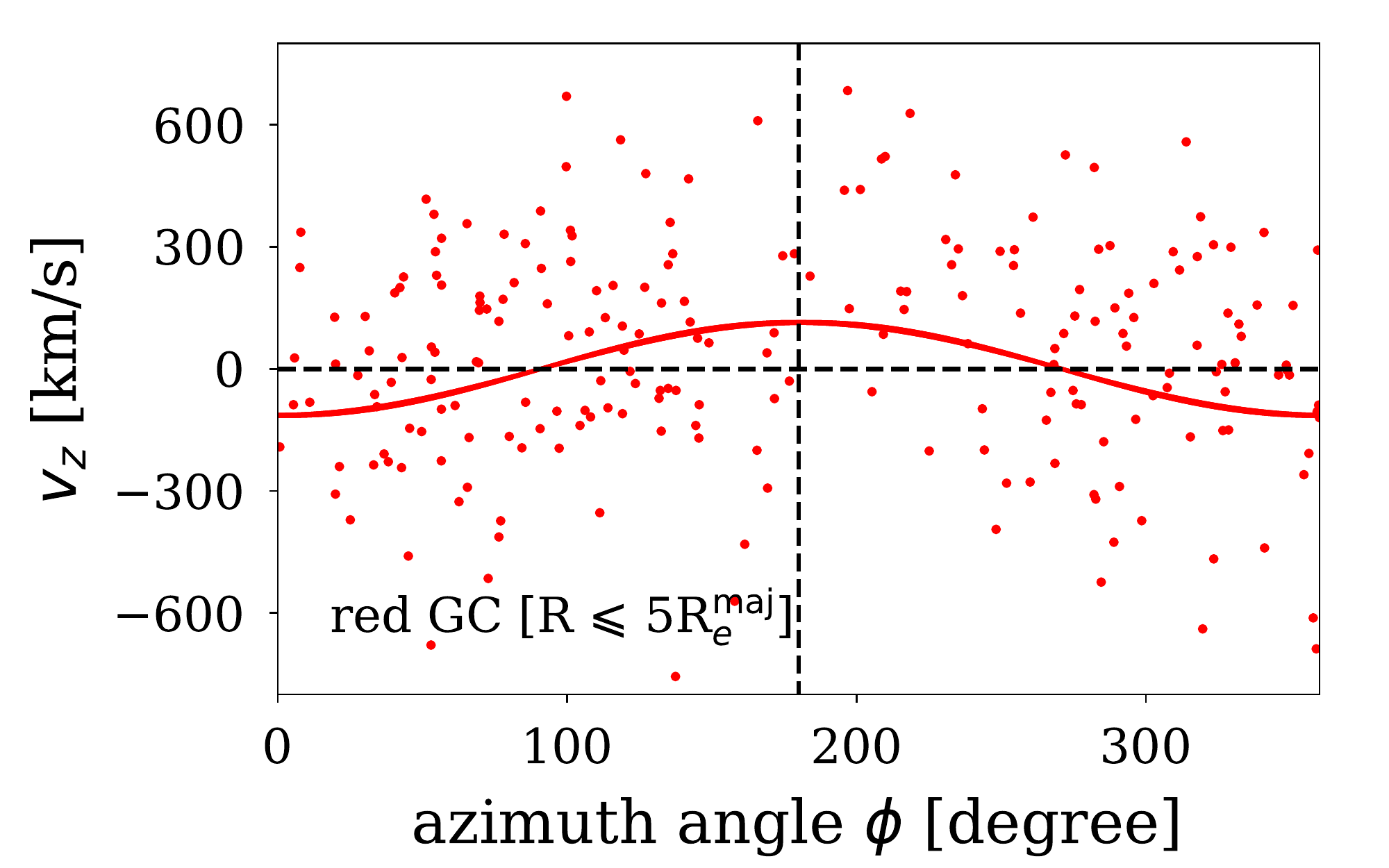}
    \caption{The LOS velocity as a function of azimuthal angle $\phi$ from the major axis. The points are the red GCs and the overlay curve is the corresponding best-fit sine curve $v=(114 \pm 6) \times {\rm sin}(\phi - (91^{\circ} \pm 3^{\circ}))\,[{\rm km\,s^{-1}}]$. The vertical dashed line marks the azimuth angle of $180^{\circ}$. }
    \label{fig:azimuth_red}
\end{figure}

As shown in Figs.~\ref{fig:GCkin} and \ref{fig:starkin}, only the red GCs show an obvious rotation around the minor axis, with $V_{\rm max}/\sigma \sim 0.4$, which is matched by our axisymmetric models. Here we check whether the rotation axis of red GCs is really aligned with the minor axis.

Fig.~\ref{fig:redGCs} shows the positions of the red GCs coloured by their LOS velocities. The plus symbols represent the red GCs coloured by the measured LOS velocity, while the size of each symbol is scaled with the absolute value of its LOS velocity. The vertical black line is the photometric minor axis of the galaxy. From a visual inspection of Fig.~\ref{fig:redGCs}, it should be apparent that the rotation direction of the red GCs is roughly around the photometric minor axis.

The variation of the LOS velocities of the red GCs along the azimuthal angle $\phi$ measured from the major axis is shown in Fig.~\ref{fig:azimuth_red}. We fit the velocities of the red GCs within $5\,R_e^{\rm maj}$ with a sinusoid $v=v_{\rm max}{\rm sin}(\phi + \phi_0)$ and obtain $\phi_0 = -91^{\circ} \pm 3^{\circ}$ and $v_{\rm max} = (114 \pm 6)\,{\rm km\,s^{-1}}$. $\phi_0 = \pm 90^{\circ}$ means rotation exactly around the minor axis. The values we obtained are consistent with minor axis rotation to within $1\,\sigma$ error. Consequently, an axisymmetric assumption is reasonable, and our model fits the rotation of the red GCs well. Red GCs may have a velocity twist outside $5\,R_e^{\rm maj}$ as shown in the velocity map of \citet{Zhang2015}, however, there are only a few red GCs outside $5\,R_e^{\rm maj}$ as can be seen in Fig.~\ref{fig:redGCs}.

Our red GCs have the same rotation axis and rotation direction as M87's planetary nebulae (PNe) \citep{Longobardi2018}. PNe have their strongest rotation in the region of $R\sim 500-1000$ arcsec \citep{Longobardi2018}, but with $V_{\rm max}/\sigma \sim 0.1$ which is smaller than we found for red GCs. 

\subsection{Velocity dispersion anisotropy}

\begin{figure}
	\includegraphics[width=\columnwidth]{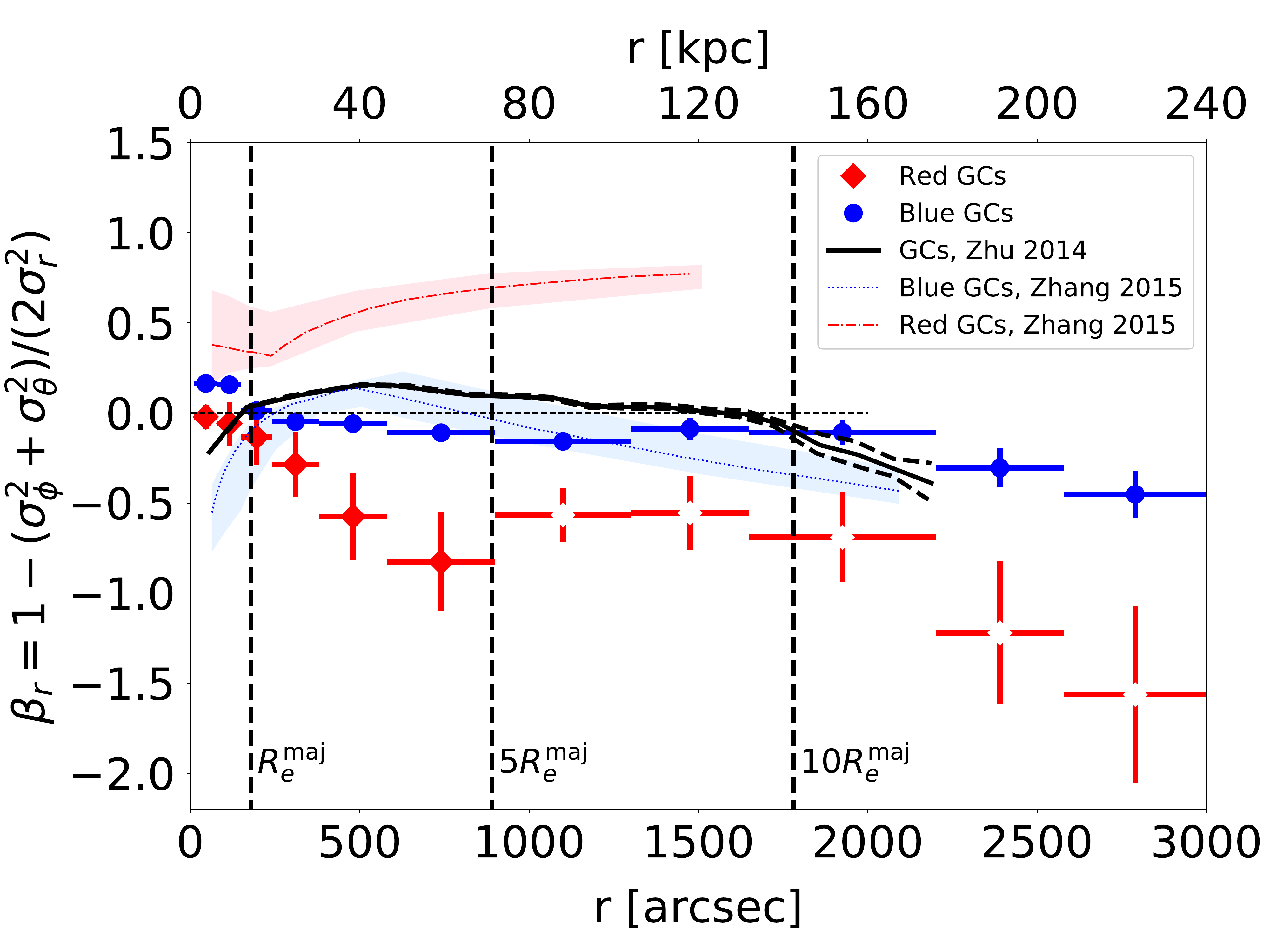}
    \caption{The velocity dispersion anisotropy $\beta_r$ profiles. The red diamonds and blue circles with error bars are the profiles of the red and blue GCs from the best-fitting models with a cored DM halo. $\beta_r$ for red GCs at $r>1000$ arcsec should have large uncertainties and we mark these points with open symbols. The error bars indicate the scatter of $\beta_r$ in the models within a $1\,\sigma$ confidence level. The dash-dotted red and dotted blue curves are the results obtained by the spherical Jeans models from \citet{Zhang2015} and the red and blue regions mark the $1\,\sigma$ confidence intervals. The solid black curve is the profile obtained by the made-to-measure method from \citet{Zhu2014} and the dashed curves are the corresponding uncertainties. Note that the profile from \citet{Zhu2014} is for all GCs. The three vertical dashed lines mark radii of $(1,5,10)\,R_e^{\rm maj}$.}
    \label{fig:betaprofile}
\end{figure}

The velocity dispersion anisotropy parameter $\beta_r = 1 - (\sigma_{\phi}^2 + \sigma_{\theta}^2)/2\sigma_r^2$ is widely used as an indication of underlying orbit distribution: $\beta_r > 0$ indicates radially biased anisotropy, $\beta_r < 0$ tangentially biased anisotropy, and $\beta_r=0$ an isotropic distribution \citep[p.294]{Binney2008}. In axisymmetric JAM models, $\beta_r$ is not used as an explicit free parameter, but we can calculate it as a function of radius from the intrinsic first and second velocity moments predicted by the model. Fig.~\ref{fig:betaprofile} shows the velocity anisotropy profiles obtained from our best-fitting models. The red diamonds and blue circles with error bars represent the velocity anisotropy for our GCs from our model with a cored DM halo. There are few red GCs at $R > 1000$ arcsec, and our model has difficulty matching the dispersion along diagonals for red GCs in this region as shown in Fig.~\ref{fig:GCkin}. Thus the anisotropy for red GCs at $r>1000$ arcsec may be larger than the errors we obtain from the model scatter. We mark these points with open symbols in the figure. The cusped model also predicts an almost identical velocity anisotropy profile and we show it in the appendix A.

According to our model, the red GC system is nearly isotropic in the very inner region, becoming significantly tangential in the outer region. The blue GC system has slightly radial velocity anisotropy in the inner region, becoming slightly tangential in the outer region. The difference between the red and blue GCs velocity anisotropy could be indicated by the velocity dispersion difference along the major and minor axes. The red GCs have a larger dispersion along the major axis, while it is the opposite for the blue GCs. This results in the different velocity anisotropy we obtained for the two populations.

The dash-dotted red and dotted blue curves in Fig.~\ref{fig:betaprofile} are the results obtained by the spherical Jeans model from \citet{Zhang2015} with the red and blue regions marking the $1\,\sigma$ confidence intervals. \citet{Zhang2015} divided GCs into red and blue population in advance by colour $g-i=0.89$ and constructed two spherical Jeans models for the red and blue GCs. 
The solid black curve is the profile obtained by the made-to-measure method from \citet{Zhu2014} and the dashed curves are the corresponding uncertainties. \citet{Zhu2014} treated GCs as one population in the model. 

\begin{figure}
	\includegraphics[width=\columnwidth]{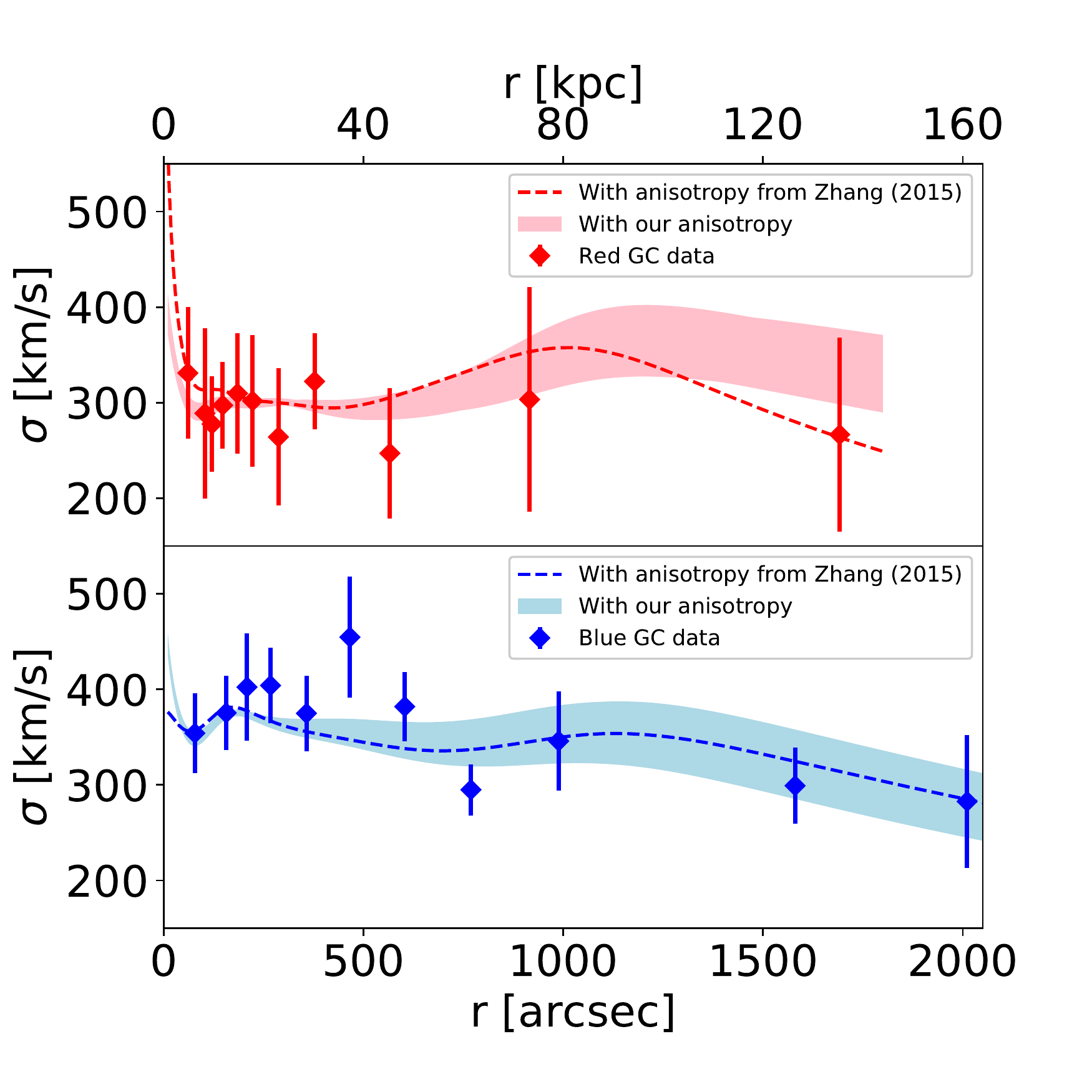}
    \caption{Comparison of the observed velocity dispersions with the predictions by the spherical Jeans models. The red and blue GCs are classified by the best-fitting chemo-dynamical model as mentioned above. The points with error bars are the observed velocity dispersion profiles of the red (top) and blue (bottom) GCs, and they are binned along the spherical radius. The pink and light-blue regions are the predictions for the velocity dispersions with $1\,\sigma$ uncertainties by the spherical Jeans models after imposing our velocity anisotropy profiles for the red and blue GCs, respectively. The red and blue dashed curves are the predictions by the spherical Jeans models with the velocity anisotropy profiles from \citet{Zhang2015}.}
    \label{fig:spherical_model}
  \end{figure}

The spherical Jeans model from \citet{Zhang2015} predicted strong radial anisotropy for the red GCs overall and slightly tangential anisotropy for the blue GCs.
\citet{Oldham2018} obtained $\beta_r \sim 0.32$ for red GCs and $\beta_r \sim 0.26$ for blue GCs which we do not show in the figure.
The velocity anisotropy profiles for the blue GCs in the literature are roughly consistent with our results within $1\,\sigma$ confidence in the outer regions beyond $1\,R_e^{\rm maj}$. For the red GCs, the radial anisotropy obtained by spherical Jeans models in the literature are not consistent with each other, and not consistent with our results.

Given the apparent inconsistency of the velocity anisotropy of the red GCs we obtained, especially when compared with that of \citet{Zhang2015}, we perform a simple test to check whether our velocity anisotropies are consistent with the data if we also make a spherical symmetry assumption.

We construct spherical Jeans models for the red GCs and blue GCs of M87, using the total mass profiles from our axisymmetric Jeans models, the velocity dispersion anisotropy $\beta_r^{\rm red}$, $\beta_r^{\rm blue}$ from our models or \citet{Zhang2015} (shown in Fig.~\ref{fig:betaprofile}), and the surface number density profiles of red and blue GCs (shown in Fig.~\ref{fig:sfprofile}) as input. We still use MGEs to describe the surface number density of each population. We impose a radial variation of $\beta_r$ in our spherical Jeans models by setting $\beta_k = \beta_r (r = \sigma_k)$ for each Gaussian $k$ with its dispersion $\sigma_k$ \citep{Cappellari2008}. We then compare the model predictions from the spherical Jeans models with the observed velocity dispersions as shown in Fig.~\ref{fig:spherical_model}. The pink and light-blue regions are the velocity dispersion predictions from the spherical Jeans models after imposing our velocity anisotropy profiles for the red and blue GCs, respectively. The red and blue dashed curves are the predictions from the spherical Jeans models using velocity anisotropy profiles from \citet{Zhang2015}.
Although the model predictions using our velocity anisotropy do not match the data for red GCs in the outmost region as well as those models with the velocity anisotropy taken from \citet{Zhang2015}, our model still generally matches the data points within $1\,\sigma$ uncertainties.
An exception is that the observed dispersion of the blue GCs at $\sim 400$ arcsec (32 kpc) is about 20 percent higher than the model prediction, and is likely due to a substructure \citep{Romanowsky2012} in the blue GCs in this region.

In a spherical Jeans model, the tangential velocity dispersion $\sigma_t$ and the radial velocity dispersion $\sigma_r$ are sensitive to $\beta_r$. For a system with a fixed mass profile $M(r)$, when $\beta_r$ increases, $\sigma_r$ increases and $\sigma_t$ decreases \citep[see equation 4-55 of][]{Binney1987}. Both $\sigma_r$ and $\sigma_t$ contribute to the velocity dispersion along the line-of-sight.
The LOS velocity dispersion does not depend strongly on $\beta_r$, as the variation of $\sigma_r$ and $\sigma_t$ partly cancel each other \citep[see equation 4-60 of][]{Binney1987}. The dependency of the LOS velocity dispersion on $\beta_r$ is affected by the tracer density profile. With the tangential anisotropy we obtained for red GCs, very different from the radial anisotropy obtained by \citet{Zhang2015}, our tests show that the dispersion obtained from the spherical Jeans models could still be consistent with the LOS velocity dispersion from the data. This indicates that the red GCs velocity anisotropy profiles from spherical models could contain large uncertainties, and perhaps the error bars in \citet{Zhang2015} have been under-estimated.
Axisymmetric Jeans models are most useful for stronger anisotropy constraints, which are provided mainly via the azimuthal information in the kinematics. We have discussed this by showing the difference in velocity dispersions along the major and minor axes in the second paragraph of this section.

\subsection{Mass profile}

\begin{figure}
	\includegraphics[width=\columnwidth]{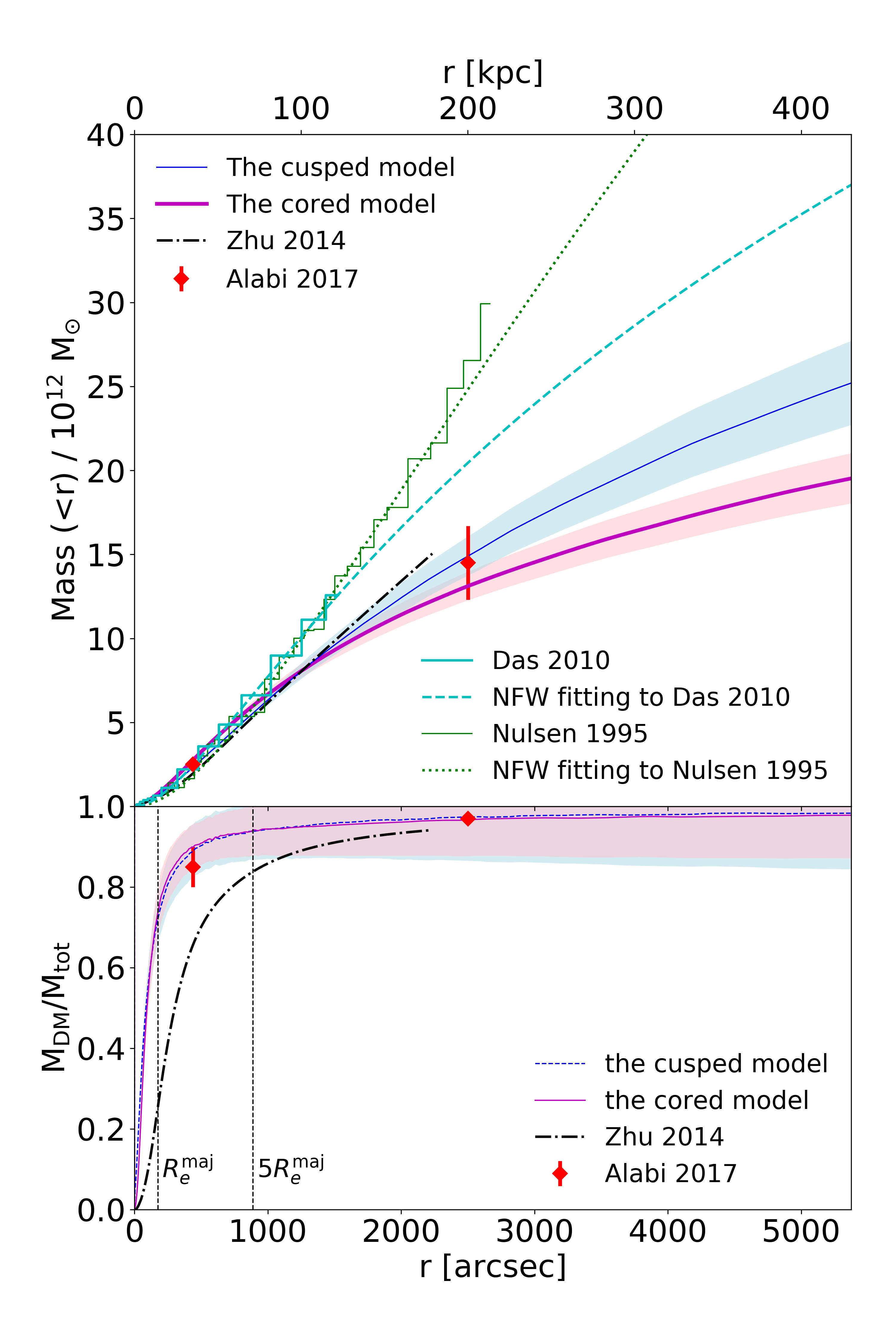}
    \caption{Total mass profile and DM fraction. \textbf{Top panel: }The total mass profiles. The solid blue, red and dash-dotted black curves represent the total mass profiles obtained by our models with a cusped DM halo, with a cored DM halo and by the made-to-measure method from \citet{Zhu2014}. The coloured regions around the solid curves represent the $1\,\sigma$ uncertainties of the total mass. The red diamonds mark the total mass obtained from \citet{Alabi2017}. The cyan step curve represents the mass profile obtained from hot gas emission based on Chandra observations \citep{Das2010}, while the green is based on ROSAT observations \citep{Nulsen1995}. The step curves represent the middle values of their measurements within the data coverage, and the corresponding dotted and dashed curves are NFW profiles fitted to the step curves. \textbf{Bottom panel: }The DM fraction. The coloured light-blue and pink regions around the solid curves represent the $1\,\sigma$ uncertainties of the DM fraction from the cusped and cored models. The red diamonds mark the DM fraction obtained from \citet{Alabi2017}. The two vertical dashes lines indicate radii of $1\,R_e^{\rm maj}, 5\,R_e^{\rm maj}$.}
    \label{fig:dm_fraction}
\end{figure}

We show in Fig.~\ref{fig:dm_fraction} the total mass profiles and the DM fractions obtained by our best-fitting models, with blue and magenta representing the cusped and the cored models. The coloured regions around the solid curves indicate the corresponding $1\,\sigma$ uncertainties of the total mass profile. The bottom panel shows the DM fraction. The light-blue and pink regions around the solid curves represent the $1\,\sigma$ uncertainties of the DM fraction from the cusped and cored models.

The total mass profiles from our two models are consistent with each other to within $1\,\sigma$ uncertainties up to $\sim$ 220 kpc. The difference in the mass profiles becomes obvious beyond 220 kpc. By comparing the observed kinematics with model predictions from the cored (Fig.~\ref{fig:GCkin}) and cusped (Fig.~\ref{fig:GCkin_cusped}) models, we see that the cored model matches the data slightly better than the cusped model beyond 150 kpc for the red GCs and the corresponding mean $\chi^2$ of the cored model is slightly lower. However, the preference of the cored model is not statistically significant due to the small number of data points at large radii. As the scale factors $\Upsilon$ obtained by the two sets of models are close, the stellar mass profiles are almost the same and the DM fraction profiles from our two models are also consistent. The DM halo contributes $(73 \pm 8)$ percent of the total mass within $1\,R_e^{\rm maj}$ (14.2 kpc), and $(94 \pm 6)$ percent within $5\,R_e^{\rm maj}$ (71.2 kpc).

With data extending to $\sim 400$ kpc, we obtain total mass values of $M_{\rm tot} = (1.95 \pm 0.17) \times 10^{13} M_{\odot}$ for the cored model and $M_{\rm tot} = (2.32 \pm 0.22) \times 10^{13} M_{\odot}$ for the cusped model. Incorporating the uncertainties in the cored and cusped models, we conclude $M_{\rm tot} = (2.16 \pm 0.38) \times 10^{13} M_{\odot}$ within $\sim 400$ kpc for M87.

The red diamonds at different radii in Fig.~\ref{fig:dm_fraction} mark the total mass and DM fraction obtained by \citet{Alabi2017}. They obtained their mass estimates using a tracer mass estimator \citep{Watkins2010} and a stellar $M/L$ ratio that accounts for stellar age variations, along with GC kinematic data from the SLUGGS survey \citep{Brodie2014} and \citet{Strader2011}. Their results are consistent with ours as may be seen from Fig.~\ref{fig:dm_fraction}. They found that the DM fractions are $(85 \pm 5)$ percent within 35 kpc, and $(97 \pm 1)$ percent within 200 kpc.

The black dash-dotted curves in Fig.~\ref{fig:dm_fraction} represent the total mass profile and DM fraction obtained by \citet{Zhu2014}. Their  mass estimates were determined using the made-to-measure method with GC kinematic data as described in Section 3.2. Comparing with our results, the total mass profile from \citet{Zhu2014} is consistent with the $1\,\sigma$ uncertainties inside $\sim$ 180 kpc, but, in the inner region,  the stellar mass profile is about twice as high as ours and the DM mass profile is lower. As shown in the bottom panel, the DM fraction from \citet{Zhu2014} is $\sim$ 26 percent within $1\,R_e^{\rm maj}$ (14.2 kpc) and $\sim$ 84 percent within $5\,R_e^{\rm maj}$ (71.2 kpc) which are significantly smaller than our inner region predictions. This can be explained by the fact that \citet{Zhu2014} assumed a constant $M/L_r$ ratio whereas we modelled with a varying $M/L$ ratio constructed by stellar population synthesis. The DM contribution is less than $17\%$ with the cusped DM and $7\%$ with the cored DM in the inner 25 arcsec ($\sim 2$ kpc). To match the total mass (mainly stellar mass) in the very inner regions, a model with a constant stellar $M/L$ ratio will require a higher overall $M/L$ ratio, which leads to more stellar mass in the outer regions ($r \sim$ a few hundred arcsec) where both stellar mass and DM are important (see Fig.~\ref{fig:ml_sbp}).

The cyan step curve in Fig.~\ref{fig:dm_fraction} represents the mass profile obtained from hot gas X-ray emissions observed by Chandra  \citep{Das2010}, and the green is based on ROSAT observations \citep{Nulsen1995}. M87 has also been observed to large radii by XMM-Newton \citep{Urban2011}. Their temperature profile of hot gas is consistent with that from the ROSAT observations, and thus similar mass profiles are expected. The step curves represent the middle values of the measurements within the data coverage, and the corresponding dotted and dashed curves are NFW profiles fitted to the step curves. The mass profile from \citet{Nulsen1995} has a large uncertainty with the lower boundary consistent our results. 
Within the data coverage (from 0 to 200 kpc), the mass obtained from X-ray emitting hot gas is $\sim 20-100$ percent larger than our mass profile obtained from GC kinematics.
There will be further discussion of X-ray mass profiles in Section \ref{sec:rjlDisc}.

\section{Discussion}
\label{sec:rjlDisc}

\subsection{Uncertainties on mass profiles}

In the early literature, it was proposed that the dark matter around M87 is associated with the whole of Virgo \citep{McLaughlin1999}, as a universal NFW halo can fit the kinematics of GCs (within 100 kpc), dwarf galaxies (100 kpc < $R$ < 2 Mpc) and X-ray gas simultaneously. The mass profile presented by \citet{McLaughlin1999} is generally consistent with the mass profile derived from X-ray emission.
However, with the increase in the numbers of GC LOS velocity measurements, it has become obvious that GCs \citep{Longobardi2018GC} and PNe \citep{Longobardi2018} have much smaller velocity dispersions than dwarf galaxies \citep{McLaughlin1999,Ko2017} in the region of $100<R<200\,{\rm kpc}$. 
A kinematically distinct intra-cluster population of GCs has been identified \citep{Longobardi2018GC}, which has a dispersion of $\sim 600\,{\rm km\,s^{-1}}$, consistent with the velocity dispersion of the 17 GCs identified in Section 3.2, as well as the dispersion of the dwarf galaxies. A similar intra-cluster component has also been identified for PNe \citep{Longobardi2018}.

The mass obtained from X-ray gas $M(<400\,{\rm kpc}) = 5.2 \times 10^{13} M_{\odot}$ (\citealt{Urban2011}; \citealt{Das2010}; \citealt{Nulsen1995}) and that of $M(<400\,{\rm kpc}) = 8.5 \times 10^{13} M_{\odot}$ from dwarf galaxy kinematics \citep{McLaughlin1999} is larger than our value of $M(<400\,{\rm kpc}) = (2.16 \pm 0.38) \times 10^{13} M_{\odot}$). Our mass is about a factor of $\sim 2.5$ smaller than that from X-ray gas, and a factor of $\sim 4$ smaller than that from dwarf galaxy kinematics.

Due to the fact that Virgo cluster may not be fully relaxed (\citealt{Urban2011}; \citealt{Binggeli1987}), M87 seems to be located at the dynamical centre of one sub-clump of galaxies. The differences in mass profiles obtained from different tracers may be caused by the unrelaxed state. Dwarf galaxies, affected by the formation of the cluster, have large velocity dispersions which are consistent with the intra-cluster GCs. The X-ray mass profile is also significantly larger than that indicated by GCs at $R>100$ kpc. It is likely that dwarf galaxies and X-ray gas are more relaxed within the gravitational potential of the outer regions of the cluster, but not in the disturbed central regions \citep{Wood2017}. Our study, however, has attempted to carefully remove the intra-cluster GCs (see Fig.~\ref{fig:groups}), and thus the remaining GCs are supposed to be well relaxed with the gravitational potential of the central galaxy (M87).

We use JAM models to describe the kinematics of stars, red GCs and blue GCs in M87. The GC systems extending to large radii are oblate and have rotation around the photometric minor axis, and thus suit the oblate assumption of JAM modelling. The stellar component of M87 could actually be triaxial with a kinematic decoupled core \citep{Emsellem2014}. Various tests show that JAM modelling could still work well for estimating the total mass distributions of triaxial and prolate systems with $\sim 20\%$ uncertainty \citep{Li2016}. The DM mass dominates in the outer regions and is mainly constrained by the GC kinematics. We presume the uncertainty in the mass profile in the outer regions caused by triaxiality of the stellar component should be minor.

Even with 894 GCs, the data are still poorly sampled in the outer regions (see Fig.~\ref{fig:groups}). We have checked that the asymmetry in velocity distribution in the outer regions is unlikely to be caused by Milky Way stars from their photometry properties. It is hard to tell if they are actually caused by low number statistics, or contamination of foreground GCs in the Virgo cluster, where the latter could potentially cause an over-estimate of velocity dispersion, thus affecting the mass profile. More data points at $R > 2000$ arcsec are needed to investigate this scenario. There is also an indication of substructure \citep{Romanowsky2012} in the GC system around $R \sim 500$ arcsec, which ideally should be cleaned before the modelling. We note that our models are not trying to match this `bump' in velocity dispersion, but mostly matched the velocity dispersion profiles in the outer regions. In this sense, the effect of this `substructure' on our results should be small. All these effects on the data may actually cause extra systematic uncertainty rather than the statistical uncertainty to our results. In this sense, we could still under-estimate the uncertainty of the masses we reported.

As discussed in Section 3.1, the MGE constructed surface density profile for red GCs may have relatively large uncertainties in the northern region at $R>1000$ arcsec. Our MGE model fits the southern side and so we may have created a too steep surface density slope. This may cause an over-estimation of the enclosed mass with fixed kinematics. In our case, there are 23 red GCs with elliptical x-axis radii $R>1000$ arcsec, including only 8 in the northern regions, thus the mass profile in the outer regions is dominated by fitting to the blue GCs (as can be seen in Fig.~\ref{fig:GCkin}, the kinematics of the blue GCs are fitted well in the outer regions, but not for the red GCs). We therefore presume that the red GCs surface density profile uncertainties have only a minor effect on the overall mass profile estimation.

We have few constraints on the dark matter inner slope, and no significant preference is found for a cored or cusped dark matter halo. A cored dark matter is preferred in \cite{Oldham2018}. Note that in their work, all components are modelled with radially anisotropic spherical models (constant $\beta_r$ for red and blue GCs, and a scaled Osipkov-Merrit profile for stars which is isotropic at $r=0$ becoming radially anisotropic with a constant value at infinity). These are strong assumptions and act to narrow down the statistical uncertainty on their mass profiles. Stronger assumptions in models can cause larger systematic uncertainties, which are usually poorly evaluated. We have allowed our velocity anisotropy and rotation parameters to be totally free in our axisymmetric models. This gives more freedom in the determination of mass profiles, and naturally causes relatively large statistical uncertainties on the gravitational potential parameters including the DM density slope $\gamma$.

We have included the $M/L$ gradient from SPS \citet{Sarzi2018}. However, the scale parameter we obtained is smaller than 1, with the stellar mass $M^{\rm dyn} / M^{\rm SPS} \sim 0.62-0.68$ regardless of the dark matter model. Note that DM contributes less than $\sim 7\%$ with the cored DM in the inner $\sim 2$ kpc, so a higher stellar mass can not be accommodated in our model. Since the stellar system of M87 is triaxial, the uncertainty in the total mass could be caused by applying a JAM model to a triaxial system. \citet{Li2017} assessed the mass uncertainty in such a case to be $\sim 20\%$, but this would not fully resolve the discrepancy here. Note that our result is consistent with \citet{Oldham2018}, who obtained a similar $M/L$ gradient independently from dynamical models, but with a systematically lower ratio than \citet{Sarzi2018}. 
Our result indicates that the $M/L_r$ from SPS \citep{Sarzi2018} might be over-estimated, especially in the very inner regions. This would cause an over-estimate of stellar mass and an underestimate of DM in the very inner regions in our model. However, overall, it should have limited impact on the whole DM profile and on the fit to the GC kinematics.

\subsection{Velocity anisotropies of GCs in other galaxies}
\begin{figure}
	\includegraphics[width=\columnwidth]{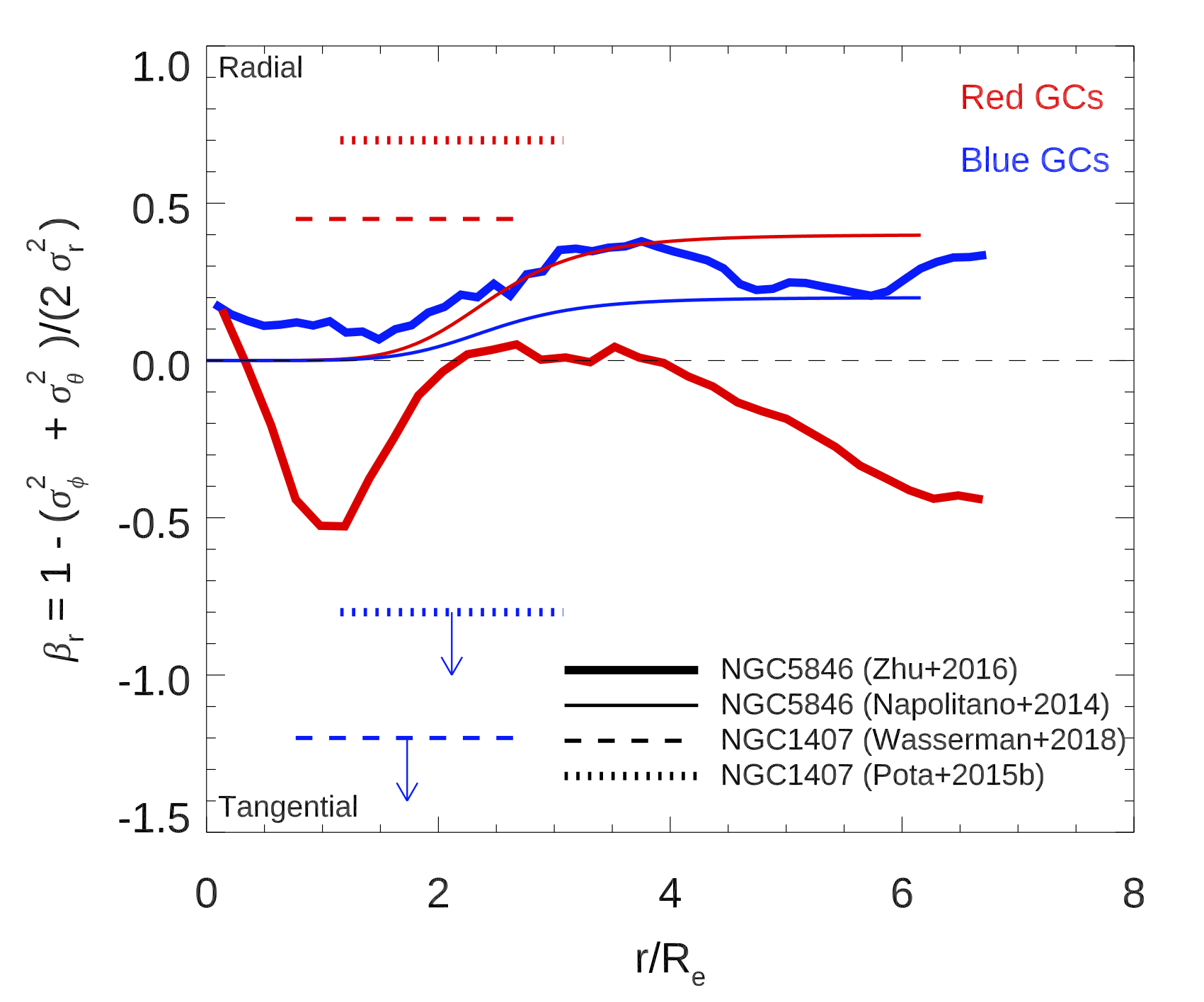}
    \caption{The velocity anisotropy profiles of red and blue GCs obtained by various two-component dynamical models in the literature: NGC 5846 from \citet{Zhu2016a} obtained by discrete chemo-dynamical JAM models and confirmed with the made-to-measure method (thick solid lines), NGC 5846 from \citet{Napolitano2014} obtained using kurtosis (thin solid lines), NGC 1407 obtained by two-component spherical Jeans models from \citet{Wasserman2018} (dashed lines) and \citet{Pota2015b} (dotted lines). The red and blue lines represent $\beta_r$ of red and blue GCs, respectively. Fig.~\ref{fig:betaprofile} shows the anisotropy profiles of M87 for comparison.}
    \label{fig:betar_compare_m87}
\end{figure}

The velocity anisotropy $\beta_r$ values obtained by different methods are not always consistent with each other. As shown in Fig.~\ref{fig:betaprofile}, we obtain tangential anisotropy for the red GCs of M87, while the two-component spherical Jeans model yields a highly radial anisotropy $\beta_r \sim 0.8$ in \citet{Zhang2015} and medium radial anisotropy $\beta_r \sim 0.3$ in \citet{Oldham2018}. Also, highly radial anisotropy was obtained for the red GCs of NGC 1407 by using a similar two-component spherical Jeans model (\citealt{Wasserman2018}; \citealt{Pota2015b}).

In Fig.~\ref{fig:betar_compare_m87}, we summarize the velocity anisotropy profiles of red and blue GCs obtained by various two-component dynamical models in the literature: NGC 5846 from \citet{Zhu2016a} obtained by discrete chemo-dynamical JAM models and confirmed with the made-to-measure method, NGC 5846 from \citet{Napolitano2014} obtained using kurtosis, NGC 1407 obtained by two-component spherical Jeans model from \citet{Wasserman2018} and \citet{Pota2015b}. 

Highly radial anisotropy of red GCs and highly tangential anisotropy of blue GCs are obtained for NGC 1407 by using a similar two-component spherical Jeans model \citep{Wasserman2018,Pota2015b} as \citet{Zhang2015} used for M87. Unlike an axisymmetric JAM model, with a spherical Jeans model, the information on anisotropy in the data itself across the 2D observational plane is actually washed out. The velocity anisotropies are obtained by the balance of the multiple tracer populations with different number density slopes and velocity dispersions. The red GCs have smaller velocity dispersions than blue GCs, both in M87 and NGC 1407, which we presume is the direct reason for obtaining radial anisotropy for red GCs and tangential anisotropy for the blue GCs through the two-component spherical Jeans models. The GC systems have rotation and different velocity dispersions along the major and minor axes, both in M87 and NGC 1407. These systems are more axisymmetric and deviate more from the spherical assumption, thus the uncertainty of the velocity anisotropy obtained by this spherical Jeans method should be large. As described in Section 5.4, we suggest that the uncertainty in velocity anisotropy obtained from the spherical Jeans modelling could be heavily under-estimated.

The velocity anisotropy profiles we obtained here for M87 by discrete chemo-dynamical JAM models are similar to those of NGC 5846 obtained by \citet{Zhu2016a}, where blue GCs are near isotropic and red GCs are tangentially anisotropic. They are consistent with the scenario that blue GCs are mostly accreted, and that red GCs, even at very large radii, could either have been born in-situ, or have been heavily disrupted on radial orbits. Red GCs have qualitatively similar rotation and velocity dispersion profiles to PNe out to $\sim 2000$ arcsec \citep{Longobardi2018}, which supports the in-situ formation history. The near isotropic blue GCs and tangential red GCs suggest that there could be significant GC disruption even at very large radii. GC disruption \citep{Vesperini2003b, Vesperini2013, Brockamp2014} is an interesting open question requiring further investigation. It may be better understood by considering the formation and disruption of GCs within cosmological simulations. \citet{Ramos2019} using the Illustris simulation \citep{Vogelsberger2014} provides a good reference on how to tag intra-cluster GCs in simulations. They found that intra-cluster GCs are expected to be highly radially biased ($\beta \geq 0.3$) at all radii within the cluster. The slightly tangential anisotropy for blue GCs could be partially caused by localized substructures, because there is a known substructure at $\sim$ 50-100 kpc \citep{Romanowsky2012} which we did not remove.

\section{SUMMARY}

In Section 1, the Introduction, we set out our objectives for our investigation. We have been successful in that we have met all of the objectives. We now have 

(1) a new mass estimate for M87 extending to $\sim 400$ kpc,

(2) estimates of the velocity dispersion anisotropy of M87's two populations of GCs, and

(3) an assessment of the formation and capture of the GC populations which is consistent with established formation/capture scenarios.

We find that the mass profiles are well-constrained by the central galaxy stellar kinematics and the GC kinematics from the innermost to the outermost region. Both cusped and cored DM halos are found to be acceptable. The total mass within $5\,R_e^{\rm maj}$ (71.2 kpc) is $(5.8 \pm 0.5) \times 10^{12} M_{\odot}$. The DM halo contributes $(73 \pm 8)$ percent of the total mass within $1\,R_e^{\rm maj}$ (14.2 kpc), and $(94 \pm 6)$ percent within $5\,R_e^{\rm maj}$ (71.2 kpc). Our estimates benefit from both the large number of GCs we 
have used and the large spatial distribution of the GCs around the central galaxy. Estimates for the central region benefit from the use of a radially varying $M/L$ ratio taken from \citet{Sarzi2018}.

With data extending to $\sim 400$ kpc, we obtain total mass values of $M_{\rm tot} = (1.95 \pm 0.17) \times 10^{13} M_{\odot}$ for the cored model and $M_{\rm tot} = (2.32 \pm 0.22) \times 10^{13} M_{\odot}$ for the cusped model. Incorporating the uncertainties in the cored and cusped models, we conclude $M_{\rm tot} = (2.16 \pm 0.38) \times 10^{13} M_{\odot}$ within $\sim 400$ kpc for M87.

The red GCs present no anisotropic dispersion bias in the very inner region, and become tangentially anisotropic as the radius increases. The blue GC system varies from slightly radially to slightly tangentially anisotropic from the inner to outer region. Considering GC formation, these anisotropy values are consistent with the scenario that blue GCs are mostly accreted. The red GCs, even at very large radii, could either have been born in-situ, or have been heavily tidally disrupted by being on radial orbits \citep{Wu2014, Brockamp2014}.

For the future, the method we have used from \citet{Zhu2016a} has performed well and we intend to use it on other galaxies with extensive GC systems.

\section*{Acknowledgements}
We thank the anonymous referee for helping to improve the paper. The models were accomplished on the "Zen" cluster at National Astronomical Observatories, Chinese Academy of Sciences (NAOC) and the "Venus" cluster at Tsinghua University. We acknowledge useful discussions with Eric Emsellem and Hongyu Li. This work is partly supported by the National Key Basic Research and Development Program of China (No. 2018YFA0404501 to SM), by the National Science Foundation of China (Grant No. 11821303, 11333003, 11390372 and 11761131004 to SM). LZ acknowledges support from Shanghai Astronomical Observatory, Chinese Academy of Sciences under grant NO. Y895201009. GvdV acknowledges funding from the European Research Council (ERC) under the European Union's Horizon 2020 research and innovation programme under grant agreement No. 724857 (Consolidator Grant ArcheoDyn). AL acknowledges the support by the Centre national d'etudes spatiales, CNES.

%%%%%%%%%%%%%%%%%%%%%%%%%%%%%%%%%%%%%%%%%%%%%%%%%%



%\bibliographystyle{mnras}

\bibliographystyle{mnras}
\bibliography{m87} 

\begin{thebibliography}{}
\makeatletter
\relax
\def\mn@urlcharsother{\let\do\@makeother \do\$\do\&\do\#\do\^\do\_\do\%\do\~}
\def\mn@doi{\begingroup\mn@urlcharsother \@ifnextchar [ {\mn@doi@}
  {\mn@doi@[]}}
\def\mn@doi@[#1]#2{\def\@tempa{#1}\ifx\@tempa\@empty \href
  {http://dx.doi.org/#2} {doi:#2}\else \href {http://dx.doi.org/#2} {#1}\fi
  \endgroup}
\def\mn@eprint#1#2{\mn@eprint@#1:#2::\@nil}
\def\mn@eprint@arXiv#1{\href {http://arxiv.org/abs/#1} {{\tt arXiv:#1}}}
\def\mn@eprint@dblp#1{\href {http://dblp.uni-trier.de/rec/bibtex/#1.xml}
  {dblp:#1}}
\def\mn@eprint@#1:#2:#3:#4\@nil{\def\@tempa {#1}\def\@tempb {#2}\def\@tempc
  {#3}\ifx \@tempc \@empty \let \@tempc \@tempb \let \@tempb \@tempa \fi \ifx
  \@tempb \@empty \def\@tempb {arXiv}\fi \@ifundefined
  {mn@eprint@\@tempb}{\@tempb:\@tempc}{\expandafter \expandafter \csname
  mn@eprint@\@tempb\endcsname \expandafter{\@tempc}}}

\bibitem[\protect\citeauthoryear{{Agnello}, {Evans}, {Romanowsky}  \&
  {Brodie}}{{Agnello} et~al.}{2014}]{Agnello2014}
{Agnello} A.,  {Evans} N.~W.,  {Romanowsky} A.~J.,   {Brodie} J.~P.,  2014,
  \mn@doi [\mnras] {10.1093/mnras/stu960}, \href
  {http://adsabs.harvard.edu/abs/2014MNRAS.442.3299A} {442, 3299}

\bibitem[\protect\citeauthoryear{{Alabi} et~al.,}{{Alabi}
  et~al.}{2017}]{Alabi2017}
{Alabi} A.~B.,  et~al., 2017, \mn@doi [\mnras] {10.1093/mnras/stx678}, \href
  {http://adsabs.harvard.edu/abs/2017MNRAS.468.3949A} {468, 3949}

\bibitem[\protect\citeauthoryear{{Amorisco}}{{Amorisco}}{2018}]{Amorisco2018}
{Amorisco} N.~C.,  2018, preprint, \href
  {http://adsabs.harvard.edu/abs/2018arXiv180200812A} {} (\mn@eprint {arXiv}
  {1802.00812})

\bibitem[\protect\citeauthoryear{{Behroozi}, {Conroy}  \&
  {Wechsler}}{{Behroozi} et~al.}{2010}]{Behroozi2010}
{Behroozi} P.~S.,  {Conroy} C.,   {Wechsler} R.~H.,  2010, \mn@doi [\apj]
  {10.1088/0004-637X/717/1/379}, \href
  {https://ui.adsabs.harvard.edu/abs/2010ApJ...717..379B} {717, 379}

\bibitem[\protect\citeauthoryear{{Binggeli}, {Tammann}  \&
  {Sandage}}{{Binggeli} et~al.}{1987}]{Binggeli1987}
{Binggeli} B.,  {Tammann} G.~A.,   {Sandage} A.,  1987, \mn@doi [\aj]
  {10.1086/114467}, \href {http://adsabs.harvard.edu/abs/1987AJ.....94..251B}
  {94, 251}

\bibitem[\protect\citeauthoryear{{Binney} \& {Tremaine}}{{Binney} \&
  {Tremaine}}{1987}]{Binney1987}
{Binney} J.,  {Tremaine} S.,  1987, {Galactic dynamics}

\bibitem[\protect\citeauthoryear{{Binney} \& {Tremaine}}{{Binney} \&
  {Tremaine}}{2008}]{Binney2008}
{Binney} J.,  {Tremaine} S.,  2008, {Galactic Dynamics: Second Edition}.
Princeton University Press

\bibitem[\protect\citeauthoryear{{Blakeslee} et~al.,}{{Blakeslee}
  et~al.}{2009}]{Blakeslee2009}
{Blakeslee} J.~P.,  et~al., 2009, \mn@doi [\apj] {10.1088/0004-637X/694/1/556},
  \href {http://adsabs.harvard.edu/abs/2009ApJ...694..556B} {694, 556}

\bibitem[\protect\citeauthoryear{{Brockamp}, {K{\"u}pper}, {Thies}, {Baumgardt}
   \& {Kroupa}}{{Brockamp} et~al.}{2014}]{Brockamp2014}
{Brockamp} M.,  {K{\"u}pper} A.~H.~W.,  {Thies} I.,  {Baumgardt} H.,   {Kroupa}
  P.,  2014, \mn@doi [\mnras] {10.1093/mnras/stu562}, \href
  {http://adsabs.harvard.edu/abs/2014MNRAS.441..150B} {441, 150}

\bibitem[\protect\citeauthoryear{{Brodie} \& {Strader}}{{Brodie} \&
  {Strader}}{2006}]{Brodie2006}
{Brodie} J.~P.,  {Strader} J.,  2006, \mn@doi [\araa]
  {10.1146/annurev.astro.44.051905.092441}, \href
  {http://adsabs.harvard.edu/abs/2006ARA%26A..44..193B} {44, 193}

\bibitem[\protect\citeauthoryear{{Brodie} et~al.,}{{Brodie}
  et~al.}{2014}]{Brodie2014}
{Brodie} J.~P.,  et~al., 2014, \mn@doi [\apj] {10.1088/0004-637X/796/1/52},
  \href {https://ui.adsabs.harvard.edu/abs/2014ApJ...796...52B} {796, 52}

\bibitem[\protect\citeauthoryear{{Cappellari}}{{Cappellari}}{2002}]{Cappellari2002}
{Cappellari} M.,  2002, \mn@doi [\mnras] {10.1046/j.1365-8711.2002.05412.x},
  \href {http://adsabs.harvard.edu/abs/2002MNRAS.333..400C} {333, 400}

\bibitem[\protect\citeauthoryear{{Cappellari}}{{Cappellari}}{2008}]{Cappellari2008}
{Cappellari} M.,  2008, \mn@doi [\mnras] {10.1111/j.1365-2966.2008.13754.x},
  \href {http://adsabs.harvard.edu/abs/2008MNRAS.390...71C} {390, 71}

\bibitem[\protect\citeauthoryear{{Cappellari} et~al.,}{{Cappellari}
  et~al.}{2013}]{Cappellari2013}
{Cappellari} M.,  et~al., 2013, \mn@doi [\mnras] {10.1093/mnras/stt562}, \href
  {http://adsabs.harvard.edu/abs/2013MNRAS.432.1709C} {432, 1709}

\bibitem[\protect\citeauthoryear{{Chen}, {C{\^o}t{\'e}}, {West}, {Peng}  \&
  {Ferrarese}}{{Chen} et~al.}{2010}]{Chen2010}
{Chen} C.-W.,  {C{\^o}t{\'e}} P.,  {West} A.~A.,  {Peng} E.~W.,   {Ferrarese}
  L.,  2010, \mn@doi [\apjs] {10.1088/0067-0049/191/1/1}, \href
  {https://ui.adsabs.harvard.edu/abs/2010ApJS..191....1C} {191, 1}

\bibitem[\protect\citeauthoryear{{C{\^o}t{\'e}}, {Marzke}  \&
  {West}}{{C{\^o}t{\'e}} et~al.}{1998}]{Cote1998}
{C{\^o}t{\'e}} P.,  {Marzke} R.~O.,   {West} M.~J.,  1998, \mn@doi [\apj]
  {10.1086/305838}, \href {http://adsabs.harvard.edu/abs/1998ApJ...501..554C}
  {501, 554}

\bibitem[\protect\citeauthoryear{{Das}, {Gerhard}, {Churazov}  \&
  {Zhuravleva}}{{Das} et~al.}{2010}]{Das2010}
{Das} P.,  {Gerhard} O.,  {Churazov} E.,   {Zhuravleva} I.,  2010, \mn@doi
  [\mnras] {10.1111/j.1365-2966.2010.17417.x}, \href
  {http://adsabs.harvard.edu/abs/2010MNRAS.409.1362D} {409, 1362}

\bibitem[\protect\citeauthoryear{{Durrell} et~al.,}{{Durrell}
  et~al.}{2014}]{Durrell2014}
{Durrell} P.~R.,  et~al., 2014, \mn@doi [\apj] {10.1088/0004-637X/794/2/103},
  \href {http://adsabs.harvard.edu/abs/2014ApJ...794..103D} {794, 103}

\bibitem[\protect\citeauthoryear{{Dutton} \& {Macci{\`o}}}{{Dutton} \&
  {Macci{\`o}}}{2014}]{Dutton2014}
{Dutton} A.~A.,  {Macci{\`o}} A.~V.,  2014, \mn@doi [\mnras]
  {10.1093/mnras/stu742}, \href
  {http://adsabs.harvard.edu/abs/2014MNRAS.441.3359D} {441, 3359}

\bibitem[\protect\citeauthoryear{{Eke} et~al.,}{{Eke} et~al.}{2004}]{Eke2004}
{Eke} V.~R.,  et~al., 2004, \mn@doi [\mnras]
  {10.1111/j.1365-2966.2004.07408.x}, \href
  {http://adsabs.harvard.edu/abs/2004MNRAS.348..866E} {348, 866}

\bibitem[\protect\citeauthoryear{{Emsellem}, {Monnet}  \& {Bacon}}{{Emsellem}
  et~al.}{1994}]{Emsellem1994}
{Emsellem} E.,  {Monnet} G.,   {Bacon} R.,  1994, \aap, \href
  {http://adsabs.harvard.edu/abs/1994A%26A...285..723E} {285, 723}

\bibitem[\protect\citeauthoryear{{Emsellem}, {Krajnovi{\'c}}  \&
  {Sarzi}}{{Emsellem} et~al.}{2014}]{Emsellem2014}
{Emsellem} E.,  {Krajnovi{\'c}} D.,   {Sarzi} M.,  2014, \mn@doi [\mnras]
  {10.1093/mnrasl/slu140}, \href
  {http://adsabs.harvard.edu/abs/2014MNRAS.445L..79E} {445, L79}

\bibitem[\protect\citeauthoryear{{Ferrarese} et~al.,}{{Ferrarese}
  et~al.}{2016}]{Ferrarese2016}
{Ferrarese} L.,  et~al., 2016, \mn@doi [\apj] {10.3847/0004-637X/824/1/10},
  \href {https://ui.adsabs.harvard.edu/abs/2016ApJ...824...10F} {824, 10}

\bibitem[\protect\citeauthoryear{{Foreman-Mackey}, {Hogg}, {Lang}  \&
  {Goodman}}{{Foreman-Mackey} et~al.}{2013}]{Foreman-Mackey2013}
{Foreman-Mackey} D.,  {Hogg} D.~W.,  {Lang} D.,   {Goodman} J.,  2013, \mn@doi
  [\pasp] {10.1086/670067}, \href
  {http://adsabs.harvard.edu/abs/2013PASP..125..306F} {125, 306}

\bibitem[\protect\citeauthoryear{{Gebhardt} \& {Thomas}}{{Gebhardt} \&
  {Thomas}}{2009}]{Gebhardt2009}
{Gebhardt} K.,  {Thomas} J.,  2009, \mn@doi [\apj]
  {10.1088/0004-637X/700/2/1690}, \href
  {http://adsabs.harvard.edu/abs/2009ApJ...700.1690G} {700, 1690}

\bibitem[\protect\citeauthoryear{{Gebhardt}, {Adams}, {Richstone}, {Lauer},
  {Faber}, {G{\"u}ltekin}, {Murphy}  \& {Tremaine}}{{Gebhardt}
  et~al.}{2011}]{Gebhardt2011}
{Gebhardt} K.,  {Adams} J.,  {Richstone} D.,  {Lauer} T.~R.,  {Faber} S.~M.,
  {G{\"u}ltekin} K.,  {Murphy} J.,   {Tremaine} S.,  2011, \mn@doi [\apj]
  {10.1088/0004-637X/729/2/119}, \href
  {http://adsabs.harvard.edu/abs/2011ApJ...729..119G} {729, 119}

\bibitem[\protect\citeauthoryear{{Geller} \& {Huchra}}{{Geller} \&
  {Huchra}}{1983}]{Geller1983}
{Geller} M.~J.,  {Huchra} J.~P.,  1983, \mn@doi [\apjs] {10.1086/190859}, \href
  {http://adsabs.harvard.edu/abs/1983ApJS...52...61G} {52, 61}

\bibitem[\protect\citeauthoryear{{Huchra} \& {Geller}}{{Huchra} \&
  {Geller}}{1982}]{Huchra1982}
{Huchra} J.~P.,  {Geller} M.~J.,  1982, \mn@doi [\apj] {10.1086/160000}, \href
  {http://adsabs.harvard.edu/abs/1982ApJ...257..423H} {257, 423}

\bibitem[\protect\citeauthoryear{{Ko} et~al.,}{{Ko} et~al.}{2017}]{Ko2017}
{Ko} Y.,  et~al., 2017, \mn@doi [\apj] {10.3847/1538-4357/835/2/212}, \href
  {https://ui.adsabs.harvard.edu/abs/2017ApJ...835..212K} {835, 212}

\bibitem[\protect\citeauthoryear{{Li}, {Li}, {Mao}, {Xu}, {Long}  \&
  {Emsellem}}{{Li} et~al.}{2016}]{Li2016}
{Li} H.,  {Li} R.,  {Mao} S.,  {Xu} D.,  {Long} R.~J.,   {Emsellem} E.,  2016,
  \mn@doi [\mnras] {10.1093/mnras/stv2565}, \href
  {https://ui.adsabs.harvard.edu/abs/2016MNRAS.455.3680L} {455, 3680}

\bibitem[\protect\citeauthoryear{{Li} et~al.,}{{Li} et~al.}{2017}]{Li2017}
{Li} H.,  et~al., 2017, \mn@doi [\apj] {10.3847/1538-4357/aa662a}, \href
  {https://ui.adsabs.harvard.edu/abs/2017ApJ...838...77L} {838, 77}

\bibitem[\protect\citeauthoryear{{Long} \& {Mao}}{{Long} \&
  {Mao}}{2010}]{Long2010}
{Long} R.~J.,  {Mao} S.,  2010, \mn@doi [\mnras]
  {10.1111/j.1365-2966.2010.16438.x}, \href
  {http://adsabs.harvard.edu/abs/2010MNRAS.405..301L} {405, 301}

\bibitem[\protect\citeauthoryear{{Longobardi}, {Arnaboldi}, {Gerhard},
  {Pulsoni}  \& {S{\"o}ldner-Rembold}}{{Longobardi}
  et~al.}{2018a}]{Longobardi2018}
{Longobardi} A.,  {Arnaboldi} M.,  {Gerhard} O.,  {Pulsoni} C.,
  {S{\"o}ldner-Rembold} I.,  2018a, \mn@doi [\aap]
  {10.1051/0004-6361/201832729}, \href
  {http://adsabs.harvard.edu/abs/2018A%26A...620A.111L} {620, A111}

\bibitem[\protect\citeauthoryear{{Longobardi} et~al.,}{{Longobardi}
  et~al.}{2018b}]{Longobardi2018GC}
{Longobardi} A.,  et~al., 2018b, \mn@doi [\apj] {10.3847/1538-4357/aad3d2},
  \href {http://adsabs.harvard.edu/abs/2018ApJ...864...36L} {864, 36}

\bibitem[\protect\citeauthoryear{{Lyubenova} et~al.,}{{Lyubenova}
  et~al.}{2016}]{Lyubenova2016}
{Lyubenova} M.,  et~al., 2016, \mn@doi [\mnras] {10.1093/mnras/stw2434}, \href
  {http://adsabs.harvard.edu/abs/2016MNRAS.463.3220L} {463, 3220}

\bibitem[\protect\citeauthoryear{{McLaughlin}}{{McLaughlin}}{1999}]{McLaughlin1999}
{McLaughlin} D.~E.,  1999, \mn@doi [\apjl] {10.1086/311860}, \href
  {http://adsabs.harvard.edu/abs/1999ApJ...512L...9M} {512, L9}

\bibitem[\protect\citeauthoryear{{Mei} et~al.,}{{Mei} et~al.}{2007}]{Mei2007}
{Mei} S.,  et~al., 2007, \mn@doi [\apj] {10.1086/509598}, \href
  {http://adsabs.harvard.edu/abs/2007ApJ...655..144M} {655, 144}

\bibitem[\protect\citeauthoryear{{Montes}, {Trujillo}, {Prieto}  \&
  {Acosta-Pulido}}{{Montes} et~al.}{2014}]{Montes2014}
{Montes} M.,  {Trujillo} I.,  {Prieto} M.~A.,   {Acosta-Pulido} J.~A.,  2014,
  \mn@doi [\mnras] {10.1093/mnras/stu037}, \href
  {http://adsabs.harvard.edu/abs/2014MNRAS.439..990M} {439, 990}

\bibitem[\protect\citeauthoryear{{Mu{\~n}oz} et~al.,}{{Mu{\~n}oz}
  et~al.}{2014}]{Munoz2014}
{Mu{\~n}oz} R.~P.,  et~al., 2014, \mn@doi [\apjs] {10.1088/0067-0049/210/1/4},
  \href {https://ui.adsabs.harvard.edu/abs/2014ApJS..210....4M} {210, 4}

\bibitem[\protect\citeauthoryear{{Murphy}, {Gebhardt}  \& {Adams}}{{Murphy}
  et~al.}{2011}]{Murphy2011}
{Murphy} J.~D.,  {Gebhardt} K.,   {Adams} J.~J.,  2011, \mn@doi [\apj]
  {10.1088/0004-637X/729/2/129}, \href
  {http://adsabs.harvard.edu/abs/2011ApJ...729..129M} {729, 129}

\bibitem[\protect\citeauthoryear{{Napolitano}, {Pota}, {Romanowsky}, {Forbes},
  {Brodie}  \& {Foster}}{{Napolitano} et~al.}{2014}]{Napolitano2014}
{Napolitano} N.~R.,  {Pota} V.,  {Romanowsky} A.~J.,  {Forbes} D.~A.,  {Brodie}
  J.~P.,   {Foster} C.,  2014, \mn@doi [\mnras] {10.1093/mnras/stt2484}, \href
  {http://adsabs.harvard.edu/abs/2014MNRAS.439..659N} {439, 659}

\bibitem[\protect\citeauthoryear{{Navarro}, {Frenk}  \& {White}}{{Navarro}
  et~al.}{1997}]{Navarro1997}
{Navarro} J.~F.,  {Frenk} C.~S.,   {White} S.~D.~M.,  1997, \mn@doi [\apj]
  {10.1086/304888}, \href {http://adsabs.harvard.edu/abs/1997ApJ...490..493N}
  {490, 493}

\bibitem[\protect\citeauthoryear{{Nulsen} \& {Bohringer}}{{Nulsen} \&
  {Bohringer}}{1995}]{Nulsen1995}
{Nulsen} P.~E.~J.,  {Bohringer} H.,  1995, \mn@doi [\mnras]
  {10.1093/mnras/274.4.1093}, \href
  {http://adsabs.harvard.edu/abs/1995MNRAS.274.1093N} {274, 1093}

\bibitem[\protect\citeauthoryear{{Oldham} \& {Auger}}{{Oldham} \&
  {Auger}}{2018}]{Oldham2018}
{Oldham} L.,  {Auger} M.,  2018, \mn@doi [\mnras] {10.1093/mnras/stx2969},
  \href {http://adsabs.harvard.edu/abs/2018MNRAS.474.4169O} {474, 4169}

\bibitem[\protect\citeauthoryear{{Oser}, {Ostriker}, {Naab}, {Johansson}  \&
  {Burkert}}{{Oser} et~al.}{2010}]{Oser2010}
{Oser} L.,  {Ostriker} J.~P.,  {Naab} T.,  {Johansson} P.~H.,   {Burkert} A.,
  2010, \mn@doi [\apj] {10.1088/0004-637X/725/2/2312}, \href
  {http://adsabs.harvard.edu/abs/2010ApJ...725.2312O} {725, 2312}

\bibitem[\protect\citeauthoryear{{Peng} et~al.,}{{Peng}
  et~al.}{2008}]{Peng2008}
{Peng} E.~W.,  et~al., 2008, \mn@doi [\apj] {10.1086/587951}, \href
  {http://adsabs.harvard.edu/abs/2008ApJ...681..197P} {681, 197}

\bibitem[\protect\citeauthoryear{{Pointecouteau}, {Arnaud}  \&
  {Pratt}}{{Pointecouteau} et~al.}{2005}]{Pointecouteau2005}
{Pointecouteau} E.,  {Arnaud} M.,   {Pratt} G.~W.,  2005, \mn@doi [\aap]
  {10.1051/0004-6361:20042569}, \href
  {http://adsabs.harvard.edu/abs/2005A%26A...435....1P} {435, 1}

\bibitem[\protect\citeauthoryear{{Pota} et~al.,}{{Pota}
  et~al.}{2015}]{Pota2015b}
{Pota} V.,  et~al., 2015, \mn@doi [\mnras] {10.1093/mnras/stv831}, \href
  {http://adsabs.harvard.edu/abs/2015MNRAS.450.3345P} {450, 3345}

\bibitem[\protect\citeauthoryear{{Prieto} \& {Gnedin}}{{Prieto} \&
  {Gnedin}}{2008}]{Prieto2008}
{Prieto} J.~L.,  {Gnedin} O.~Y.,  2008, \mn@doi [\apj] {10.1086/591777}, \href
  {http://adsabs.harvard.edu/abs/2008ApJ...689..919P} {689, 919}

\bibitem[\protect\citeauthoryear{{Ramos-Almendares}, {Sales}, {Abadi},
  {Doppel}, {Muriel}  \& {Peng}}{{Ramos-Almendares} et~al.}{2019}]{Ramos2019}
{Ramos-Almendares} F.,  {Sales} L.~V.,  {Abadi} M.~G.,  {Doppel} J.~E.,
  {Muriel} H.,   {Peng} E.~W.,  2019, arXiv e-prints, \href
  {https://ui.adsabs.harvard.edu/abs/2019arXiv190611921R} {p. arXiv:1906.11921}

\bibitem[\protect\citeauthoryear{{Romanowsky}, {Strader}, {Brodie}, {Mihos},
  {Spitler}, {Forbes}, {Foster}  \& {Arnold}}{{Romanowsky}
  et~al.}{2012}]{Romanowsky2012}
{Romanowsky} A.~J.,  {Strader} J.,  {Brodie} J.~P.,  {Mihos} J.~C.,  {Spitler}
  L.~R.,  {Forbes} D.~A.,  {Foster} C.,   {Arnold} J.~A.,  2012, \mn@doi [\apj]
  {10.1088/0004-637X/748/1/29}, \href
  {https://ui.adsabs.harvard.edu/abs/2012ApJ...748...29R} {748, 29}

\bibitem[\protect\citeauthoryear{{R{\"o}ttgers}, {Naab}  \&
  {Oser}}{{R{\"o}ttgers} et~al.}{2014}]{Rottgers2014}
{R{\"o}ttgers} B.,  {Naab} T.,   {Oser} L.,  2014, \mn@doi [\mnras]
  {10.1093/mnras/stu1762}, \href
  {http://adsabs.harvard.edu/abs/2014MNRAS.445.1065R} {445, 1065}

\bibitem[\protect\citeauthoryear{{Sarzi}, {Spiniello}, {La Barbera},
  {Krajnovi{\'c}}  \& {van den Bosch}}{{Sarzi} et~al.}{2018}]{Sarzi2018}
{Sarzi} M.,  {Spiniello} C.,  {La Barbera} F.,  {Krajnovi{\'c}} D.,   {van den
  Bosch} R.,  2018, \mn@doi [\mnras] {10.1093/mnras/sty1092}, \href
  {http://adsabs.harvard.edu/abs/2018MNRAS.478.4084S} {478, 4084}

\bibitem[\protect\citeauthoryear{{Starkenburg} et~al.,}{{Starkenburg}
  et~al.}{2009}]{Starkenburg2009}
{Starkenburg} E.,  et~al., 2009, \mn@doi [\apj] {10.1088/0004-637X/698/1/567},
  \href {http://adsabs.harvard.edu/abs/2009ApJ...698..567S} {698, 567}

\bibitem[\protect\citeauthoryear{{Strader} et~al.,}{{Strader}
  et~al.}{2011}]{Strader2011}
{Strader} J.,  et~al., 2011, \mn@doi [\apjs] {10.1088/0067-0049/197/2/33},
  \href {http://adsabs.harvard.edu/abs/2011ApJS..197...33S} {197, 33}

\bibitem[\protect\citeauthoryear{{Syer} \& {Tremaine}}{{Syer} \&
  {Tremaine}}{1996}]{Syer1996}
{Syer} D.,  {Tremaine} S.,  1996, \mn@doi [\mnras] {10.1093/mnras/282.1.223},
  \href {http://adsabs.harvard.edu/abs/1996MNRAS.282..223S} {282, 223}

\bibitem[\protect\citeauthoryear{{Urban}, {Werner}, {Simionescu}, {Allen}  \&
  {B{\"o}hringer}}{{Urban} et~al.}{2011}]{Urban2011}
{Urban} O.,  {Werner} N.,  {Simionescu} A.,  {Allen} S.~W.,   {B{\"o}hringer}
  H.,  2011, \mn@doi [\mnras] {10.1111/j.1365-2966.2011.18526.x}, \href
  {http://adsabs.harvard.edu/abs/2011MNRAS.414.2101U} {414, 2101}

\bibitem[\protect\citeauthoryear{{Vesperini} \& {Zepf}}{{Vesperini} \&
  {Zepf}}{2003}]{Vesperini2003b}
{Vesperini} E.,  {Zepf} S.~E.,  2003, \mn@doi [\apjl] {10.1086/375313}, \href
  {https://ui.adsabs.harvard.edu/abs/2003ApJ...587L..97V} {587, L97}

\bibitem[\protect\citeauthoryear{{Vesperini}, {Zepf}, {Kundu}  \&
  {Ashman}}{{Vesperini} et~al.}{2003}]{Vesperini2003}
{Vesperini} E.,  {Zepf} S.~E.,  {Kundu} A.,   {Ashman} K.~M.,  2003, \mn@doi
  [\apj] {10.1086/376688}, \href
  {http://adsabs.harvard.edu/abs/2003ApJ...593..760V} {593, 760}

\bibitem[\protect\citeauthoryear{{Vesperini}, {McMillan}, {D'Antona}  \&
  {D'Ercole}}{{Vesperini} et~al.}{2013}]{Vesperini2013}
{Vesperini} E.,  {McMillan} S. L.~W.,  {D'Antona} F.,   {D'Ercole} A.,  2013,
  \mn@doi [\mnras] {10.1093/mnras/sts434}, \href
  {https://ui.adsabs.harvard.edu/abs/2013MNRAS.429.1913V} {429, 1913}

\bibitem[\protect\citeauthoryear{{Vogelsberger} et~al.,}{{Vogelsberger}
  et~al.}{2014}]{Vogelsberger2014}
{Vogelsberger} M.,  et~al., 2014, \mn@doi [\mnras] {10.1093/mnras/stu1536},
  \href {https://ui.adsabs.harvard.edu/abs/2014MNRAS.444.1518V} {444, 1518}

\bibitem[\protect\citeauthoryear{{Wasserman} et~al.,}{{Wasserman}
  et~al.}{2018}]{Wasserman2018}
{Wasserman} A.,  et~al., 2018, \mn@doi [\apj] {10.3847/1538-4357/aad236}, \href
  {http://adsabs.harvard.edu/abs/2018ApJ...863..130W} {863, 130}

\bibitem[\protect\citeauthoryear{{Watkins}, {Evans}  \& {An}}{{Watkins}
  et~al.}{2010}]{Watkins2010}
{Watkins} L.~L.,  {Evans} N.~W.,   {An} J.~H.,  2010, \mn@doi [\mnras]
  {10.1111/j.1365-2966.2010.16708.x}, \href
  {https://ui.adsabs.harvard.edu/abs/2010MNRAS.406..264W} {406, 264}

\bibitem[\protect\citeauthoryear{{Watkins}, {van de Ven}, {den Brok}  \& {van
  den Bosch}}{{Watkins} et~al.}{2013}]{Watkins2013}
{Watkins} L.~L.,  {van de Ven} G.,  {den Brok} M.,   {van den Bosch} R.~C.~E.,
  2013, \mn@doi [\mnras] {10.1093/mnras/stt1756}, \href
  {http://adsabs.harvard.edu/abs/2013MNRAS.436.2598W} {436, 2598}

\bibitem[\protect\citeauthoryear{{Wood} et~al.,}{{Wood}
  et~al.}{2017}]{Wood2017}
{Wood} R.~A.,  et~al., 2017, \mn@doi [\apj] {10.3847/1538-4357/aa8723}, \href
  {https://ui.adsabs.harvard.edu/abs/2017ApJ...847...79W} {847, 79}

\bibitem[\protect\citeauthoryear{{Wu}, {Gerhard}, {Naab}, {Oser},
  {Martinez-Valpuesta}, {Hilz}, {Churazov}  \& {Lyskova}}{{Wu}
  et~al.}{2014}]{Wu2014}
{Wu} X.,  {Gerhard} O.,  {Naab} T.,  {Oser} L.,  {Martinez-Valpuesta} I.,
  {Hilz} M.,  {Churazov} E.,   {Lyskova} N.,  2014, \mn@doi [\mnras]
  {10.1093/mnras/stt2415}, \href
  {http://adsabs.harvard.edu/abs/2014MNRAS.438.2701W} {438, 2701}

\bibitem[\protect\citeauthoryear{{Zhang} et~al.,}{{Zhang}
  et~al.}{2015}]{Zhang2015}
{Zhang} H.-X.,  et~al., 2015, \mn@doi [\apj] {10.1088/0004-637X/802/1/30},
  \href {http://adsabs.harvard.edu/abs/2015ApJ...802...30Z} {802, 30}

\bibitem[\protect\citeauthoryear{{Zhu} et~al.,}{{Zhu} et~al.}{2014}]{Zhu2014}
{Zhu} L.,  et~al., 2014, \mn@doi [\apj] {10.1088/0004-637X/792/1/59}, \href
  {http://adsabs.harvard.edu/abs/2014ApJ...792...59Z} {792, 59}

\bibitem[\protect\citeauthoryear{{Zhu} et~al.,}{{Zhu} et~al.}{2016}]{Zhu2016a}
{Zhu} L.,  et~al., 2016, \mn@doi [\mnras] {10.1093/mnras/stw1931}, \href
  {http://adsabs.harvard.edu/abs/2016MNRAS.462.4001Z} {462, 4001}

\bibitem[\protect\citeauthoryear{{de Zeeuw} et~al.,}{{de Zeeuw}
  et~al.}{2002}]{deZeeuw2001}
{de Zeeuw} P.~T.,  et~al., 2002, \mn@doi [\mnras]
  {10.1046/j.1365-8711.2002.05059.x}, \href
  {https://ui.adsabs.harvard.edu/\#abs/2002MNRAS.329..513D} {329, 513}

\bibitem[\protect\citeauthoryear{{van den Bosch} \& {van de Ven}}{{van den
  Bosch} \& {van de Ven}}{2009}]{Remco2009}
{van den Bosch} R.~C.~E.,  {van de Ven} G.,  2009, \mn@doi [\mnras]
  {10.1111/j.1365-2966.2009.15177.x}, \href
  {http://adsabs.harvard.edu/abs/2009MNRAS.398.1117V} {398, 1117}

\makeatother
\end{thebibliography}

%\begin{thebibliography}{99}
%\end{thebibliography}

%%%%%%%%%%%%%%%%%%%%%%%%%%%%%%%%%%%%%%%%%%%%%%%%%%

%%%%%%%%%%%%%%%%% APPENDICES %%%%%%%%%%%%%%%%%%%%%

\appendix

\section{Cusped Dark Matter Models}

In the main text of the paper, we adopt a cored dark matter profile. In this appendix, we show the results from a cusped model. Most results are unchanged (for more discussion, see Section 5.1).

\begin{figure}
	\includegraphics[width=\columnwidth]{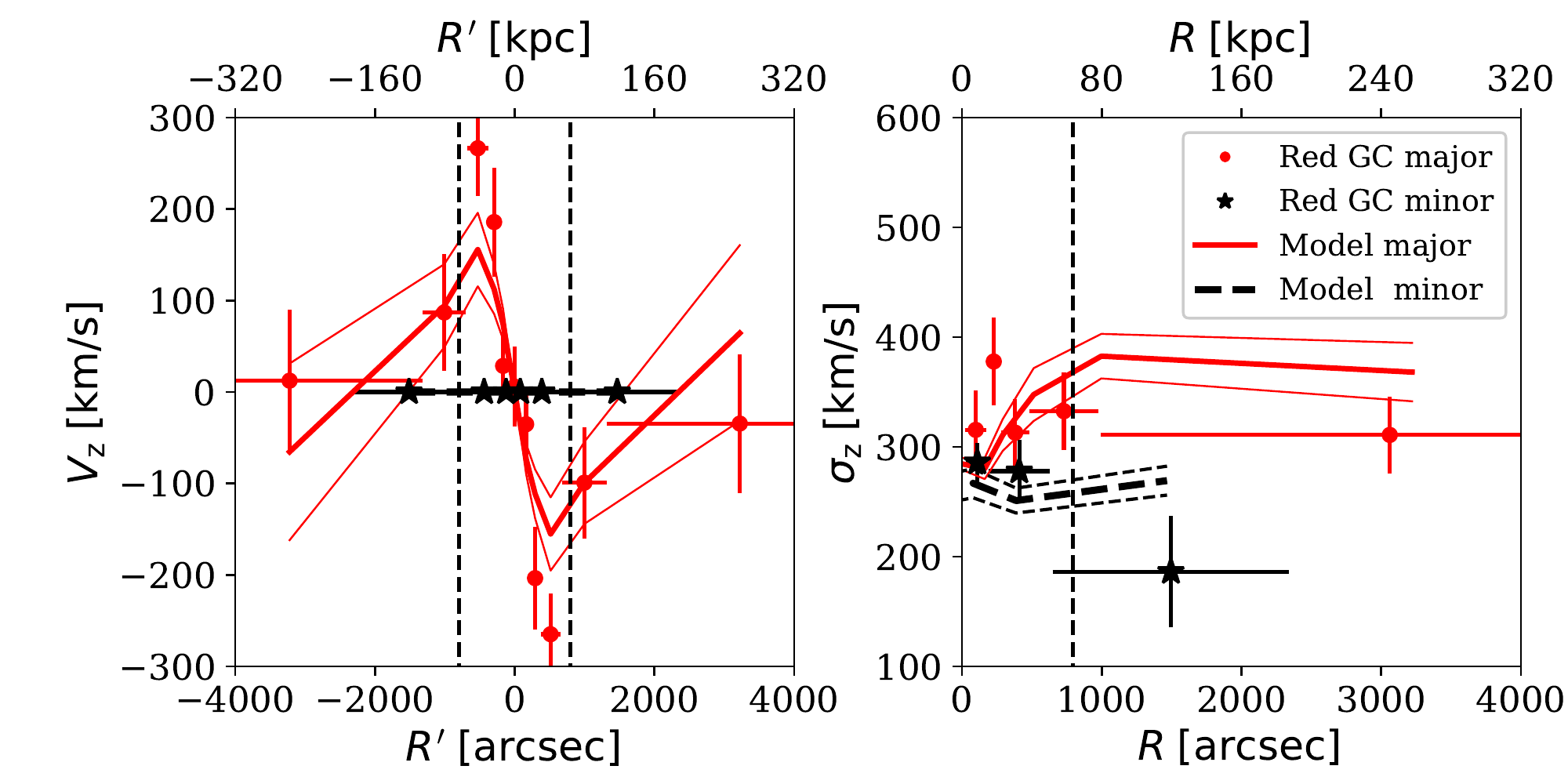}
	\includegraphics[width=\columnwidth]{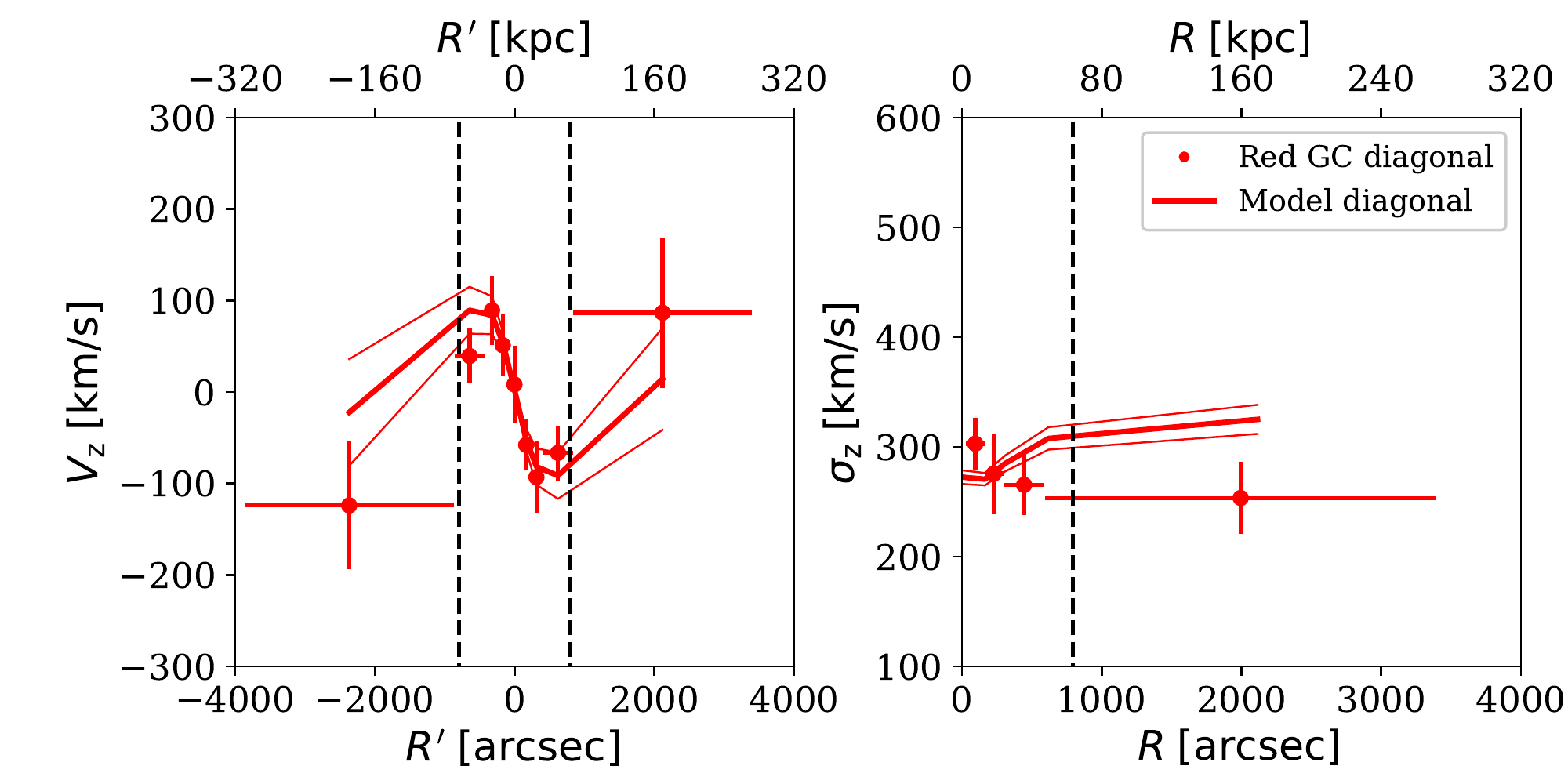}
	\includegraphics[width=\columnwidth]{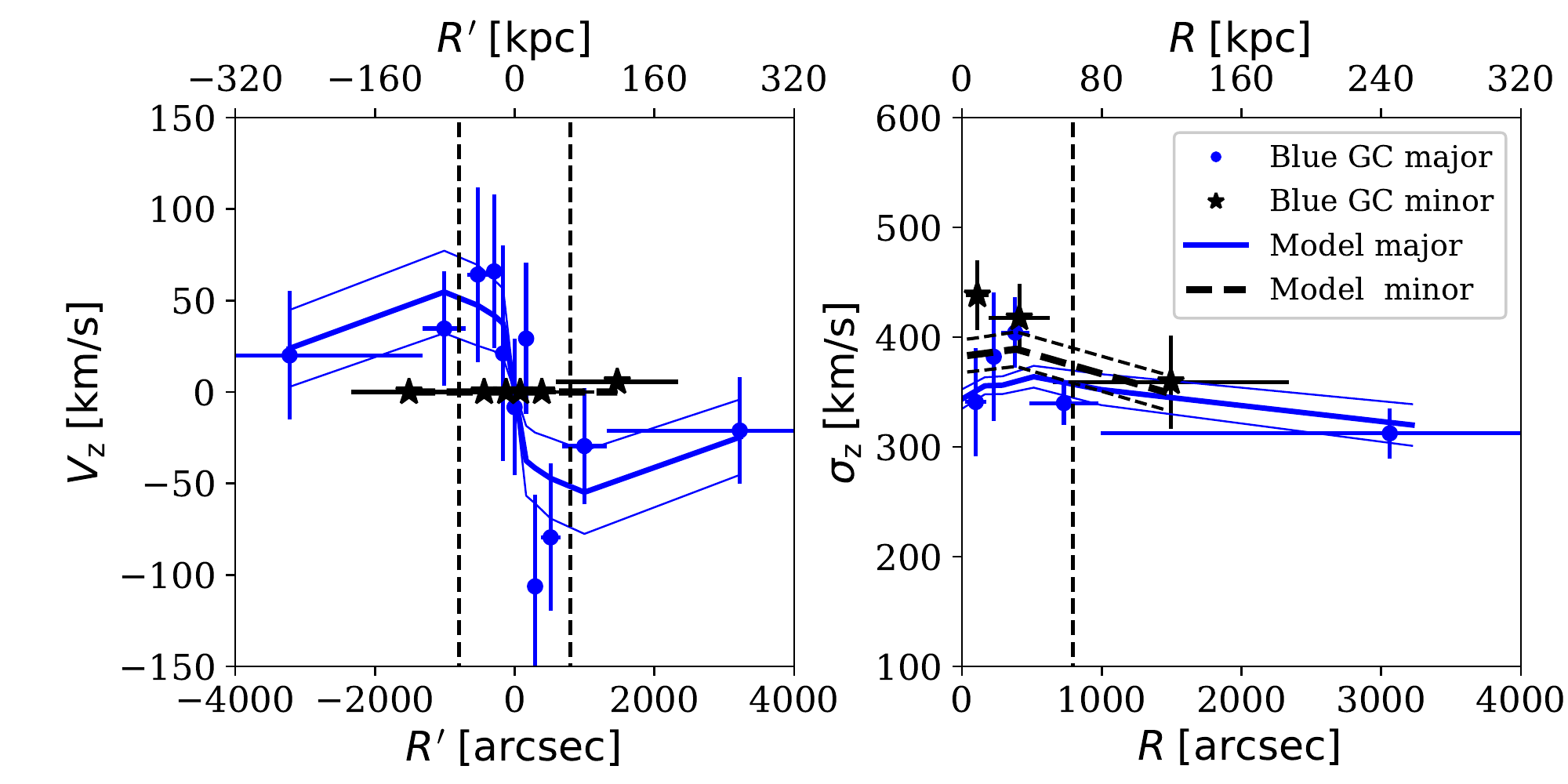}
	\includegraphics[width=\columnwidth]{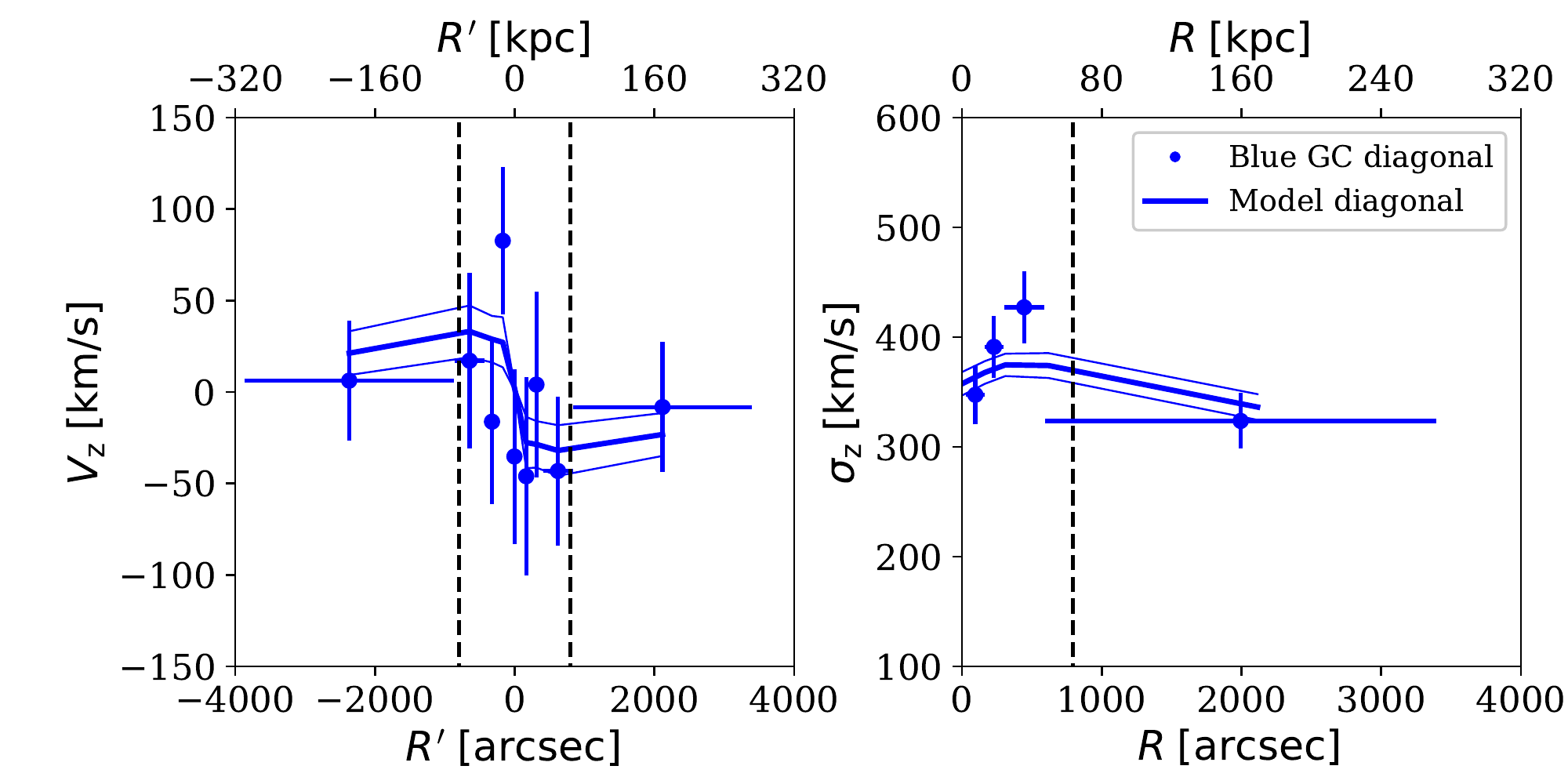}
    \caption{Comparison of the observed GC data with model predictions of the best-fitting cusped model. Fig.~\ref{fig:GCkin} shows the corresponding cored model.}
    \label{fig:GCkin_cusped}
\end{figure}

\begin{figure}
	\includegraphics[width=\columnwidth]{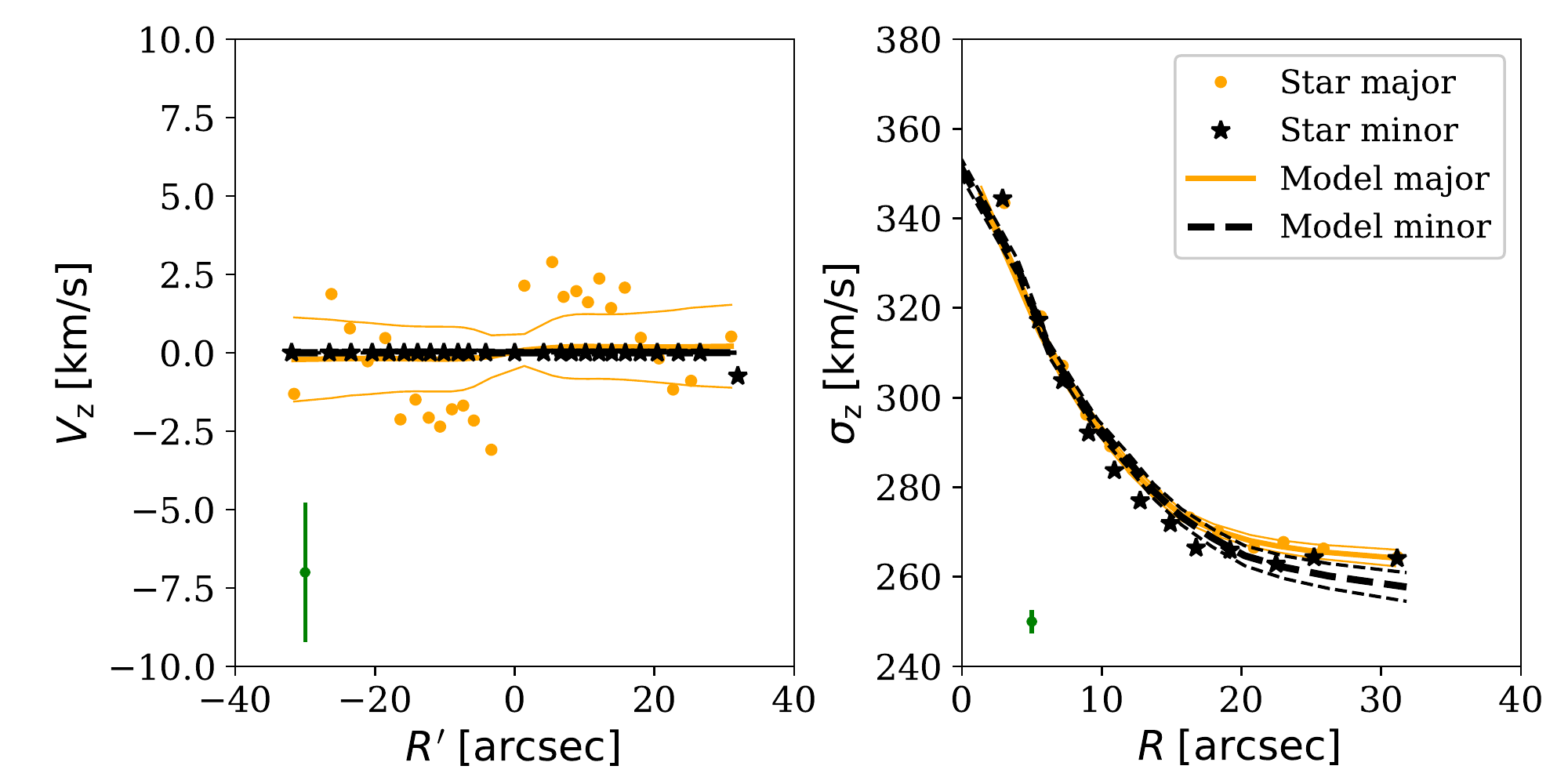}
	\includegraphics[width=\columnwidth]{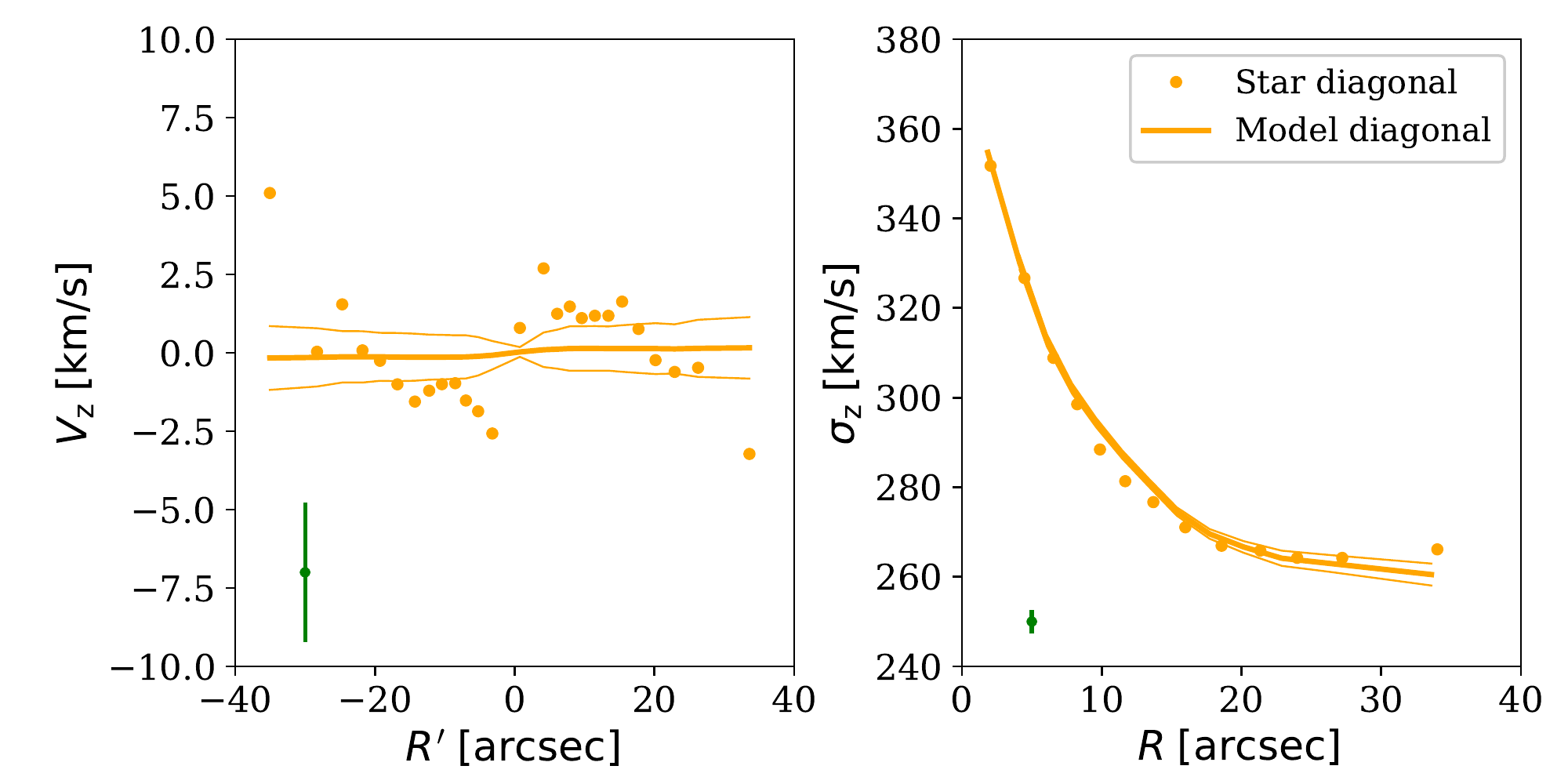}
    \caption{Comparison of the stellar data with model predictions of the best-fitting cusped model. Fig.~\ref{fig:starkin} shows the corresponding cored model.}
    \label{fig:starkin_cusped}
\end{figure}

\begin{figure}
	\includegraphics[width=\columnwidth]{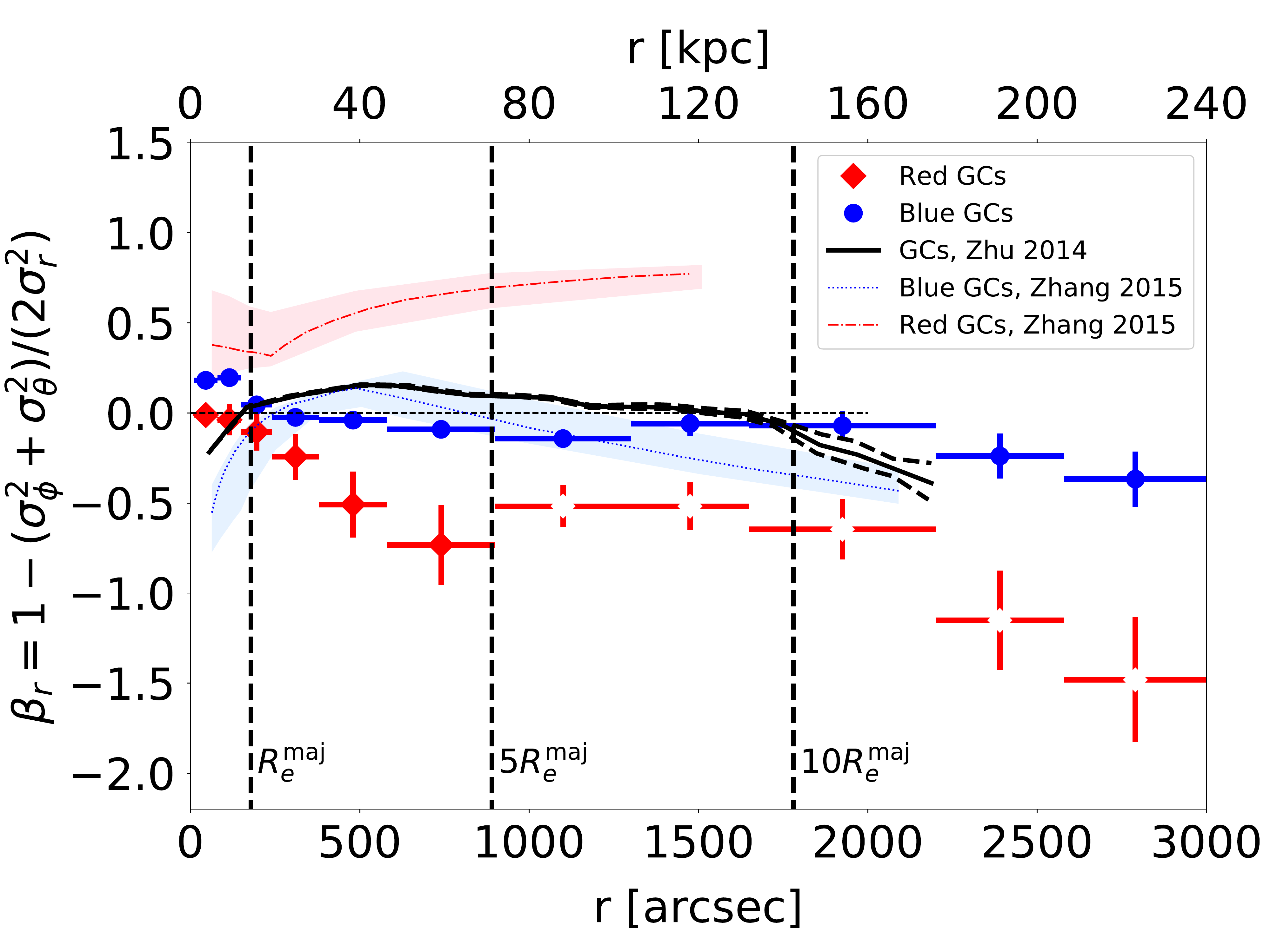}
    \caption{The velocity dispersion anisotropy $\beta_r$ profiles from the cusped model. $\beta_r$ for red GCs at $r>1000$ arcsec should have large uncertainties and we mark these points with open symbols. Fig.~\ref{fig:betaprofile} shows the corresponding cored model.}
    \label{fig:beta_cusped}
\end{figure}

Figs.~\ref{fig:GCkin_cusped} and \ref{fig:starkin_cusped} show the comparison of the observed data with the model predictions. The data and model predictions are binned in the same way as in Fig.~\ref{fig:binscheme}. The best-fitting cusped model matches the GC data with a mean $\chi^2$ of 1.15 for the red population and that of 0.51 for the blue population. The mean $\chi^2$ of the central galaxy directly from the model is 0.33. The velocity dispersion anisotropy profile obtained by the cusped model is shown in Fig.~\ref{fig:beta_cusped}.

\section{MGE fitting to the GC populations}
\begin{figure*}
	\includegraphics[width=2\columnwidth]{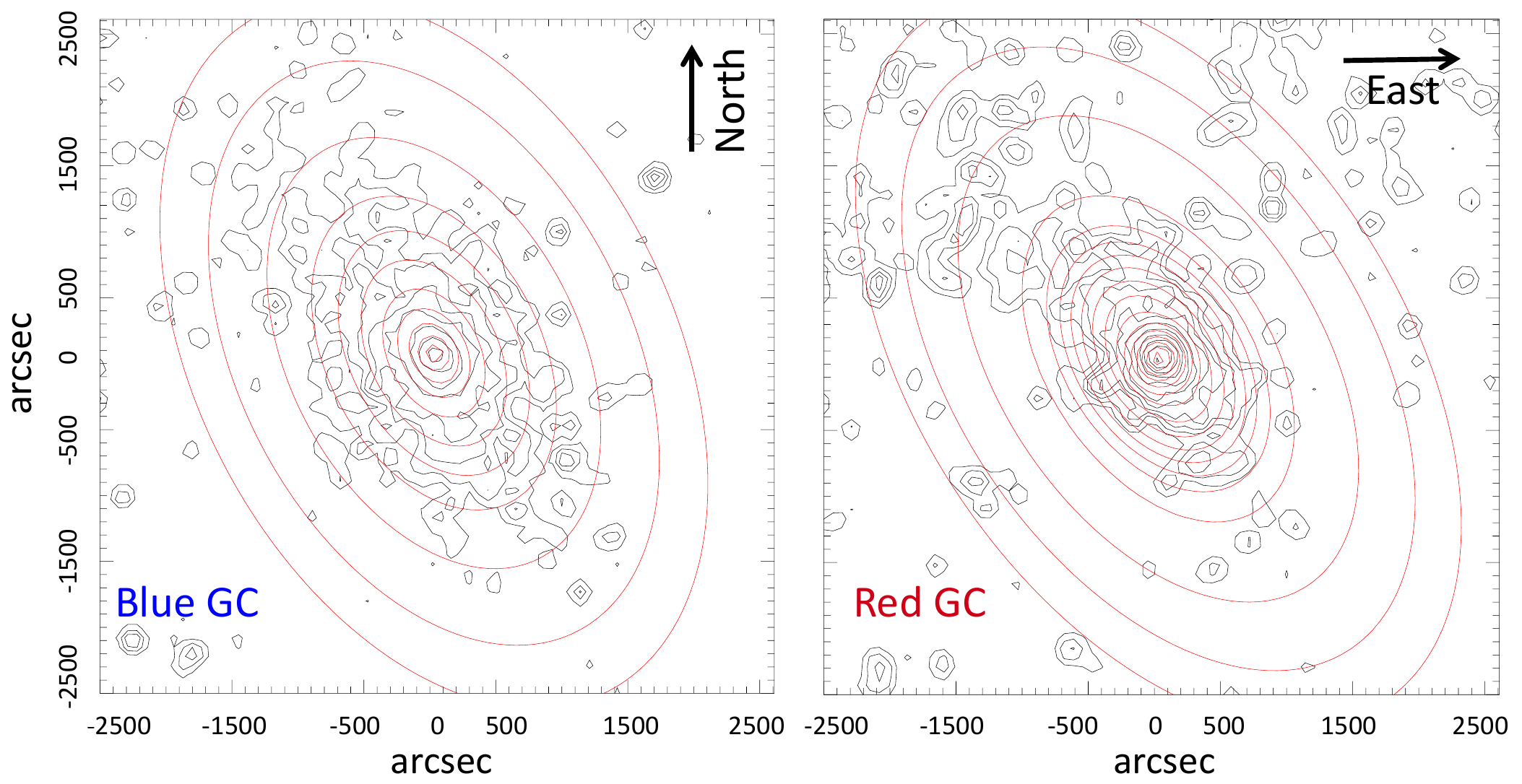}
    \caption{The 2D MGEs fitting to the surface number density maps of blue (left) and red (right) GCs. The black contours are the smoothed surface number density, the red ellipses are the best-fitting MGE models. Notice that for the red GCs, there is an asymmetric structure in the north at $R > 1000$ arcsec, which our model does not match. Instead, our model matched the south side of the data.}
    \label{fig:sd2D}
  \end{figure*}

\begin{figure*}
	\includegraphics[width=1\columnwidth]{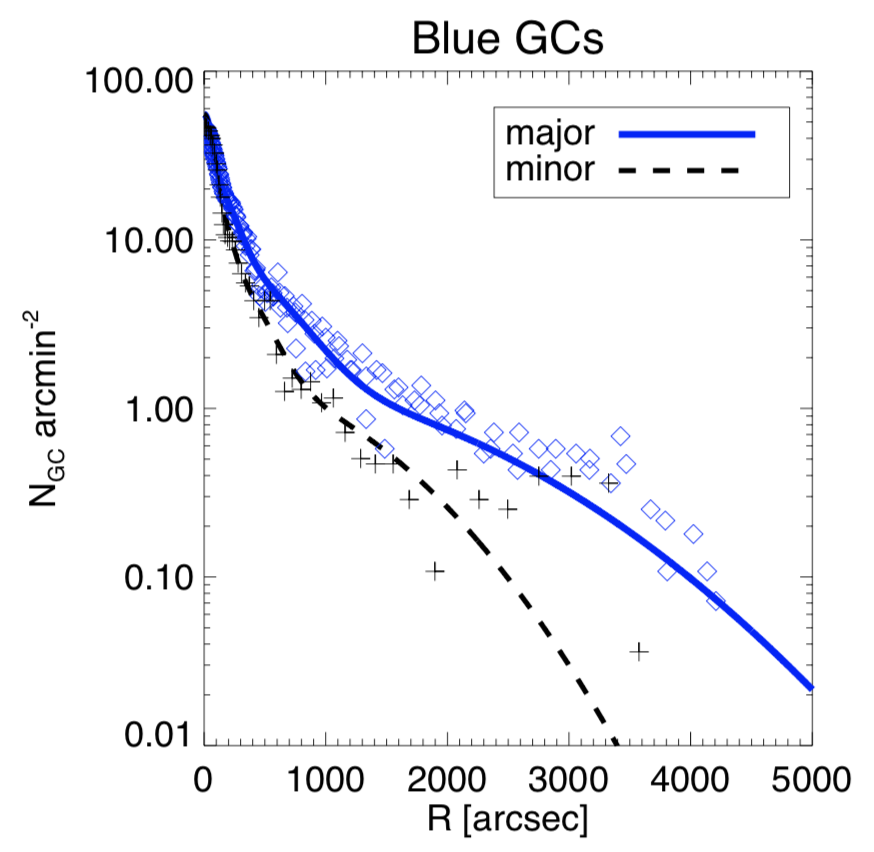}
	\includegraphics[width=1\columnwidth]{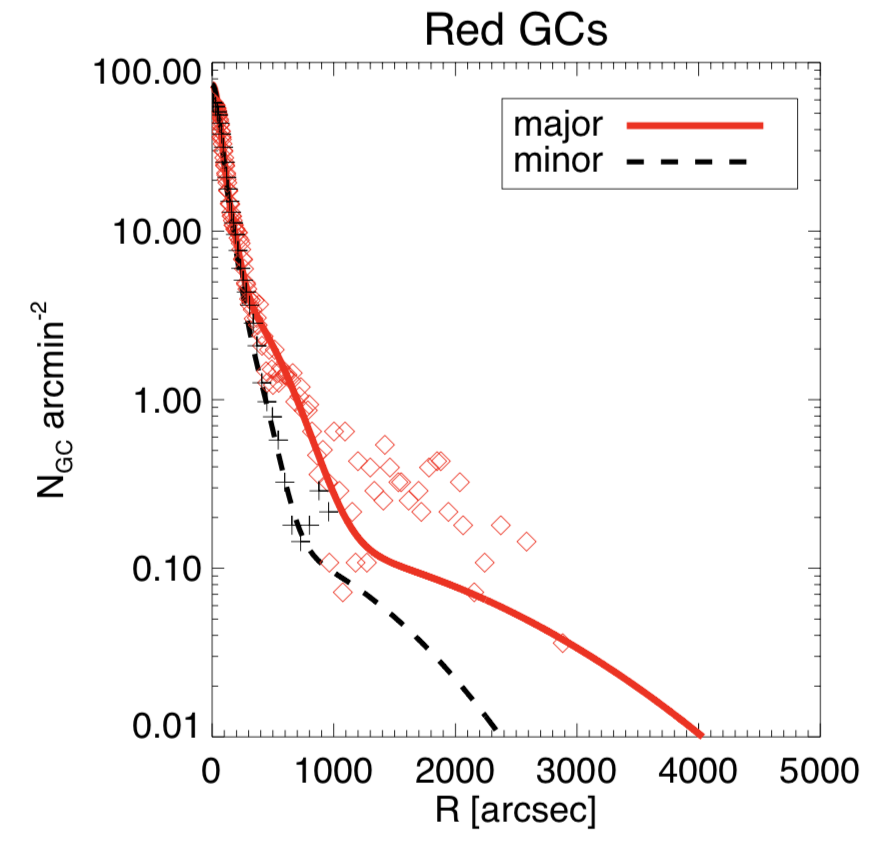}
    \caption{The diamonds and pluses are the surface brightness along the major and minor axes constructed from the photometric data. The solid and dashed curves are the MGE fits along the major and minor axes, respectively.}
    \label{fig:sd1D}
  \end{figure*}

In Fig.~\ref{fig:sd2D}, we show the 2D MGEs fitted to the surface number density maps of the blue (left) and red (right) GCs. The black contours are the smoothed surface number density, the red ellipses are the best-fitting MGE models.
In Fig.~\ref{fig:sd1D}, we show how well the MGE models match the surface density profiles along the major and minor axes.  

\section{Cored dark matter model with free inclination}
\begin{figure*}
	\includegraphics[width=2\columnwidth]{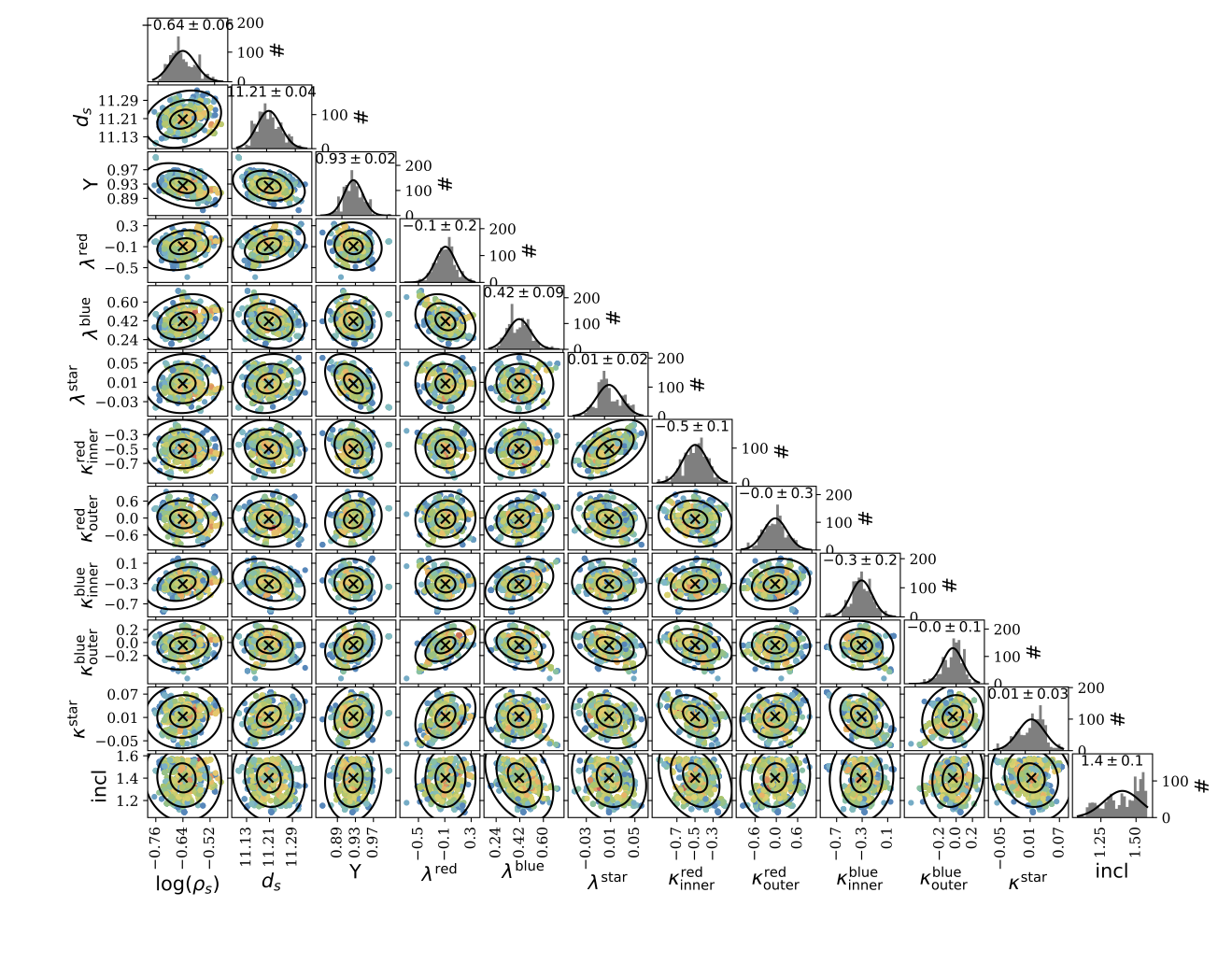}
    \caption{The posterior marginalized distributions for the cored model using a free inclination angle. The preferred inclination angle is $90^{\circ}$, but has large uncertainties. The rest of the parameters converge very similarly to the model with inclination angle fixed at $90^{\circ}$ (Fig.~\ref{fig:dbns}).}
    \label{fig:dbns_incl}
\end{figure*}

We created a model with a cored DM halo and leaving the inclination angle $\vartheta$ free. The best-fitting model parameters are shown in Fig.~\ref{fig:dbns_incl}. The model prefers $\vartheta = 90^{\circ}$, but has large uncertainties. There is a weak degeneracy between the inclination angle and the velocity anisotropy $\beta_z$ for the blue GCs. $\beta_z$ tends to be larger when the system is more face-on. This degeneracy is intrinsic in JAM models constrained by 2D kinematic maps, and the degeneracy is stronger for systems with positive $\beta_z$ \citep{Cappellari2008}. However, overall, the degeneracy is weak in our case, and we obtained similar results for the cored model with either the inclination angle free or fixed with $\vartheta= 90^{\circ}$.
%%%%%%%%%%%%%%%%%%%%%%%%%%%%%%%%%%%%%%%%%%%%%%%%%%

% Don't change these lines
\bsp	% typesetting comment
\label{lastpage}
\end{document}